\documentclass[fleqn,usenatbib]{mnras}

\usepackage{newtxtext,newtxmath}

\usepackage[T1]{fontenc}

\DeclareRobustCommand{\VAN}[3]{#2}
\let\VANthebibliography\thebibliography
\def\thebibliography{\DeclareRobustCommand{\VAN}[3]{##3}\VANthebibliography}

\usepackage{graphicx}	
\usepackage{amsmath}	
\usepackage{ulem}
\usepackage[dvipsnam]{xcolor}

\newcommand{\ergs}{{\rm erg\,s^{-1}}}
\newcommand{\kms}{{\rm km\,s^{-1}}}
\newcommand{\tff}{t_{\rm ff}}
\newcommand{\tcc}{t_{\rm cc}}
\newcommand{\tsp}{t_{\rm sp}}
\newcommand{\vwind}{v_{\rm w}}
\newcommand{\nc}{\Bar{n}_{\rm c}}
\newcommand{\MachWind}{\mathcal{M}_{\rm w}}
\newcommand{\kmin}{k_{\rm min}}
\newcommand{\pwind}{p_{\rm w}}
\newcommand{\cc}{{\rm cm^{-3}}}
\newcommand{\cswind}{c_{\rm s,w}}

\newcommand{\pow}[1]{10^{#1}}
\newcommand{\Pout}{P_{\rm w}}
\newcommand{\nwind}{n_{\rm w}}
\newcommand{\fst}{1^{\rm st}}
\newcommand{\snd}{2^{\rm nd}}
\newcommand{\trd}{3^{\rm rd}}
\newcommand{\fth}{4^{\rm th}}
\newcommand{\tkh}{t_{\rm KH}}
\newcommand{\tdrag}{t_{\rm drag}}
\newcommand{\pc}{{\rm pc}}
\newcommand{\kpc}{{\rm kpc}}
\newcommand{\sigmav}{\sigma_V}
\newcommand{\alphavir}{\alpha_{\rm vir}}
\newcommand{\SFRff}{{\rm SFR_{ff}}}
\newcommand{\Mach}{\mathcal{M}}
\newcommand{\Msun}{{\rm M_{\odot}}}
\newcommand{\Mdot}{\Dot{M}_{\rm OF}}
\newcommand{\ROF}{R_{\rm OF}}
\newcommand{\curl}[1]{\boldsymbol{\nabla}\times\boldsymbol{#1}}
\newcommand{\tcool}{t_{\rm cool}}
\newcommand{\Mdotavg}{\Dot{M}_{\rm OF, avg}}
\newcommand{\del}[2]{\frac{\partial #1}{\partial #2}}

\title[Cloud - AGN-wind interaction]{Probing the role of self-gravity in clouds impacted by AGN-driven winds}

\author[A Mandal et al.]{Ankush Mandal,$^{1}$\thanks{E-mail: \href{mailto:ankushm@iucaa.in}{ankushm@iucaa.in}} Dipanjan Mukherjee,$^{1}$\thanks{E-mail: \href{mailto:dipanjan@iucaa.in}{dipanjan@iucaa.in}} Christoph Federrath$^{2,3}$, Geoffrey V.~Bicknell$^{2}$, \newauthor Nicole P.~H.~Nesvadba$^{4}$ and Andrea Mignone$^{5}$
\\
$^{1}$Inter-University Centre for Astronomy and Astrophysics, Pune-411007, India\\
$^{2}$Research School of Astronomy and Astrophysics, Australian National University, Canberra, ACT 2611, Australia\\
$^{3}$ARC Centre of Excellence for Astronomy in Three Dimensions (ASTRO-3D), Canberra ACT 2601, Australia\\
$^{4}$Universit\'e de la C\^ote d’Azur, Observatoire de la C\^ote d’Azur, CNRS, Laboratoire Lagrange, Bd de l’Observatoire, CS 34229,06304 Nice cedex 4, France \\
$^{5}$Dipartimento di Fisica, Universit\'a di Torino via Pietro Giuria 1 (I-10125) Torino, Italy}

\date{}

\pubyear{2024}

\begin{document}
\label{firstpage}
\pagerange{\pageref{firstpage}--\pageref{lastpage}}
\maketitle

\begin{abstract}
The impact of winds and jet-inflated bubbles driven by active galactic nuclei (AGN) are believed to significantly affect the host galaxy's interstellar medium (ISM) and regulate star formation. To explore this scenario, we perform a suite of hydrodynamic simulations to model the interaction between turbulent star-forming clouds and highly pressurised AGN-driven outflows, focusing on the effects of self-gravity.
Our results demonstrate that the cloudlets fragmented by the wind can become gravitationally bound, significantly increasing their survival time. While external pressurisation leads to a global collapse of the clouds in cases of weaker winds ($\pow{42}-\pow{43}~\ergs$), higher-power winds ($\pow{44}-\pow{45}~\ergs$) disperse the gas and cause localised collapse of the cloudlets. We also demonstrate that a kinetic energy-dominated wind is more efficient in accelerating and dispersing the gas than a thermal wind with the same power. The interaction can give rise to multi-phase outflows with velocities ranging from a few 100 to several 1000~$\kms$. The mass outflow rates are tightly correlated with the wind power, which we explain by an ablation-based mass-loss model. Moreover, the velocity dispersion and the virial parameter of the cloud material can increase by up to one order of magnitude through the effect of the wind. Even though the wind can suppress or quench star formation for about 1 Myr during the initial interaction, a substantial number of gravitationally bound dense cloudlets manage to shield themselves from the wind's influence and subsequently undergo rapid gravitational collapse, leading to an enhanced star formation rate (SFR).

\end{abstract}

\begin{keywords}
ISM -- self-gravity -- AGN -- outflows -- hydrodynamics -- star formation
\end{keywords}



\section{Introduction}\label{sec:introduction}
The feedback from active galactic nuclei (AGNs) on the overall evolution of their host galaxies is thought to be a dominant mechanism in galaxy evolution theory \citep{Silk_1998,Fabian_2012}.
It is postulated that the large-scale outflow in the form of `jets' from the AGN heats up the intra-cluster medium and stops the cooling flow towards the centre of the cluster, therefore regulating the star-forming fuel \citep{McNamara_2007,McNamara_2012}.
Indeed, in recent cosmological simulations, it is necessary to include various models of feedback from the AGN by injecting thermal or kinetic energy \citep{Springel_2005,Schaye_2015,Weinberger_2017,Dave_2019,Schaye_2023}, in order to regulate star formation in massive galaxies and reproduce various observed scaling relations, including the luminosity functions and the $M_{\rm BH}-\sigma$ relation \citep[see the recent review][and references therein]{Vogelsberger_2020}.

While modern cosmological simulations successfully replicate the statistical characteristics and the redshift evolution of galaxies, they lack the ability to predict the feedback's impact on individual host galaxies due to challenges in accurately modelling the multi-phase interstellar medium (ISM) and associated small-scale physics, i.e., these processes are included as sub-grid recipes in cosmological simulations.
There is increasing observational evidence that wind and young radio jets originating from the central AGN significantly affect the host galaxy's ISM by driving multiphase outflows, which expel gas from the central region, driving turbulence, and potentially diminishing the star-forming fuel \citep{Nesvadba_2010,Harrison_2014,Garcia-Burillo_2014,Fluetsch_2019,Almeida_2022,Girdhar_2022,Leftley_2024}. This is also demonstrated by dedicated hydrodynamic simulations \citep{Sutherland_2007,Wagner_2012,Mukherjee_2018,Mandal_2021,Meenakshi_2022b,Tanner_2022}. 
These phenomena have a direct impact on the star formation activity inside the host as demonstrated by several observational studies where it has been found that some galaxies hosting radio-loud AGN show a lower star formation rate (SFR) compared to main-sequence galaxies, which follow the standard Kennicutt-Schmidt relation \citep{Schmidt_1959,Schmidt_1963,Kennicutt_1998a,Kennicutt_1998b} between gas-mass and SFR surface density \citep{Ogle_2007,Ogle_2010,Nesvadba_2010,Nesvadba_2011,Nesvadba_2021,Alatalo_2014,Alatalo_2015,Lanz_2016}.

Conversely, the over-pressurized winds/jet can cause significant compression of the ISM and may trigger collapse to rapidly form stars  \citep{Silk_2005,Gaibler_2012,Zubovas_2014a,Dugan_2017b,Mukherjee_2018}. Observational evidence also supports this hypothesis, where compact radio jets or quasar winds are found to enhance star formation activity \citep{Bicknell_2000,Inskip_2008,Zinn_2013,Kalfountzou_2014,Salome_2015,Salome_2017,Lacy_2017}.

Advancements in observational techniques and improved modelling of star formation physics within hydrodynamic simulations are beginning to shed light on the distinction between 'negative' and 'positive' feedback from AGN. Recent observations indicate the coexistence of both types of feedback within a single system \citep{Cresci_2018,Shin_2019}. Indeed, recent hydrodynamical simulations of jet-ISM interactions have revealed that while jet-inflated bubbles globally reduce star formation by enhancing turbulence, they can cause local regions of enhanced SFR due to the compression near the nuclear region \citep{Mandal_2021,Felize_2023}.
Thus, how AGN-driven outflows affect star formation is a complex competition between various phenomena on different scales.

However, a complete understanding of star formation as well as the survivability of the dense gas subjected to powerful AGN outflows remains elusive without the effect of the self-gravity of the gas. 
With typical densities around $100~\cc$ \citep{Miville_2017}, star-forming giant molecular clouds (GMC) have a freefall timescale of a few Myr, which is comparable to or shorter than the typical duration of AGN episodes, lasting between $10-100~{\rm Myr}$ \citep[e.g.,][]{Marconi_2004}.
Moreover, the presence of self-gravity can increase/prolong the survival time of the clouds, when faced with strong outflows from AGN, by making them dense and compact, effectively shielding them from erosion caused by the outflow and/or AGN radiation.
Conversely, the fragmentation induced by self-gravity can give rise to numerous smaller cloudlets that may be susceptible to evaporation or entrainment by the hot wind/jet cocoon, leading to the formation of multiphase outflows, which may regulate the available fuel to form stars.

Therefore, the significance of self-gravity at the cloud level can influence how AGN-driven winds/jet cocoons impact the host galaxy on larger scales.
Thus, well-resolved simulations modelling the interaction between AGN-driven winds and individual star-forming clouds may offer a supplementary perspective to both observations and global-scale simulations.
After all, the ultimate fate of the clouds depends on small-scale processes.
Additionally, the results from these small-scale studies are important for building better sub-resolution prescriptions of different mechanisms in global (galaxy and cosmological) simulations.

In this study, we revisit the classical `cloud-crushing' problem \citep{Klein_1994} with the help of a suite of three-dimensional (3D), self-gravitational hydrodynamics simulations in the context of the interaction between AGN-driven winds/jet cocoons and star-forming clouds. There have been extensive studies of the effects of external shocks or winds on clouds in various different contexts, with a primary emphasis on supersonic winds/shocks from galactic winds \citep[e.g.,][]{Klein_1994,Nakamura_2006,Pittard_2009,Pittard_2010,Scannapieco_2015,Gronke_2018,Banda_2016,Cottle_2018,Cottle_2020,Banda_2018,Banda_2019,Banda_2020,Banda_2021}.
However, wind-driven bubbles or jet-cocoons are known to be highly pressurized during the energy-driven phase \citep[e.g.,][]{Begelman_1989,Wagner_2011}, and thus can be subsonic depending on the density of the wind, while also exhibiting extreme velocities of up to tens of thousands of $\kms$.
Nonetheless, only a limited number of studies have taken into account the parameters of shocks/winds (e.g., density, velocity, and pressure) which can reach extremes comparable to those generated by AGN-jet cocoons or quasar winds \citep{Mellema_2002,Fragile_2004,Cooper_2009,Dugan_2017a,Cottle_2018}.

\citet{Mellema_2002} examined the influence of a radio jet cocoon on uniform spherical and elliptical clouds using 2D simulations. 
They identified 3 significant phases in the evolution \citep{Klein_1994} of these interactions: (i) the initial impact between the blast wave and the cloud, (ii) compression induced by the thermal and ram pressure of the wind, and (iii) fragmentation of the cloud. 
Additionally, their work revealed that in the presence of radiative cooling, the growth rate of the Kelvin-Helmholtz instability was highly suppressed and the mixed gas fraction was considerably reduced (less than 1\% of the original cloud mass), implying a slow evaporation process and a prolonged lifetime of the cloudlets. 
\citet{Cooper_2009} reached a similar conclusion and further demonstrated that when dealing with a fractal cloud, the fragmentation induced by the wind is more pronounced compared to a uniform cloud structure.
The fragmented cloudlets are compressed to high densities, which can then cool very efficiently and survive for a much longer period. 
While they discussed the potential role of self-gravity in this scenario, self-gravity was not included explicitly in the simulations.

Considering the effect of external pressurisation on a spherical cloud that may result from AGN winds or jet cocoons, \citet{Zubovas_2014a} demonstrated that the pressure confinement triggers the collapse of the cloud, leading to an enhanced SFR.
Using adaptive mesh refinement (AMR), self-gravitating simulations of the interaction between more realistic AGN-driven winds and Bonnor-Ebert spheres (resembling star-forming cores), \citet{Dugan_2017a} also arrived at a similar conclusion. 
Additionally, they identified a threshold ram pressure of the wind, above which the cloud will be destroyed before significant amounts of star formation can take place.
Another recent study by \citet{Li_2020} concluded that, if the cloud is initially Jeans' unstable, the interaction will eventually enhance the collapse of the cloud in the presence of self-gravity.
While these studies have individually delved into various significant aspects and distinct physical processes, there is a lack of a comprehensive global perspective that incorporates all the crucial parameters and physics.

This study seeks to extend previous research on the interaction between AGN-driven outflows with more realistic fractal star-forming interstellar clouds, including radiative cooling and self-gravity.
We explore a wide parameter space, systematically varying parameters including the wind power, the average cloud density, the internal fractal density distribution within the cloud, and whether the wind is primarily dominated by kinetic or thermal energy. 
This approach enables us to investigate diverse facets of the cloud's evolution.
The paper is organised as follows. 
In Sec.~\ref{sec:method}, we describe the simulation method and the choice of initial conditions. 
In Sec.~\ref{sec:result}, we present the main results of this study. We discuss the implications of this work in Sec.~\ref{sec:discussion}. 
Finally, in Sec.~\ref{sec:conclusion}, we summarize and conclude.

\section{Method}\label{sec:method}
\subsection{Simulation code}\label{sec:setup}
The numerical simulations presented in this study are performed using the grid-based code \textsc{Pluto v4.4} \citep{Mignone_2007,Mignone_2012} in 3D ($x,y,z$) Cartesian geometry. We use the HLLC Riemann solver \citep{Gurski_2004,Li_2005} along with a piecewise parabolic reconstruction scheme \citep[PPM;][]{Colella_1984} for solving the self-gravitating hydrodynamic (HD) equations:
\begin{gather}
      \frac{\partial\rho}{\partial t} 
    + \boldsymbol{\nabla}\cdot(\rho\boldsymbol{v}) = 0, \label{eq:mass_conservation} \\
    \frac{\partial(\rho\boldsymbol{v})}{\partial t} 
    + \boldsymbol{\nabla}\cdot\left[\rho\boldsymbol{v}\boldsymbol{v} + p\boldsymbol{I}\right] = \rho \boldsymbol{g}, \label{eq:momentum_conservation} \\
    \frac{\partial E_t}{\partial t} 
    + \boldsymbol{\nabla}\cdot\left[\left(E_t + p \right)\boldsymbol{v} )\right] = \rho\boldsymbol{v}\cdot\boldsymbol{g} - \left(\frac{\rho}{\mu m_{\rm H}}\right)^2\Lambda(T), \label{eq:energy_conservation} \displaybreak\\
    \frac{\partial(\rho C)}{\partial t} + \boldsymbol{\nabla}\cdot{\rho C \boldsymbol{v}} = 0 \label{eq:tracer}, \\
    \nabla^2\Phi = 4\pi G \rho, \label{eq:poisson}
\end{gather}
where $\rho$ is the mass density, $\boldsymbol{v}$ is the velocity, $p$ is the thermal pressure, $C$ is a Lagrangian scalar used to track gas in different components (i.e., cloud and wind), $\boldsymbol{g}=-\boldsymbol{\nabla}\Phi$ (where $\Phi$ is the gravitational potential) is the acceleration due to gravity, and $E_t$ is the total energy density given by,
\begin{equation}\label{eq:etot}
    E_t = \rho e + \frac{\rho\boldsymbol{v}^2}{2}.
\end{equation}
The above equations are closed by an ideal gas equation of state (EOS):
\begin{equation}\label{eq:EOS}
    p = (\gamma - 1)\rho e,
\end{equation}
where we consider $\gamma=5/3$ throughout. The energy conservation equation (Eq.~\ref{eq:energy_conservation}) also includes the cooling term ($\Lambda$) in order to account for the radiative losses, discussed in Sec.~\ref{sec:cooling}. 
Time evolution of Eq.~\eqref{eq:mass_conservation}-\eqref{eq:tracer} is performed using a $3^{\rm rd}$-order Runge-Kutta time-stepping scheme. 
For simulations involving gravity, we incorporate the self-gravity module\footnote{The self-gravity patch for the PLUTO code is publicly available at \url{https://bitbucket.org/mankush/pluto-4.4-self-gravity-patch}} developed by \citet{Mandal_2023}, which employs a Runge-Kutta-Legendre-based Poisson solver coupled to a V-cycle multigrid algorithm to solve the Poisson equation for the gravitational potential. 
Details of the numerical implementation of the self-gravity module are presented in \citet{Mandal_2023}. 
We also solve for the potential in the non-self-gravitating runs but do not couple the gravitational acceleration terms to the hydrodynamics (Eq.~\ref{eq:momentum_conservation} and \ref{eq:energy_conservation}). In this way, we calculate the gravitational potential-related quantities for these simulations and compare them with the corresponding self-gravitating runs.

\subsection{Computational domain}
\label{sec:domain}
In our simulations, we employ a uniform Cartesian domain ($x\mbox{-}y\mbox{-}z$) with a physical range of $-50~\mathrm{pc}\leq x\leq 150~\mathrm{pc}$, $-50~\mathrm{pc}\leq y\leq 50~\mathrm{pc}$, and $-50~\mathrm{pc}\leq z\leq 50~\mathrm{pc}$. 
The domain is discretized into a grid with $1024\times512\times512$ cells, resulting in a computational cell size (resolution) of $0.195$~pc.
In order to study the effect of the numerical resolution, we also perform two lower-resolution simulations with a grid size of $512\times256\times256$ and $256\times128\times 128$, respectively, and present the results in Appendix~\ref{sec:resolution_study}.
The fractal clouds (see Sec.~\ref{sec:cloud_setup}) are initially positioned at the origin (0,0,0) of the domain.
The clouds possess a core radius of 25~pc and an envelope of width 5~pc. Therefore, the core radius of the cloud is resolved with $\sim 128$ cells, this being adequate to capture the overall evolution \citep{Klein_1994,Fujita_2009,Banda_2018}. 
The wind is launched from the $y-z$ boundary at $x=-50$~pc in the positive $x$-direction.  
The details of the cloud setup are described in Sec.~\ref{sec:cloud_setup} and the wind parameters and their injection in Sec.~\ref{sec:wind_parameters} and Sec.~\ref{sec:BC} respectively.

\subsection{Cloud initialisation}
\label{sec:cloud_setup}
Several theoretical and observational studies have shown that the density structures in the turbulent molecular cloud are well described by a log-normal distribution function \citep{Vazquez-Semadeni_1994,Passot_1998,Federrath_2010a,Price_2011,Federrath_2012,Federrath_Ban_2015,Kritsuk_2017,Mandal_2020}. 
Hence, we model the cloud in our simulation such that the probability distribution function (PDF) in terms of the logarithmic density $s=\ln(\rho/\rho_0)$ (where $\rho_0$ is the mean density of the cloud) is given by a lognormal distribution:
\begin{equation}\label{eq:density_PDF}
    P(s) = \frac{1}{\sqrt{2\pi\sigma_s^2}}\exp\left[-\frac{(s-s_0)^2}{2\sigma_s^2}\right].
\end{equation}
Here $s_0$ and $\sigma_s$ are the mean and dispersion of the logarithmic density fluctuation, which from normalisation constraints $(\int e^s P(s) ds = 1)$ are related by $s_0=-\sigma_s^2/2$ \citep{Ostriker_2001,Li_2003,Federrath_2012}.

The log-normal density field of the cloud is constructed using the pyFC library\footnote{\url{https://www2.ccs.tsukuba.ac.jp/Astro/Members/ayw/code/pyFC}}, which generates a periodic random scalar field in 3D Cartesian space from a given PDF and power-law spectrum, $D(k)\propto k^{-\beta}$, in Fourier space, where $k$ is the dimensionless wavenumber. 
The two-point fractal distribution is characterised by the slope of the power-law ($\beta$), the Nyquist limit $k_{\rm max}$ and a lower cutoff wavenumber, $\kmin$, which corresponds to the largest spatially-correlated scale ($\lambda_{\max} \approx L/\kmin$) for a positive value of $\beta$.
For a given value of $\kmin$, the largest size of the perturbations or `cloudlets' is $r_{{\rm cloudlets},\kmin} \approx L/(2\kmin)$, where $L$ is the size of the periodic box \citep{Lewis_2002,Sutherland_2007,Wagner_2012}.
In this study, we have set the value of $\beta$ to be 1.66 for all the clouds. This falls in the range of cloud density spectral indices for supersonic turbulence\citep[e.g., see][]{Federrath_2009,Federrath_2013}. The $\kmin$ value is primarily set to 3 ($\lambda_{\rm max} \approx 20~\pc$) for most of the clouds while varying the wind properties to investigate their effects. 
Nevertheless, we also explore different values of $\kmin=$ 1, 3, 6, and 10, while keeping the wind parameters identical. This allows us to examine the impact of variations in the density distribution within the cloud.

In addition to the fractal density distribution of the cloud, we also initialise a Gaussian random field for each component of the velocity with zero mean and 1-D velocity dispersion of $\sigma_v$. 
For a particular cloud setup, the values of $\kmin$ and $\beta$ for the velocity field are the same as the log-normal density field. 
We choose the value of $\sigma_v$ for most of the simulations (except for the cloud with lower mean density) such that the 3D velocity dispersion ($\sigmav=\sqrt{3}\sigma_v$) of the cloud is $\sim 8\,\kms$, which is typical of the observed velocity dispersion of GMCs in the Milky Way and nearby galaxies on this scale \citep[e.g.,][]{Hughes_2010,Hughes_2013,Miville_2017,Faesi_2018}.

We set the mean number density of the fractal cloud to $200~\cc$, which gives us an initial virial parameter ($\alphavir = 2E_{\rm kin}/|E_{\rm grav}$) of $\sim 0.9$, again typical for star-forming clouds \citep{Miville_2017,Faesi_2018}.

The standard deviation of the density PDF in Eq.~\eqref{eq:density_PDF} of our clouds is calculated using the well-established relation:
\begin{equation}\label{eq:sigma_s}
    \sigma_s \approx \left[\ln\left(1+b^2\Mach^2\right)\right]^{1/2}
\end{equation}
which connects the standard deviation of the log-density ($\sigma_s$) and the turbulent Mach number ($\Mach = \sigmav/c_{\rm s, rms}$, where $c_{\rm s, rms}$ is the root mean square sound speed) \citep{Federrath_2008,Federrath_2010a,Federrath_2012}. 
The parameter $b$ is the driving parameter of turbulence, which represents the ratios of energies in solenoidal and compressive modes turbulence and varies between 1/3 (purely solenoidal) and 1 (purely compressive), respectively \citep{Federrath_2008}. 
Here we set $b=0.4$ for a mixed mode of turbulence driving \citep{Federrath_2010a}, as often observed in different environments \citep{Federrath_2016,Menon_2021,Sharda_2022,Dhawalikar_2022,Gerrard_2023}. 
The resulting value of $\sigma_s$ then serves as the input parameter in the pyFC routine which generates the fractal density field.

For the cloud density, we create a $310^3$-sized log-normal data cube with a mean of 1, which corresponds to a physical domain of $(x,y,z)\in[-30~\pc,30\pc]$\footnote{for this choice, the scale largest correlated density structure for a given value of $\kmin$ is $\lambda_{{\rm max},\kmin} \approx (60/\kmin)~\pc$}, following the method described above.
The density cube is then multiplied by the desired mean density of the cloud and a spherical volume of a radius of $30~\pc$ is extracted from the cube.
The sphere is then tapered with a radially decreasing function in order to ensure a smooth transition between the cloud's edge and the ambient medium:
\begin{equation}
    \rho_{\rm c}(r) = \rho_{\rm a} + \frac{\rho_{\rm cube}(r)}{\cosh{\left[\left(\frac{r}{r_{\rm core}}\right)^8\right]}},
\end{equation}
where $\rho_{\rm a}$, $\rho_{\rm cube}$ and $\rho_{\rm c}$ are the ambient density, the original density values in the fractal data cube and the final density for the cloud material that is used for the simulations, respectively.
Here, $r_{\rm core} = 25~\pc$ is the core radius of the cloud where the density remains the same as the density cube, along with an envelope with $5~\pc$ radially decreasing density.
Finally, the turbulent cloud is placed at the origin $(0,0,0)$ of the computational domain using a tri-linear interpolation scheme.

We also perform the same procedure for each velocity component where a $310^3$-sized Gaussian random data cube is generated for each component of the velocity field and mapped into the computational grid in a similar way to the density initialization.
The remaining portion of the computational domain is initialised with a static ambient medium having a density of $0.1~\cc$ and a temperature of $10^6~{\rm K}$, while the cloud is set to be in pressure equilibrium with this ambient medium initially.

\subsection{Wind parameters}\label{sec:wind_parameters}
In this study, our main objective is to examine how AGN-driven ``fast'' winds or jet-inflated cocoons affect star-forming complexes within host galaxies. 
Therefore, we focus on parameters typical of such scenarios, particularly considering the pressure and velocity ranges associated with these AGN-driven processes. 
Numerous observational studies have demonstrated that the velocities of ionized winds \citep{Rupke_2013,Harrison_2014,Genzel_2014,Brusa_2015,Carniani_2015,Bischetti_2017} and broad absorption line (BAL) winds \citep{Korista_2008,Moe_2009,Borguet_2013} driven by the central AGN on kiloparsec scales range from a few hundred to several thousand $\kms$ \citep[see][for a review]{Fiore_2017}. 
In this work, we consider the cloud to be located at $\sim 1~\kpc$ away from the AGN and the velocity of the injected winds at ($\vwind$) this distance are in the range of 400 to 4000 $\kms$, with a total (kinetic + thermal) wind power ($\Pout$) ranging from $\pow{42}$ to $\pow{45}~\ergs$. From theoretical and numerical investigations, these types of winds are found to be very hot and highly pressurised during the energy-conserving phase with pressure ranging from $10^{-10}$ to $10^{-7}~{\rm dyne\,cm^{-2}}$ depending on the wind power \citep[e.g.,][]{Begelman_1989,Sutherland_2007,Wagner_2011,Mukherjee_2016,Mukherjee_2018,Richings_2018a,Costa_2020}.    
Within these parameter ranges the thermal energy of the winds dominates the total energy budget, making them subsonic.
Interestingly, the self-similar solution describing an expanding bubble propelled by a central source, as proposed by \citet{Weaver_1977}, aligns with these pressure values. 
For an AGN-driven bubble with a given injected power ($\Pout$), expanding in a spatially homogeneous ambient medium of density $\rho_{\rm a}$, the pressure ($\pwind$) at a distance $R_{\rm w}$ from the central source is expressed as
\begin{gather}
    \pwind = \frac{7}{25}\left(\frac{125}{154\pi}\right)^{2/3} \rho_{\rm a}^{1/3} \Pout^{2/3} R_{\rm w}^{-4/3},\nonumber\\
    \begin{split}\label{eq:wind_pressure_norm}
        \approx 5.5\times 10^{-10} \left(\frac{n_{\rm a}}{0.1~\cc}\right)^{1/3}\left(\frac{\Pout}{10^{43}~\ergs}\right)^{2/3}\times \\ \left(\frac{R_{\rm w}}{\kpc}\right)^{-4/3}~{\rm dyne\,cm^{-2}}.
    \end{split}
\end{gather}
In this study, we set the pressure of the wind from the above equation (Eq.~\ref{eq:wind_pressure_norm}) for $R_{\rm w}\sim 1~\kpc$, which yields values in range $\pwind \sim 10^{-10}-10^{-8}~{\rm dyne\,cm^{-2}}$. The pressure is kept constant in time at the injection region, unlike the true \citet{Weaver_1977} wind solution. We discuss the implications of this choice in later paragraphs.
Moreover, the velocity of the bubble's forward shock at $R_{\rm w} = 1~\kpc$, as given by \citet{Weaver_1977},
\begin{gather}
    \vwind = \left(\frac{243}{3850\pi}\right)^{1/5} \rho_{\rm a}^{-1/5} \Pout^{1/5} t^{-2/5},\nonumber\\
    \approx 850 \left(\frac{n_{\rm a}}{0.1~\cc}\right)^{-1/3}\left(\frac{\Pout}{10^{43}~\ergs}\right)^{1/3} \left(\frac{R_{\rm w}}{\kpc}\right)^{-2/3}~\kms, \label{eq:wind_velocity_norm}
\end{gather}
also predicts values within the considered velocity ranges $\sim 400-4000~\kms$. In Table~\ref{tab:sim_list}, we list the wind parameters for each simulation.

We also consider a wind of power $\pow{45}\,\ergs$ whose velocity is set to $15,\!000\,\kms$ and whose pressure is such that it is supersonic. The total energy, in this case, is dominated by the kinetic energy (see Table~\ref{tab:sim_list}).
This kind of extreme outflow parameter can appear in the scenario where a cloud directly lies along the path of an AGN jet.
Additionally, this choice allows us to explore the effect of thermal vs.~kinetic (subsonic vs.~supersonic) winds of the same power in an otherwise identical cloud setup.

In order to completely specify the wind state, the wind density is calculated from energy conservation, i.e., the total energy flux of the thermal and kinetic components is equal to the injected power of the wind,
\begin{equation}\label{eq:wind_energy_conservation}
    \left(\frac{1}{2}\rho_{\rm w}\vwind^2 + \frac{\pwind}{\gamma -1}\right)4\pi R_{\rm w}^2\vwind = \Pout.
\end{equation}
Therefore, for the given values of wind power ($\Pout$), velocity ($\vwind$) and pressure ($\pwind$), the density ($n_{\rm w}$) of the wind is given by
\begin{equation}\label{eq:wind_density_norm}
    \begin{split}
        \nwind \approx 3\times 10^{-2}\left(\frac{\vwind}{10^3~\kms}\right)^{-2}\left[5.57\left(\frac{\Pout}{10^{43}~\ergs}\right) \right.\\
        \left. \times\left(\frac{\vwind}{10^3~\kms}\right)^{-1}\left(\frac{R_{\rm w}}{\kpc}\right)^{-2} - \left(\frac{\pwind}{10^{-10}~{\rm dyne\,cm^{-2}}}\right)\right]~\cc,
    \end{split}
\end{equation}
which is typically $n_{\rm w} \approx 0.01~\cc$ for the wind powers with corresponding velocity and pressure considered in this study. Therefore, we set the density of the wind to $\nwind=0.01~\cc$ for all the simulations. With these wind parameters, the mass outflow rate over a surface of $4\pi R_{\rm w}^2$ at $R_{\rm w} = 1~\kpc$ lies within the range of $\sim 0.7-7~{\mathrm{M}_{\odot}\,{\rm yr^{-1}}}$, which is typical of the mass outflow rates of ionized winds found in observations \citep[e.g., see][and references therein]{Fiore_2017}.

Although we set the pressure and velocity of the winds with different powers assuming that the cloud is located at $R_{\rm w} = 1~\kpc$ from the AGN, from Eq.~\eqref{eq:wind_pressure_norm} and \eqref{eq:wind_velocity_norm}, it becomes apparent that similar values of $\pwind$ and $\vwind$ can be achieved for a wind of different power when considering an alternative $R_{\rm w}$ value. 
For instance, $\pwind$ and $\vwind$ are very similar for winds with $\Pout = \pow{44}~\ergs$ at $R_{\rm w} = 1~\kpc$ and $\Pout=\pow{42}~\ergs$, but at $R_{\rm w} = 0.1~\kpc$.
Therefore, the wind parameters considered in this study not only reflect the strength of the wind at a particular location ($\sim 1~\kpc$) for varying wind powers, but can also be mapped to different locations relative to the central AGN for a particular wind power.

It is essential to emphasise that \citet{Weaver_1977} solution is only used to set the reference values of wind parameters, which are kept constant throughout the simulation. 
In reality, the radius of the bubble ($R_{\rm w}$) increases with time as it expands. Therefore, the wind solutions are time dependent. Ideally, one should consider a self-consistent evolution of the bubble with time for the wind injection. Additionally, the bubble solution of an AGN-driven wind solution consists of several distinct internal structures, i.e., the forward shock, shocked ambient medium, shocked wind and wind material \citep[e.g.,][]{Weaver_1977,King_2003,Zubovas_2012,FGQ_2012,Costa_2014}, which are not generally easy to implement in local box simulations like ours. Moreover, as our main focus is to investigate the effect of a steady wind whose parameters are similar to AGN-driven outflows on kpc scales on star-forming complexes, we adopt the simplistic wind injection model. Importantly, the main purpose of this study is to investigate the impact of the wind on clouds with and without self-gravity. Thus, while details of the wind modelling are simplified, we can still make a meaningful comparison, as clouds in the simulation with and without self-gravity face the exact same wind.

\subsection{Cooling}\label{sec:cooling}
In order to account for the energy losses due to radiative cooling (see Eq.~\ref{eq:energy_conservation}), we use the non-equilibrium cooling function calculated using \textsc{Mappings V} code \citep{Sutherland_2017,Sutherland_2018}.
This code utilizes a comprehensive database of atomic data to self-consistently compute the optically thin cooling rate for various gas phases, including cold neutral, warm neutral, partially ionized, and fully ionized gas in the temperature range $10^2-10^9\,{\rm K}$.
For temperatures exceeding $\sim 10^9\,{\rm K}$, the cooling function is extended by assuming bremsstrahlung emission \citep[e.g.,][]{Krause_2007,Mukherjee_2018}.
We also impose a lower temperature threshold of $T_{\rm floor} = 100\,{\rm K}$, below which the cooling is turned off and the temperature of any cell falling below $T_{\rm floor}$ is enforced to stay at $T_{\rm floor}$.
No additional heating terms, such as the galactic UV background or ionizing photons (UV, soft and hard X-ray) from the AGN, are included. 
In our simulations, the gas that can cool below $10^{4}~{\rm K}$ is sufficiently dense ($n_{\rm c} > 500~\cc$). 
In this density range, the high-density cores can be safely assumed to be self-shielded from the external UV and soft X-ray photons \citep{Rahmati_2013,Meenakshi_2022a}. 
Conversely, due to the very small photo-absorption cross-section of hard X-ray photons ($E\gtrsim 20~{\rm keV}$, primarily due to K-shell ionization of Fe and Ni ions) \citep{Band_1990,Verner_1993}, the cloudlets remain optically thin, resulting in negligible heating. 
Through explicit radiative-transfer calculations using \textsc{Cloudy} \citep{Ferland_2017} with the AGN spectral energy distribution (SED) as used in \citet{Meenakshi_2022a}, we confirm that ionization within the dense region ($n_{\rm c} > 500~\cc$) by an external radiation field (AGN and UV background) is insignificant, even for an AGN with a bolometric luminosity of $L_{\rm bol} = 10^{45}~\ergs$, where radiation can only penetrate to a depth of $\lesssim 0.5~\pc$ from the illuminated surface. 
This depth is even smaller for lower luminosities or higher densities. 
Therefore, the impact of ionising radiation can be safely disregarded.

\begin{table*}
    \caption{Initial conditions of all the simulations in this study.}
    \begin{tabular}{|c|c|c|c|c|c|c|c|c|c|c|}
       \hline
       Name  & $\Pout$ $^{(1)}$ & $\vwind$ $^{(2)}$ & $\pwind$ $^{(3)}$ & $\MachWind$ $^{(4)}$ & $\nc$ $^{(5)}$& Self-gravity? & $\kmin$ $^{(6)}$ & $\tcc$ $^{(7)}$ & $\tff$ $^{(8)}$ & $\tkh$ $^{(9)}$\\
         & ($\ergs$) & ($\kms$) & (${\rm dyne\,cm^{-2}}$) &  & ($\cc$) & & & (Myr) & (Myr) & (Myr)\\
       \hline
       \hline
       C\_no\_wind\_k3 & - & - & - & - & $200$ & No & 3 & - & 3.65 & -\\
       GC\_no\_wind\_k3 & - & - & - & - & $200$ & yes & 3 & - & 3.65 & - \\
       \hline
       C42\_k3 & $\pow{42}$ & $400$ & $1.32 \times \pow{-10}$ & $0.286$ & $200$ & No & 3 & 8.64 & 3.65 & 4.65\\
       GC42\_k3 & $\pow{42}$ & $400$ & $1.32 \times \pow{-10}$ & $0.286$ & $200$ & Yes & 3 & 8.64 & 3.65 & 4.65\\[0.2cm]
       
       C43\_k3 & $\pow{43}$ & $1000$ & $5.25 \times \pow{-10}$ & $0.329$ & $200$ & No & 3 & 3.46 & 3.65 & 1.86\\
       GC43\_k3 & $\pow{43}$ & $1000$ & $5.25 \times \pow{-10}$ & $0.329$ & $200$ & Yes & 3 & 3.46 & 3.65 & 1.86\\[0.2cm]
       
       C44\_k3 & $\pow{44}$ & $1800$ & $2.95 \times \pow{-9}$ & $0.277$ & $200$ & No & 3 & 1.92 & 3.65 & 1.03\\
       GC44\_k3 & $\pow{44}$ & $1800$ & $2.95 \times \pow{-9}$ & $0.277$ & $200$ & Yes & 3 & 1.92 & 3.65 & 1.03\\[0.2cm]
       
       C45\_k3 & $\pow{45}$ & $4000$ & $1.32 \times \pow{-8}$ & $0.285$ & $200$ & No & 3 & 0.86 & 3.65 & 0.46\\
       GC45\_k3 & $\pow{45}$ & $4000$ & $1.32 \times \pow{-8}$ & $0.285$ & $200$ & Yes & 3 & 0.86 & 3.65 & 0.46\\
       \hline
       GC43\_k1 & $\pow{43}$ & $1000$ & $5.25 \times \pow{-10}$ & $0.329$ & $200$ & Yes & 1 & 3.46 & 3.65 & 5.58\\
       GC43\_k6 & $\pow{43}$ & $1000$ & $5.25 \times \pow{-10}$ & $0.329$ & $200$ & Yes & 6 & 3.46 & 3.65 & 0.93\\
       GC43\_k10 & $\pow{43}$ & $1000$ & $5.25 \times \pow{-10}$ & $0.329$ & $200$ & Yes & 10 & 3.46 & 3.65 & 0.56\\
       \hline
       GC43\_uniform & $\pow{43}$ & $1000$ & $5.25 \times \pow{-10}$ & $0.329$ & $200$ & Yes & - & 3.46 & 3.65 & 4.61\\
       \hline
       GC43\_k3\_low & $\pow{43}$ & $1000$ & $5.25 \times \pow{-10}$ & $0.329$ & $20$ & Yes & 3 & 1.09 & 11.55 & 0.59\\
       \hline
       GC45\_k3\_kinetic & $\pow{45}$ & $15000$ & $1.00 \times \pow{-10}$ & $12.16$ & $200$ & Yes & 3 & 0.23 & 3.65 & 0.13\\
       \hline
       \hline 
    \end{tabular}
    \\
    \begin{flushleft}
        (1) Power of the expanding bubble.
        (2) Speed of the wind material.
        (3) Pressure of the wind.
        (4) Mach number of the wind defined as $\MachWind = \vwind/\cswind$, where $\cswind$ is the sound speed inside the wind.
        (5) Mean number density of the cloud. 
        (6) The minimum normalized wavenumber ($\kmin \equiv k$) of the density distribution of the cloud.
        (7) Cloud-crushing timescale.
        (8) Freefall timescale.
        (9) Growth timescale for the Kelvin-Helmholtz instabilities.
    \end{flushleft}
    \label{tab:sim_list}
\end{table*}

We create a table of the cooling rate $\Lambda$ as a function of temperature in the range of $10^2-10^{10}\,{\rm K}$ assuming solar metallicities \citep{Asplund_2009}. The values are then interpolated to every computational cell at runtime and are treated as a source term in Eq.~\eqref{eq:energy_conservation}. 
The equation is solved using a fractional step formalism, where the hydrodynamic evolution and source step are solved separately through operator splitting.
The energy losses from radiative cooling are computed by integrating the internal energy equation
\begin{equation}
    \frac{\partial (\rho e)}{\partial t} = - \left(\frac{\rho}{\mu m_{\rm H}}\right)^2\Lambda(T),
\end{equation}
using an adaptively-chosen, explicit or semi-implicit Embedded Runge-Kutta method, depending on the ``stiffness'' of the equation \citep[see e.g., ][]{Tesileanu_2008}.

We note that we do not account for the explicit density dependence of the cooling function ($\Lambda$), which becomes significant in gas with $T \lesssim 10^4~{\rm K}$ and $n \gtrsim 10^4~\cc$ in the case for non-equilibrium cooling. 
In our simulations, however, the initial temperature of gas with $n > 10^3~\cc$ is below the floor temperature of $T < 100~{\rm K}$, where radiative cooling is turned off. 
Additionally, the shock-heated gas in our simulations rarely reaches the density and temperature ranges where the density dependence of the cooling curve would be significant. 
Therefore, the simplification of the cooling curve as a function of temperature only does not affect our results. 
Moreover, our simulations do not include molecular cooling, which might be a dominant mechanism below $T = 1000\,{\rm K}$, and completely dominates the cooling process below $100\,{\rm K}$ \citep[e.g.,][]{Goldsmith_1978}.
Therefore, for more accurate modelling of gas cooling and star formation, one must include the low-temperature processes of the cold molecular phase \citep[][]{Koyama_2000,Vazquez-Semadeni_2007,Glover_2010}.
Nonetheless, as we impose a temperature floor of $100\,{\rm K}$, the temperature does not fall below this range. Hence, the absence of molecular cooling does not affect the result significantly.

\subsection{Wind injection and boundary conditions}\label{sec:BC}
We approximate the AGN-driven spherically symmetric wind at a distance of $1~\kpc$ from the central AGN as a planar wind on the scale of the cloud, propagating in the positive $x$-direction. 
The initial position of the forward shock is set at $x=-40~\mathrm{pc}$ and the domain within $-50~\pc\leq x \leq-40~\pc$ is initialised with the wind state.
The wind is constantly injected from the left boundary along the $x$-direction using an inflow boundary condition, where we populate the fluid quantities (pressure, velocity and density) in the ghost cells with the wind properties as listed in Table~\ref{tab:sim_list}.

\begin{figure*}
    \centerline{
    \def\arraystretch{1.0}
    \setlength{\tabcolsep}{0.0pt}
        \begin{tabular}{lcr}
            \includegraphics[width=0.5\linewidth]{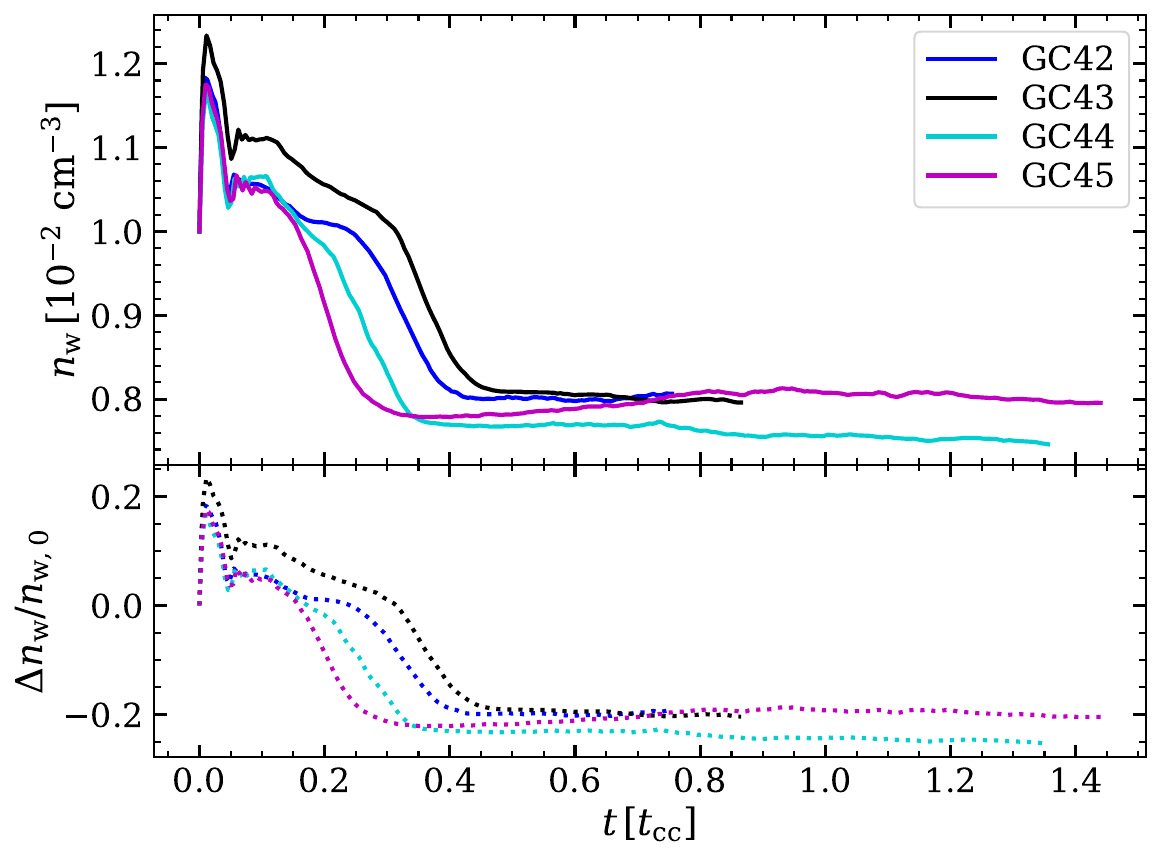} &
            \includegraphics[width=0.5\linewidth]{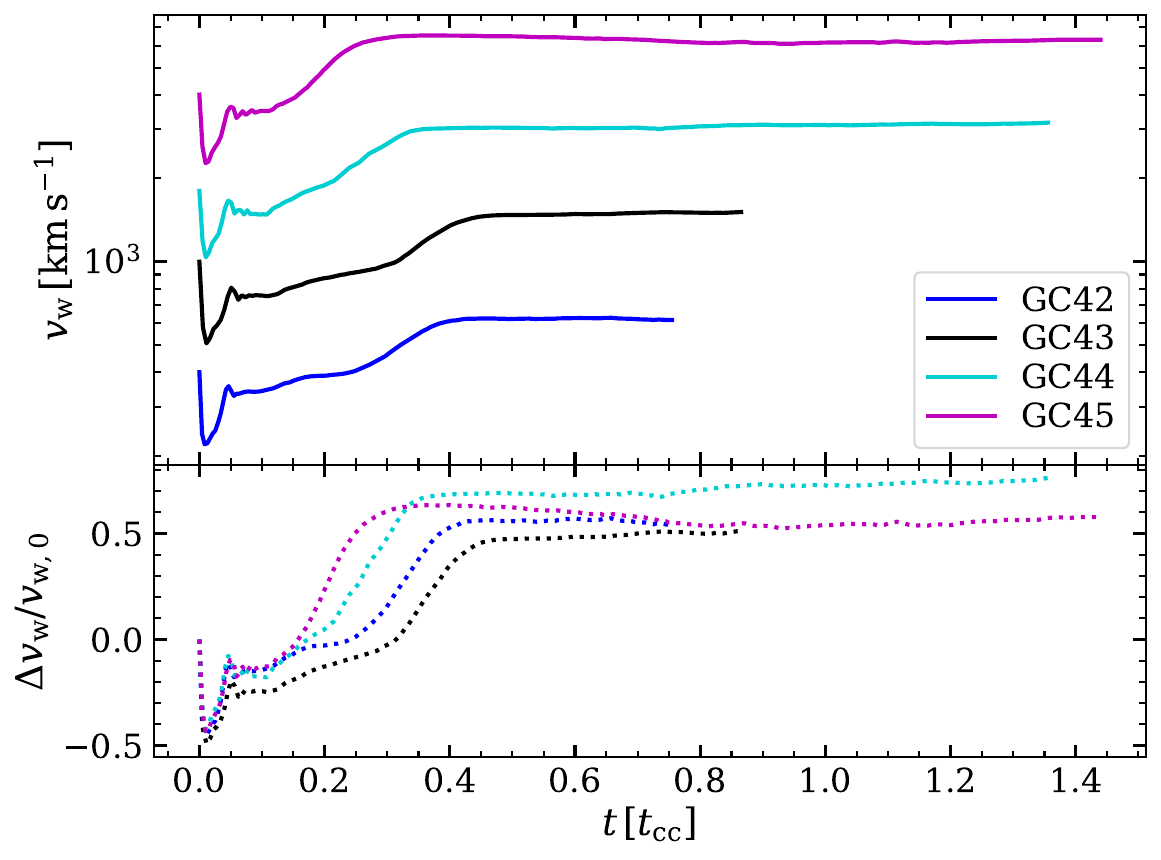} \\
            \includegraphics[width=0.5\linewidth]{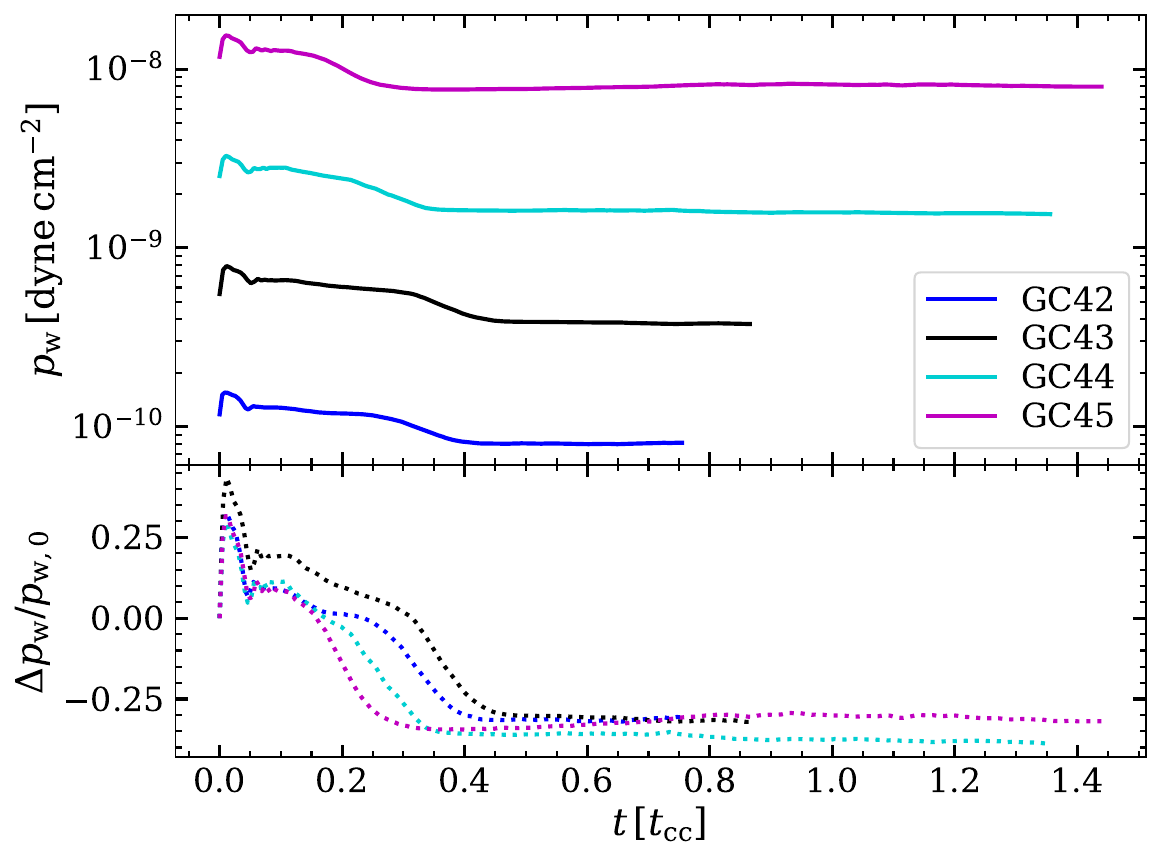} &
            \includegraphics[width=0.5\linewidth]{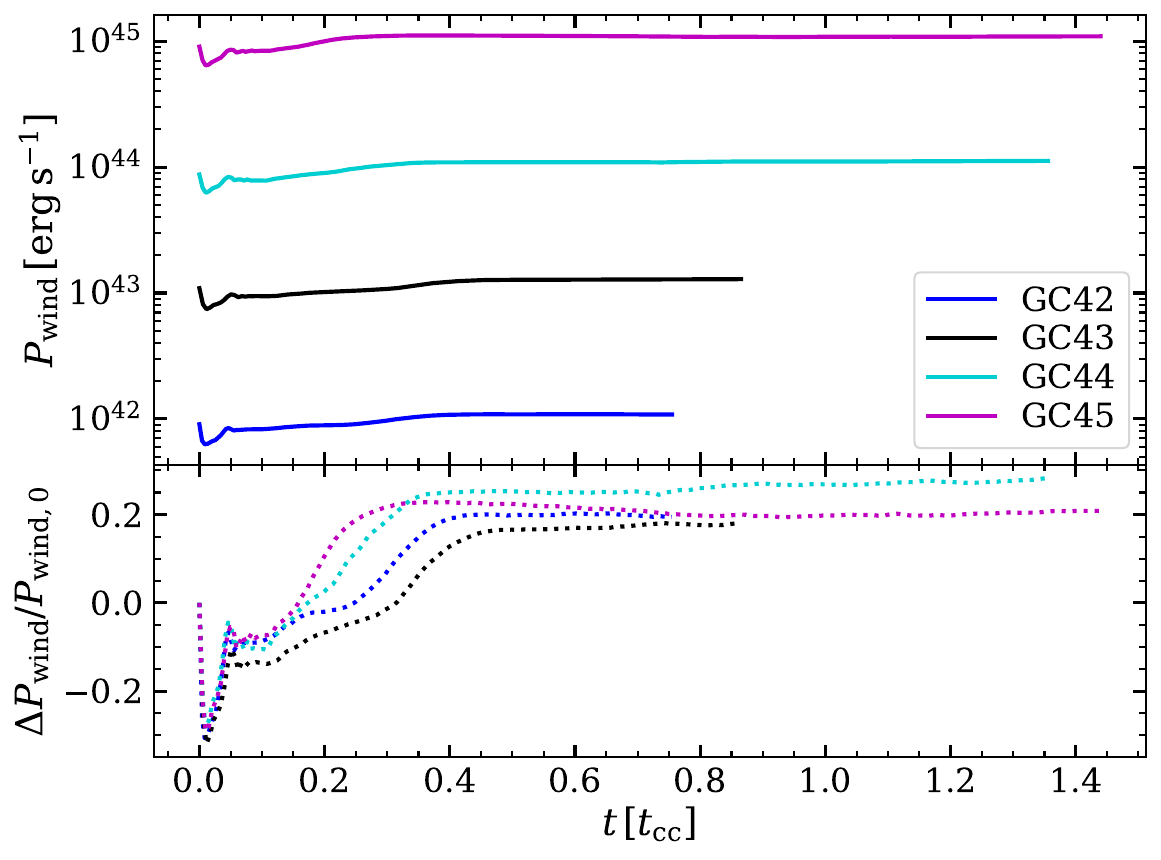}
        \end{tabular}}
        \caption{Time evolution of the wind parameters, i.e., density (top-left), velocity (top-right), pressure (bottom-left) and power (bottom-right) in the simulations with different power as indicated in the legend. The bottom section of each panel shows the fractional deviation of the parameter from the initial values.}
        \label{fig:input_parameter_simulations}
\end{figure*}

\begin{table*}
    \caption{The saturated values of the wind parameters for the subsonic cases. The subscripts with $0$ refer to the intended values of the wind parameters as tabulated in Table~\ref{tab:sim_list}, while `sat' subscripts are the saturated values of the parameters after $\gtrsim 0.4~\tcc$.}
    \begin{tabular}{c|c|c|c|c|c|c|c|c|c|c}
        \hline
        $P_{\rm w,0}$ & $P_{\rm w,sat}$ & $n_{\rm w,0}$ & $n_{\rm w,sat}$ & $v_{\rm w,0}$ & $v_{\rm w,sat}$ & $p_{\rm w,0}$ & $p_{\rm w,sat}$ & $\mathcal{M}_{\rm w,0}$ & $\mathcal{M}_{\rm w,sat}$ & $t_{\rm cc, sat}$\\
        $(\ergs)$ & $(\ergs)$& $(\cc)$ & $(\cc)$ & $(\kms)$ & $(\kms)$ & ${\rm (dyne~cm^{-2})}$ & ${\rm (dyne~cm^{-2})}$ &  &  & (Myr) \\
        \hline
        \hline
        $\pow{42}$ & $1.09\times 10^{42}$ & $10^{-2}$ & $8.01\times 10^{-3}$ & 400 & 623 & $1.32\times 10^{-10}$ & $8.04\times 10^{-11}$ & 0.286 & 0.621 & 6.20\\
        $\pow{43}$ & $1.28\times 10^{43}$ & $10^{-2}$ & $8.06\times 10^{-3}$ & 1000 & 1481 & $5.25\times 10^{-10}$ & $3.84\times 10^{-10}$ & 0.329 & 0.678 & 2.60 \\
        $\pow{44}$ & $1.10\times 10^{44}$ & $10^{-2}$ & $7.62\times 10^{-3}$ & 1800 & 3075 & $2.95\times 10^{-9}$ & $1.60\times 10^{-9}$ & 0.277 & 0.671 & 1.29\\
        $\pow{45}$ & $1.09\times 10^{45}$ & $10^{-2}$ & $7.98\times 10^{-3}$ & 4000 & 6281 & $1.32\times 10^{-8}$ & $8.02\times 10^{-9}$ & 0.285 & 0.625 & 0.62\\
        \hline
        \hline
    \end{tabular}
    \label{tab:saturated_values}
\end{table*}

However, one complication of subsonic inflows is that only two out of three characteristic waves enter the domain, while the third leaves. 
This implies that only two out of three primitive variables (ideally density-pressure or density-velocity pairs for well-poised conditions, e.g., see Sec.~19.3 of \citealp{Laney_1998}) can be specified physically at the boundary, with the remaining variable set numerically from the interior solution \citep{Thompson_1990,LeVeque_2002}. 
Yet, a difficulty arises due to the finite size of the computational domain: the outgoing wave, which should freely exit the domain, experiences numerical reflection at the boundary, thereby reducing the accuracy of the interior solution \citep{Majda_1975}. 
Therefore, the value of the third primitive variable should be chosen to allow the wave to exit the domain with minimal reflection, a condition commonly referred to as the non-reflecting boundary condition (NRBC), which itself is a broad area of research \citep{Engquist_1977,Engquist_1979,Hedstrom_1979,Bayliss_1980}.

However, in our simulations, we do not include such a treatment to minimize boundary reflections. 
As a result, the subsonic winds injected at the ghost-cell layers, do not emerge with the same parameters (density, velocity, pressure, Mach, and power) as tabulated in Table~\ref{tab:sim_list}. 
Fig.~\ref{fig:input_parameter_simulations} depicts the time evolution of the wind density (top-left), velocity (top-right), pressure (bottom-left), and power (bottom-right) for the subsonic cases, as indicated in the legend. 
The values are calculated by taking the average over a $y\mbox{-}z$ slice at $i = 1$, i.e., the first interior cell from the $x$-left boundary. We observe that all wind parameters experience an initial transient phase when the wind starts to progress through the stationary ambient medium, sweeping up the material. 
However, after $t\gtrsim 0.4~\tcc$, the parameters saturate to constant values, which are different from the injected values at the ghost zone. 
In Table~\ref{tab:saturated_values}, we tabulate the injected values of the wind parameters and their saturated values.
Nonetheless, the deviation of the parameters from the intended values is almost similar for all the powers.
The wind velocities exhibit the most significant deviation, reaching $\lesssim 75\%$ higher than the intended values. Conversely, pressure values decrease by $\lesssim 40\%$, therefore increasing the Mach number of the winds (see Table~\ref{tab:saturated_values}), which are still subsonic. 
However, as the winds are dominated by thermal energy, the deviations of the wind power from the intended values are less, increased by $\lesssim 30 \%$. 

Nevertheless, since our main objective in this study is to investigate the interaction between a cloud and winds possessing parameters akin to AGN-driven winds at specific powers, the qualitative outcomes among winds of varying powers, as well as the impact of self-gravity, remain unaffected.
Moreover, the saturated values of the wind parameters, i.e., density, velocity, pressure and power (see Table~\ref{tab:saturated_values}) fall well within the desired AGN wind parameter space, as laid out in Sec.~\ref{sec:wind_parameters}.
Furthermore, implementing non-reflecting boundary conditions at the inflow boundary does not ensure that the wind will emerge with the intended power, as one of the three primitive variables remains unconstrained and is set numerically. 
Indeed, as demonstrated in Appendix~\ref{sec:NRBC}, even though the physically specified primitive variables (density and velocity) remain constant, the pressure and power of the wind deviate significantly ($\sim 90\%$) from the intended value, exceeding that of the simulation using the wind injection method employed in this study.

Except for the inflow boundary, all other boundaries of the computational domain are set to diode boundary conditions -- the modification of the outflow boundary condition that prevents inflow into the domain so that gas can only leave the computational domain. 
We employ isolated boundary conditions for the gravitational potential ($\Phi$) on all sides of the domain, which is calculated using a multipole expansion of the density distribution up to order $l=4$ \citep[see Appendix C of][for details]{Mandal_2021}.


\subsection{Simulations and naming convention}\label{sec:IC}
In this study, we conduct a total of 13 three-dimensional (3D) simulations (including 10 simulations with self-gravity) covering a large parameter space of both the wind and cloud.
An additional two simulations (with and without self-gravity) are performed without wind in order to ascertain the effects of winds. 
For all these simulations, we initialize the cloud with an initial virial parameter, $\alpha_{\rm vir, 0} = 0.9$, and using this value, we determine the initial mean cloud density and velocity dispersion.

In our fiducial simulations, the mean density of the cloud ($\nc$) is set to $200\,\cc$ with $\kmin=3$ and an initial velocity dispersion of $8\,\kms$ to achieve $\alpha_{\rm vir, 0} = 0.9$.
Therefore, the mass of the standard cloud in our simulation is $M_{\rm c} = 2.873\times 10^{5}\,\Msun$, which is typical of the average mass of GMCs in the Milky Way and nearby galaxies \citep{Hughes_2013,Miville_2017}.
To investigate the effect of a similar power wind on a cloud with lower density,
we consider one simulation with a lower mean density of $20\,\cc$ and velocity dispersion of $2.5\,\kms$.
The fiducial wind in our simulation is thermal energy-dominated.
The wind parameters and the minimum wavelength of the cloud are varied as discussed in Sec.~\ref{sec:cloud_setup} and \ref{sec:wind_parameters}.
The initial conditions and the parameters along with the name for all the simulations are listed in Table~\ref{tab:sim_list}

It is useful to clarify the naming convention of the simulations.
For example, in GC43\_k3, `G' implies self-gravity is present and `C' stands for cooling, which is common in all the simulations.
The number `43' represents the total power of the wind of $\pow{43}\,\ergs$. the label `k3' represents the minimum wavenumber of the cloud;  in this case $k_{\rm min}=3$.
We explicitly added the label `\_low' to `GC43\_k3' for the case of the lower mean density cloud with wind power of $\pow{43}\,\ergs$.
Similarly, `\_kinetic' is added to `GC45\_k3' to indicate that the wind of power $\pow{45}\,\ergs$ is kinetic energy dominated.
If not specified, the default minimum wavenumber of the cloud is $\kmin=3$.

\subsection{Relevant timescales}\label{sec:timescale}
There are various dynamical timescales involved in the problem that we consider in this study and the absolute values for different simulations are tabulated in Tab.~\ref{tab:sim_list}.
\begin{enumerate}
    \item The shock-passing time ($\tsp$), i.e., the approximate time the initial shock takes to sweep over the whole cloud \citep{Klein_1994,Fragile_2004}:
    \begin{equation}\label{eq:t_sp}
        \tsp \approx \frac{2R_{\rm c}}{\vwind},
    \end{equation}
    where $R_{\rm c}$ is the radius of the cloud and $\vwind$ is the velocity of the wind.

    \item The cloud-crushing time ($\tcc$), the typical time in which the shock will compress the cloud \citep{Fragile_2004}:
    \begin{equation}\label{eq:t_cc}
        \tcc \approx \frac{R_{\rm c}}{v_{\rm st}} \approx \chi^{1/2}\frac{R_{\rm c}}{v_w},
    \end{equation}
    where $v_{\rm st}$ is the velocity of the transmitted shock into the cloud and $\chi = \rho_{\rm c}/\rho_{\rm w}$ is the density contrast between the cloud and wind.

    \item The freefall timescale:
    \begin{equation}\label{eq:t_ff}
        \tff = \sqrt{\frac{3\pi}{32 G \rho_{\rm c}}},
    \end{equation}

    \item The cooling timescale \citep{Klein_1994,Mellema_2002,Fragile_2004}:
    \begin{equation}\label{eq:t_cool}
        \tcool = \frac{3k_{\rm B} n_{\rm c}\langle T_{\rm c}\rangle}{2n_{\rm c}^2\Lambda(T)},
    \end{equation}
    where $k_{\rm B}$ is the Boltzmann constant, $n_{\rm c}$ and $\langle T_{\rm c}\rangle$, the average initial number density and temperature of the cloud. For the fiducial simulations with an initial mean number density of $n_{\rm c} = 200~\cc$ and mean temperature of $\langle T_{\rm c} \rangle \approx 3\times 10^3~{\rm K}$, as estimated from the pressure equilibrium condition with the ambient medium, the typical cooling timescale is of the order of $3 \times 10^2~{\rm yr}$, which is significantly shorter than relevant other timescales (see Tab.~\ref{tab:sim_list}). However, it is essential to recognize that this estimate relies on average temperature and density values, whereas the fractal nature of the clouds in the simulations introduces a wide range of temperatures due to the inhomogeneous density distribution. Therefore, the cooling timescale can exhibit significant variations across different regions within the cloud.

    \item The drag timescale ($\tdrag$), by which the cloud is accelerated to a similar velocity as that of the wind \citep{Klein_1994,Fragile_2004}:
    \begin{equation}\label{eq:t_drag}
        \tdrag \approx \chi^{1/2}\tcc
    \end{equation}

    \item The time ($\tkh$) for the Kelvin-Helmholtz (KH) instability to grow for $\chi \gg 1$ \citep{Chandrasekhar_1961}:
    \begin{equation}\label{eq:t_KH}
        \tkh \sim \frac{\chi^{1/2}}{k_{\rm KH} v_{\rm rel}} \approx \frac{4}{3}\frac{r_{\rm cloudlet, \kmin}\chi^{1/2}}{\vwind}
    \end{equation}
    where $k_{\rm KH}$ and $v_{\rm rel}$ are the wavenumber of the  KH perturbations and the relative velocity between the post-shock background and cloud. For $\chi \gg 1$, the relative velocity is approximately equal to the post-shock velocity of the background, i.e., $v_{\rm rel}\approx 3\vwind/4$. We set $k_{\rm KH}\sim 1/r_{\rm cloudlet, \kmin}$ because even though the instabilities at the smallest wavelengths grow more rapidly, the most detrimental wavelengths are those approximately equal to the cloudlet radius  \citep[e.g., see][]{Klein_1994,Poludnenko_2002}.
\end{enumerate}

\section{Results}\label{sec:result}

\subsection{Morphological evolution of the cloud}
\label{sec:evolution}

\begin{figure}
    \centering
    \includegraphics[width=\linewidth]{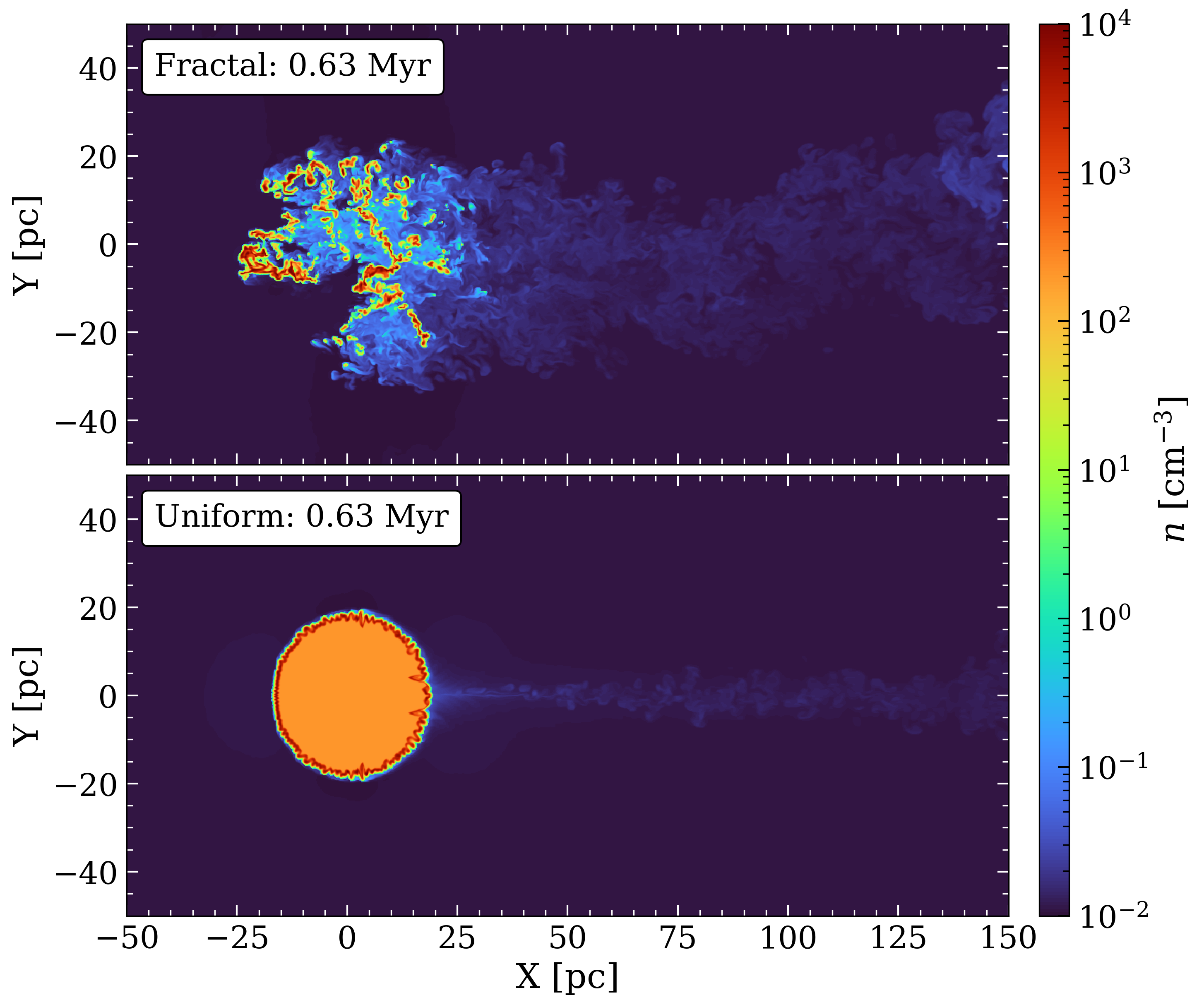}
    \caption{Number density slice through the $z=0$ plane for simulations with a fractal with $\kmin=3$ (GC43\_k3, top) and uniform (GC43\_uniform, bottom) cloud at 0.63~Myr. The wind power in both the simulation is $\pow{43}~\ergs$. The wind readily penetrates the fractal cloud through low-density inter-cloudlet channels (top panel), leading to the formation of numerous small and dense cloudlets. On the other hand, in the simulation featuring a uniform spherical cloud (bottom panel), the initial compression forms a highly dense shell, which cools efficiently. This shell slows down the transmitted shock into the cloud, significantly diminishing the wind's impact on the overall cloud.}
    \label{fig:slice_fractal_vs_uniform}
\end{figure}
\subsubsection{General features}
\label{sec:general_evolution}
The evolution of a cloud impacted by a highly pressurized wind can be divided into three phases. 
Initially, as the wind comes into contact with the cloud, it initiates an internal shock propagating from the cloud's surface toward its centre. 
The force exerted by the wind's ram pressure causes compression within the cloud, leading to a rise in the average density of the cloud material.
The log-normal density distribution of the clouds results in many high-density fragments (cloudlets) and low-density channels inside the fractal clouds. 
This aids the wind to propagate deeper into the clouds, resulting in a stronger interaction with the cloud and deeper penetration when compared to cases with a uniform spherical cloud, as demonstrated in Fig.~\ref{fig:slice_fractal_vs_uniform} \citep[also see][]{Cooper_2009,Schneider_2017,Banda_2018}.

Subsequently, as the wind continues to flow downstream wrapping around the cloud, a shear layer forms at the wind-cloud interface, leading to the onset of Kelvin-Helmholtz (KH) instability. 
As a result, the outer layer of the cloud gets stripped, mixed with the wind, and funnelled downstream where it condensates (middle left panel of Fig.~\ref{fig:proj_diff_model}), giving rise to a series of cold and dense cloudlets that form a trailing tail behind the main cloud. 
During this phase, if no cooling mechanism is present, the energy transport from the wind to the cloud in the form of thermal energy should increase the temperature of the cloud material, which would result in an expansion of the cloud, making the cloud more prone to destruction \citep{Orlando_2005,Cooper_2009}.
However, with the presence of radiative cooling, the compressed cloudlets cool efficiently, leading to a cessation of internal pressure support, contraction, and the formation of denser cloudlets. 
This contraction enhances the density contrast ($\chi$) between the cloudlets and the surrounding hot shear flow, consequently reducing the growth rate of KH instabilities (Eq.~\ref{eq:t_KH}). 
Therefore, in the presence of radiative cooling, the cloudlets are relatively protected from instabilities induced by the shear flow compared to a non-radiative scenario \citep{Mellema_2002,Cooper_2009,Scannapieco_2015,Banda_2021}.

While the presence of radiative cooling reduces the growth rate of instabilities,  it does not provide complete protection.
Therefore, the wind-induced shear flow along the cloudlet surface still causes the entrainment of cloud material, albeit to a lesser extent compared to a non-radiative scenario \citep{Cooper_2009,Banda_2021}. 
Additionally, the direct momentum transfer from the wind accelerates the cloudlets.
Combined with the shredded cloudlets from the main cloud due to the KH instability, these fragments persistently contribute to the elongation of the cloud, eventually leaving the computational domain (bottom-left panel of Fig.~\ref{fig:proj_diff_model}).

Although the general evolutionary stages of a cloud impacted by a highly pressurized wind are similar for all cases, the corresponding time scales and the overall evolution can depend on several factors which are discussed in subsequent sections.


\subsubsection{Gravity vs.~no-gravity}
\label{sec:diff_model}
\begin{figure*}
    \centering
    \includegraphics[width=\linewidth]{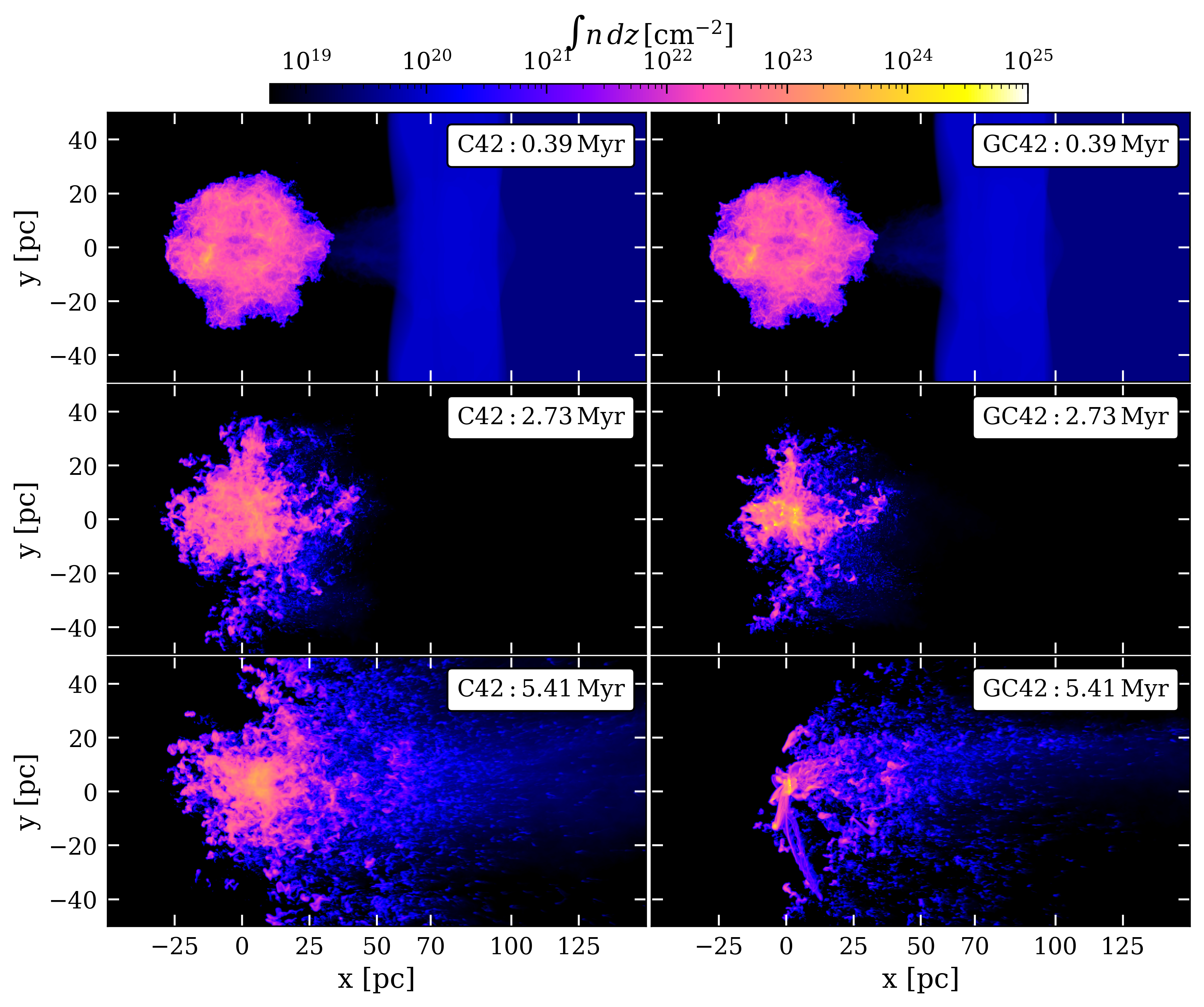}
    \caption{Time evolution (row-wise) of the projected number density along the $z$-axis of the simulations without self-gravity (left) and with self-gravity (right). The power of the wind for both simulations is $\pow{42}\,\ergs$. It is evident that in the absence of self-gravity (left), the cloud undergoes expansion after the initial compression. Whereas in the self-gravity case (right) global collapse takes place.}
    \label{fig:proj_diff_model}
\end{figure*}
Here, we examine the effect of self-gravity on the morphological evolution of the cloud by comparing it with a simulation that shares identical initial conditions and wind strength but lacks self-gravity.
In Fig.~\ref{fig:proj_diff_model}, we show the column density map in the $x\mbox{-}y$ plane of the simulations both without (C42\_k3; left panels) and with (GC42\_k3; right panels) self-gravity at three different times.
The power of the wind is $\pow{42}\,\ergs$.

During the initial compression phase, the mean density of the cloud in both cases (with and without self-gravity) increases.
After the initial shock-induced fragmentation phase, in the absence of self-gravity, the pressure-confined fragments undergo the highest attainable compression due to the transmitted shock into the cloudlets (middle-left panel of Fig.~\ref{fig:proj_diff_model}). 
Once this compression limit is reached, the density of the clumps cannot increase any further. 
Hence, the cloud as a whole commences disintegration, resulting in cloud expansion (bottom-left panel).

We point out that the expansion is not a result of adiabatic heating in the absence of radiative cooling, as outlined in various studies \citep{Banda_2018,Banda_2019}. 
Instead, the expansion occurs because the low-density channels within the fractal cloud facilitate the infiltration of high-velocity wind material, which induces turbulence and vorticity within the cloud \citep{Klein_1994}, leading to the transfer of momentum from the wind to the cloud material. 
This momentum transfer is what ultimately causes the cloud to expand.

\begin{figure*}
    \centering
    \includegraphics[width=\linewidth]{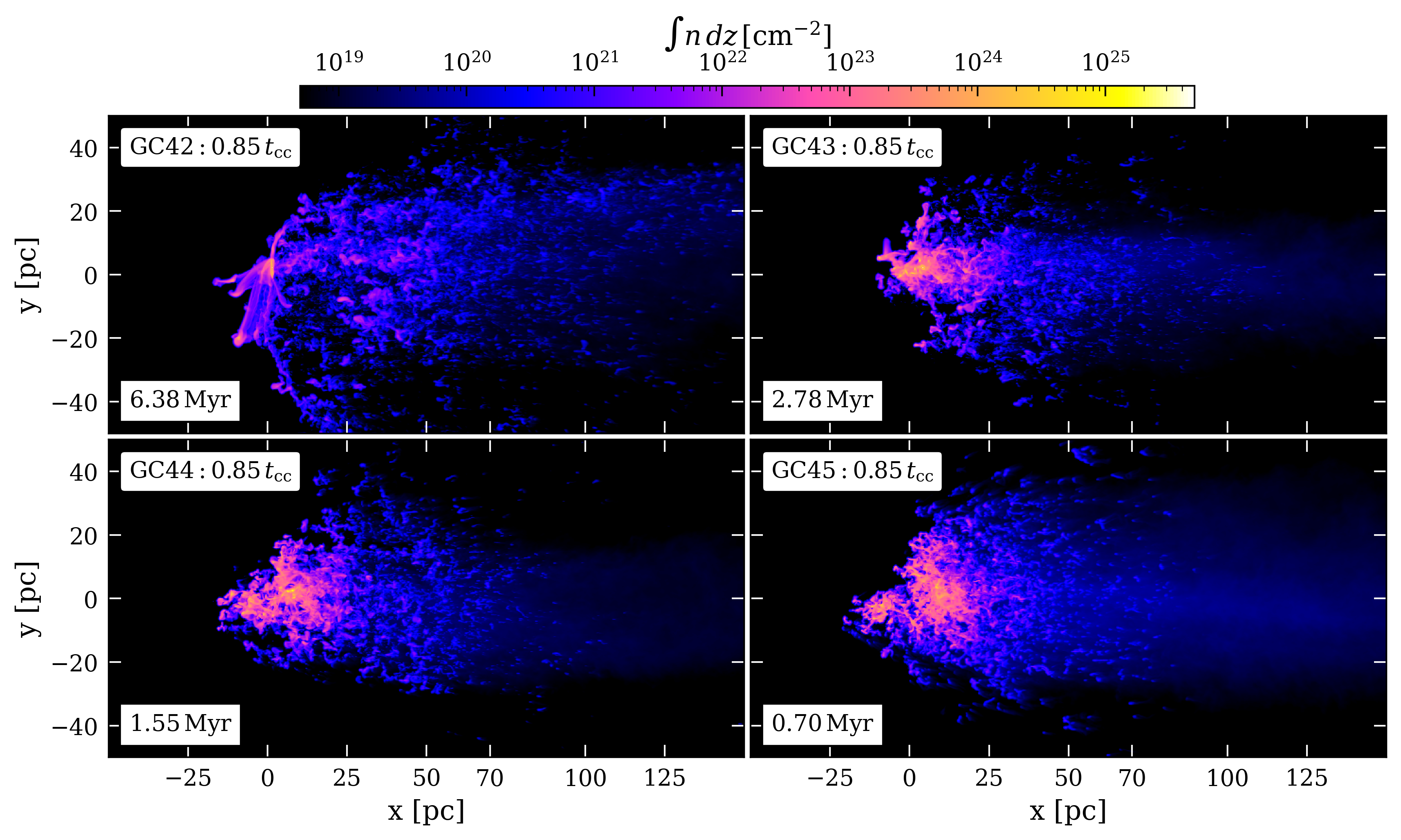}
    \caption{Projected number density along the $z$-direction at $t = 0.85\,\tcc$ of simulations with self-gravity and different wind power: $\pow{42}$ (top-left), $\pow{43}$ (top-right), $\pow{44}$ (bottom-left) and $\pow{45}\,\ergs$ (bottom-right). In the unit of the Kelvin-Helmholtz timescale, this time corresponds to $t=1.51~\tkh$. The corresponding physical time for each simulation is shown in the bottom-left corner of each panel. Interesting to note that, as wind power increases, the cloud becomes more disrupted. While in low-power cases, the cloud experiences global collapse, the higher-power winds cause localised collapses of the dense cloudlets, simultaneously still disrupting the cloud.}
    \label{fig:proj_diff_power}
\end{figure*}

In contrast, when self-gravity is present, the shock-compressed cloudlets become gravitationally bound and start to accrete material from the surroundings (middle-right panel of Fig.~\ref{fig:proj_diff_model}), thereby attaining significantly higher density and mass compared to the case without self-gravity. 
Additionally, when acting in an inhomogeneous medium, self-gravity causes additional local fragmentation of the cloud, which creates many low-density channels for the wind to propagate inside the cloud, resulting in a stronger interaction. 
The increased average cloud density produced by compression deepens the gravitational potential, resulting in a more compact and tightly bound cloud structure. 
As a result, the interplay between the gravitational pull and the strength of the wind-cloud interaction becomes the pivotal factor in deciding whether the cloud will experience a runaway collapse or will be disintegrated by the force of the wind.

For a low-power wind, e.g., $\pow{42}\,\ergs$ as shown in Fig.~\ref{fig:proj_diff_model}, the momentum transfer from the wind to the cloud is significantly lower and therefore less gas is pushed out of the potential well of the cloud (which becomes deeper with time as more gas is accreted) compared to a high-power wind.
Therefore, the wind triggers a runaway collapse of the cloud in the presence of self-gravity as can be seen from the bottom-right panel of Fig.~\ref{fig:proj_diff_model}. 
In contrast, in the simulation without self-gravity, the cloud expands (bottom-left panel of Fig.~\ref{fig:proj_diff_model}) and eventually will be elongated and destroyed by the wind.


\subsubsection{Effect of varying wind power}
\label{sec:diff_power}
Here we show the morphological differences of the cloud in simulations in the presence of self-gravity with different wind velocities and pressure, which are the proxy for the power of the wind driven by AGN. Fig.~\ref{fig:proj_diff_power} illustrates the column density map on the $x\mbox{-}y$ plane of simulations with wind power $\pow{42}$ (GC42\_k3), $\pow{43}$ (GC43\_k3), $\pow{44}$ (GC44\_k3) and $\pow{45}\,\ergs$ (GC45\_k3) at $t=0.85\,\tcc$. 
The corresponding physical time is also shown in each panel.

For the lower power cases (GC42\_k3 and GC43\_k3), the initial freefall time is shorter than the growth timescale for the KH instability due to the lower wind velocity ($\tkh \propto 1/\vwind$), which can further be reduced by the dramatic increase of density through compression.
In this scenario, the cloud as a whole undergoes runaway gravitational collapse before any instability has a chance to act. 
This is clearly visible in the top-left panel, where the cloud in GC42\_k3 has collapsed by this time, becoming very compact and highly dense.
Nonetheless, due to the continuous ablation, the loosely bound outer layer of the cloud has been gradually stripped away by the wind, giving rise to a tail in the direction of the wind. 
In GC43\_k3 (top-right), the cloud is undergoing collapse but at a somewhat lower rate than in the $\pow{42}\,\ergs$ case due to the stronger wind and comparatively quicker growth of shear instabilities (such as KH instability). 
Thus, even though the cloud's central region is experiencing collapse, the wind's influence extends to a larger section of the cloud, resulting in a more extended structure compared to the $\pow{42}\,\ergs$ case.

\begin{figure*}
    \centering
    \includegraphics[width=\linewidth]{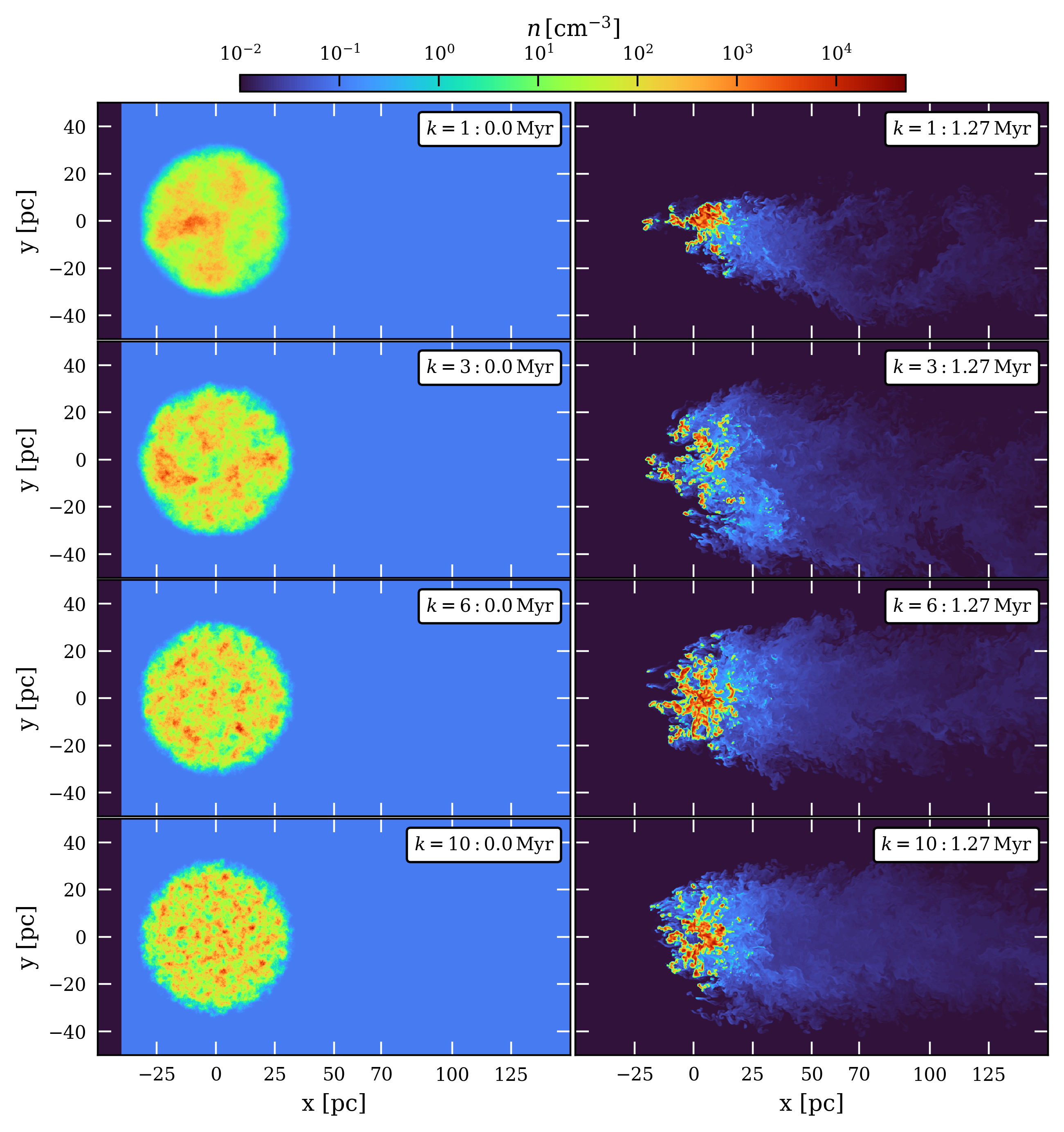}
    \caption{Number density slice through the $z=0$ plane of self-gravity simulations with $\Pout=\pow{43}\,\ergs$ and different cloud porosity: $\kmin = 1$ ($\fst$ row), 3 ($\snd$ row), 6 ($\trd$ row) and 10 ($\fth$ row). The left column shows the initial number density slice of each simulation. The right column represents the distribution at 1.27~Myr, which corresponds to $0.38\,\tcc$ and $0.33\,\tff$.}
    \label{fig:slice_diff_k}
\end{figure*}

The cloud disruption timescales in GC44\_k3 (bottom-left) and GC45\_k3 (bottom-right) are much shorter than the free-fall time due to increased velocity and ram pressure (see Tab.~\ref{tab:sim_list}).
Therefore, a high-power wind induces much stronger turbulence and percolation of gas inside the inhomogeneous cloud, preventing the runaway collapse globally, compared to a low-power wind.
Thus, in this scenario, the cloud is ablated and disrupted significantly before the influence of gravity becomes important.
However, even in the case of a strong wind, the small cloudlets formed by the fragmentation become gravitationally bound and shielded from the external wind, by forming a high-density post-shock outer layer both due to compression and radiative cooling, preventing the wind material from infiltrating these cloudlets.
Therefore, in the absence of any opposing mechanism, very high-density cloudlets collapse locally, while the wind ablates comparatively low-density material, shaping the cloud into an elongated and extended structure of cloudlets.

However, it is important to note that the density contrast between the cloud and wind material in these simulations is very high ($\chi \sim 2\times\pow{4}$), which results in a very high value of the drag timescale ($\tdrag \sim \sqrt{\chi}\tcc$), which is the time by which the cloudlets are expected to attain a similar velocity as the wind.
Therefore, even in scenarios of high power, where the cloud's overall collapse is impeded by the wind, a considerable portion of the fragmented cloudlets persists and remains gravitationally bound to the central potential well due to the inability to attain escape velocity, owing to the large drag timescale.


\subsubsection{Dependence on cloud fractal wavenumber}
\label{sec:diff_k}
The maximum size ($\lambda_{\rm max} \approx L/2\pi\kmin$) of individual cloudlets inside the cloud is parameterized by the minimum wavenumber $\kmin$.
Therefore, a higher value of $\kmin$ results in a larger number of small cloudlets as well as low-density inter-cloudlet channels.
As discussed earlier, the interaction between the wind and a fractal cloud is influenced by the extent to which the wind material can permeate the medium between individual clumps separated by low-density cloud material, in addition to its impact on the fractal surfaces of the cloud.
Thus, one can anticipate that in the case of a higher value of $\kmin$ with a greater abundance of low-density channels, it is easier for the wind to percolate through these channels into the cloudlets and mix with cloud material, thereby transferring energy and momentum.
Additionally, due to the reduced size of the cloudlets in this case, KH instabilities grow faster ($\tkh \propto 1/k$) in a cloud with lower $\kmin$.
Consequently, it is expected that a cloud with a higher $\kmin$ value would undergo more pronounced disruption.

Hence, in order to investigate the impact of the cloud fractal wavenumber, we simulate wind-cloud interaction for clouds with varying values of $\kmin$ ($ = 1, 3, 6, 10$), while keeping the mass and mean density of the clouds fixed. 
The simulations are initiated with self-gravity and a wind of power $\pow{43}\,\ergs$.
In Fig.~\ref{fig:slice_diff_k}, we show the number density slice\footnote{Here we show the density slices instead of the column depth, as presented earlier, to better demonstrate the percolation of the wind within the clouds for different values of the $\kmin$ parameter.} through the $z=0$ plane for these four cases (row-wise) at 0~Myr (left) and 1.27~Myr (right). 

If we first consider the $\kmin=1$ case (`k1', first row), the cloud contains approximately one big cloudlet with increasing density towards its centre, surrounded by low-density outer layers, which are rapidly eroded by the wind through ablation.
Nevertheless, owing to its larger size and the added shielding from compression and cooling, the shock is unable to penetrate deep into the clump, at least not at early times compared to the other $\kmin$ cases. 
Furthermore, external over-pressurization triggers the gravitational collapse of the clump to form a compact, gravitationally bound structure.

In the $\kmin=3$ case (`k3', second row), the cloud is initially composed of many small, dense clumps separated by wide inter-clump channels. 
These channels help the high-velocity wind material to percolate into the cloud, transferring momentum. 
This results in more fragmentation of the cloud, preventing the global collapse, unlike in the $\kmin=1$ case.

Clouds with smaller scale density perturbations as in the $\kmin=6$ (`k6') and $10$ (`k10') cases (third and fourth row), the numbers of cloudlets are much higher, and the inter-cloud channels are also consequently narrower. 
During the initial interaction, the swept-up material by the wind creates a dense layer near the interaction area, blocking the entrance of these narrow channels. This prevents the wind from dispersing the cloudlets efficiently, resulting in the accumulation near the centre.
Thus, external over-pressurization triggers the global collapse of the cloud in these cases similar to the `k1' case.
However, a striking difference between the 'k1' case and the $\kmin=6,10$ cases is evident. 
In the former, cloudlets form through fragmentation and stripping of the large clump. 
In contrast, in the 'k6' and 'k10' cases, cloudlets were initially seeded by the fractal generator itself at the initialisation of the simulation and they remain organized in a more spherically symmetric manner at later times.

\begin{figure*}
    \centering
    \includegraphics[width=\linewidth]{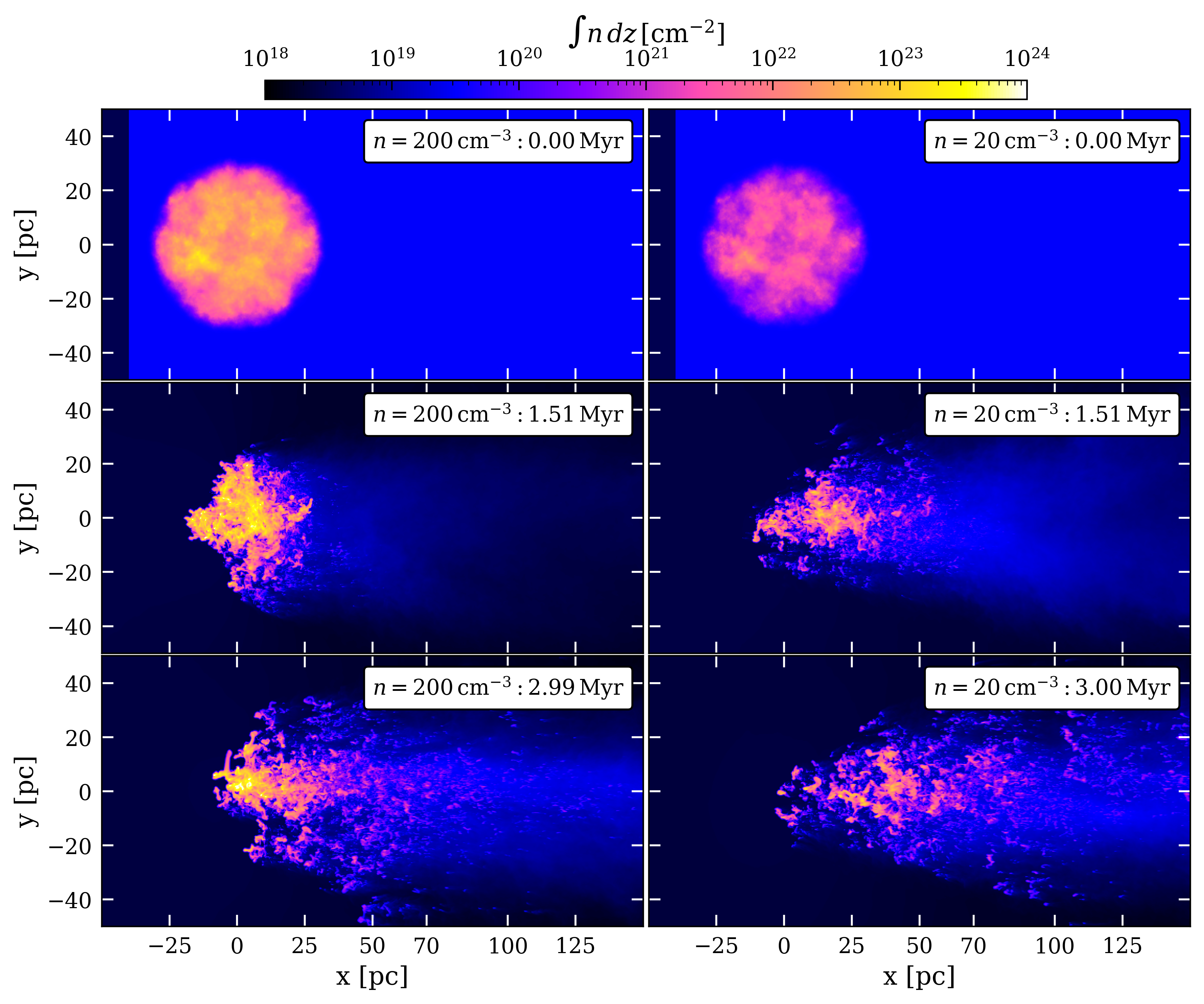}
    \caption{Time evolution of projected number density along the $z$-direction of self-gravity simulations with mean cloud density of $200\,\cc$ (left) and $20\,\cc$ (right). The power of the outflow for both simulations is $\pow{43}\,\ergs$. Here, the lower-density cloud gets quickly disrupted by the wind before self-gravity becomes important compared to the higher-density cloud where we observe a significantly higher fraction of high-density gas.}
    \label{fig:diff_density}
\end{figure*}
%


\subsubsection{Dependence on the mean cloud density}
\label{sec:diff_density}
In order to understand the influence of the density contrast between the cloud and wind, we consider two simulations including self-gravity with identical wind power of $\pow{43}\,\ergs$, but two different values of the initial mean cloud density, namely, $\nc = 200\,\cc$ ($\chi=2\times 10^4$, GC43\_k3) and $20\,\cc$ ($\chi = 2\times 10^3$, GC43\_k3\_low).
Fig.~\ref{fig:diff_density} shows the column density map on the $x\mbox{-}y$ plane of the simulation with $\nc = 200\,\cc$ (left column) and $20\,\cc$ (right column) at different times (row-wise).
We observed that the cloud in GC43\_k3 undergoes gravitational collapse due to the compression from the wind. 
However, in the case of $\nc=20$, the value of $\chi$ is one order of magnitude lower, resulting in a much smaller cloud-crushing time ($\sim 1\,{\rm Myr}$) and drag time ($\sim 44\,{\rm Myr}$). 
Hence, the wind rapidly disintegrates the cloud in the $\nc=20\,\cc$ case, giving rise to a filamentary structure before the gravitational force has a chance to significantly impact the evolution. 
Additionally, due to the relatively shorter drag time, the fragmented cloudlets attain a sufficiently high velocity to effectively overcome the gravitational potential and get dispersed.

\begin{figure*}
    \centering
    \includegraphics[width=\linewidth]{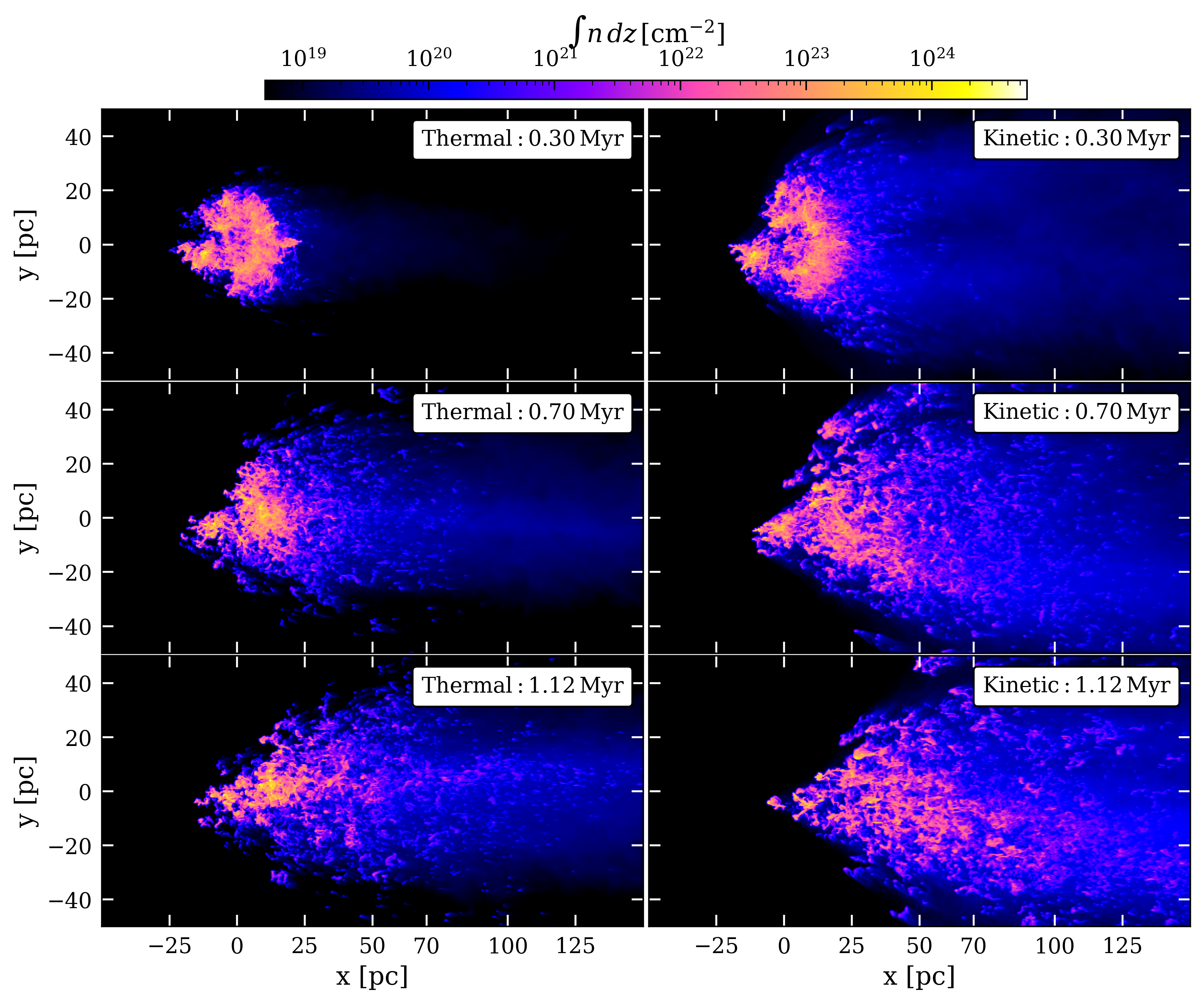}
    \caption{Time evolution (row-wise) of the projected number density along the $z$-direction of self-gravity simulations with thermal (left) and kinetic (right) winds of the same power ($\pow{45}\,\ergs$). The cloud impacted by the kinetic wind is more extended and elongated. Also, there exists a higher fraction of low-density mixed gas (blue colour) in the kinetic wind case.}
    \label{fig:proj_ThermKin}
\end{figure*}
%


\subsubsection{Thermal wind vs.~kinetic wind}\label{sec:wind_kin}
Until now, the energy budget of the wind -- specifically, whether the energy carried by the wind is primarily thermally dominated (subsonic) or kinetically dominated (supersonic) -- has received limited attention in the context of the traditional `cloud-crushing' problem.
Most of the previous studies focus on the impact of a supersonic wind on the cloud, which generally holds true for galactic or starburst-driven winds.
However, as indicated by various theoretical and numerical investigations \citep{King_2003,King_2011,Zubovas_2012,FGQ_2012,Costa_2014}, the progression of AGN-driven winds can encompass various evolutionary stages—such as pressure-dominated and kinetic-energy dominated phases—contingent upon diverse parameters, including the black hole mass, the launch velocity of the wind at the accretion scale, ambient density profile, and various cooling mechanisms, among others \citep[e.g.,][]{FGQ_2012}.
Thus, it is useful to investigate how different kinds of wind, despite having the same power, affect the evolution of the cloud, as pressure and momentum couple differently to the hydrodynamics.

Thus, to examine the effect of thermal vs.~kinetic wind on the evolution of the cloud, we consider two different simulations with the same initial cloud configuration and wind power of $\pow{45}\,\ergs$, but varying wind properties. 
One simulation involves a thermal wind (GC45\_k3), characterized by a Mach number ($\MachWind$) of 0.28 (as used in the previous simulation comparisons in this study so far), while the other initialises a kinetic wind (GC45\_k3\_kinetic, see Tab.~\ref{tab:sim_list}), with a higher velocity ($15000~\kms$) but lower pressure ($10^{-10}~{\rm dyne\,cm^{-2}}$) such that total power is $\pow{45}~\ergs$, resulting in $\MachWind = 12.16$.
Fig.~\ref{fig:proj_ThermKin} displays the column density map in the $x\mbox{-}y$ plane of the simulation with the thermal wind (left column) and the kinetic wind (right column) at different times (row-wise).  
Evidently, the cloud impacted by the kinetic wind undergoes a higher fraction of ablation across all stages of its evolution, compared to the thermal wind's effect.
We observe a significantly larger amount of gas in the low-density tail in the kinetic wind case, which has been stripped from the original cloud by the wind.
Although possessing the same power, the differences in the effect of the thermal and kinetic wind are due to the fact that cloud material is primarily entrained and accelerated by the direct momentum transfer from the wind compared to the $PdV$ work \citep{Wagner_2012,Wagner_2013,Nayakshin_2014}. 
With the thermal wind containing less kinetic energy and consequently, lower ram pressure compared to the kinetic energy-dominated wind, the mixing and acceleration of the cloudlets are significantly lower for the thermal wind case. 
In contrast, due to higher ram pressure, the direct momentum transfer is much higher for the case of the kinetic wind, leading to increased turbulence and vorticity, and causing the cloud to expand kinetically. 
As a result, the kinetic wind, despite possessing the same power, is expected to exhibit more destructive behaviour than a thermal wind with equivalent power.


\subsection{Cloud dynamics}\label{sec:dynamics}
In this section, our main emphasis lies in examining how various initial conditions influence the dynamical evolution of the cloud, i.e., mass loss, gas turbulence, cloud elongation, acceleration, etc.
In order to extract a particular quantity ($\Psi$) for the cloud material from the whole simulation domain, we use the definition of the mass-weighted volume average of $\Psi$ by \citet{Banda_2018},
\begin{equation}\label{eq:def_mass_weight}
    \left\langle\Psi\right\rangle = \frac{\int \Psi \rho C dV}{\int \rho C dV},
\end{equation}
where $\rho$ is the density and $C$ is the cloud tracer (a passive scalar that traces cloud material, i.e., if a cell contains only cloud material then $C=1$; if the cell contains no cloud material $C=0$, such that any mixture of cloud and non-cloud material in a cell can be represented by $C$). 
To exclude significant fractions of wind and hot mixed material from the cloud, we impose a threshold value on the temperature, i.e., only cells with $T<\pow{4}\,{\rm K}$ are used in Eq.~\eqref{eq:def_mass_weight}. 
This gives us a reliable estimate of a particular quantity of cold gas embedded in a hot wind.

\begin{figure}
    \centering
    \includegraphics[width=\linewidth]{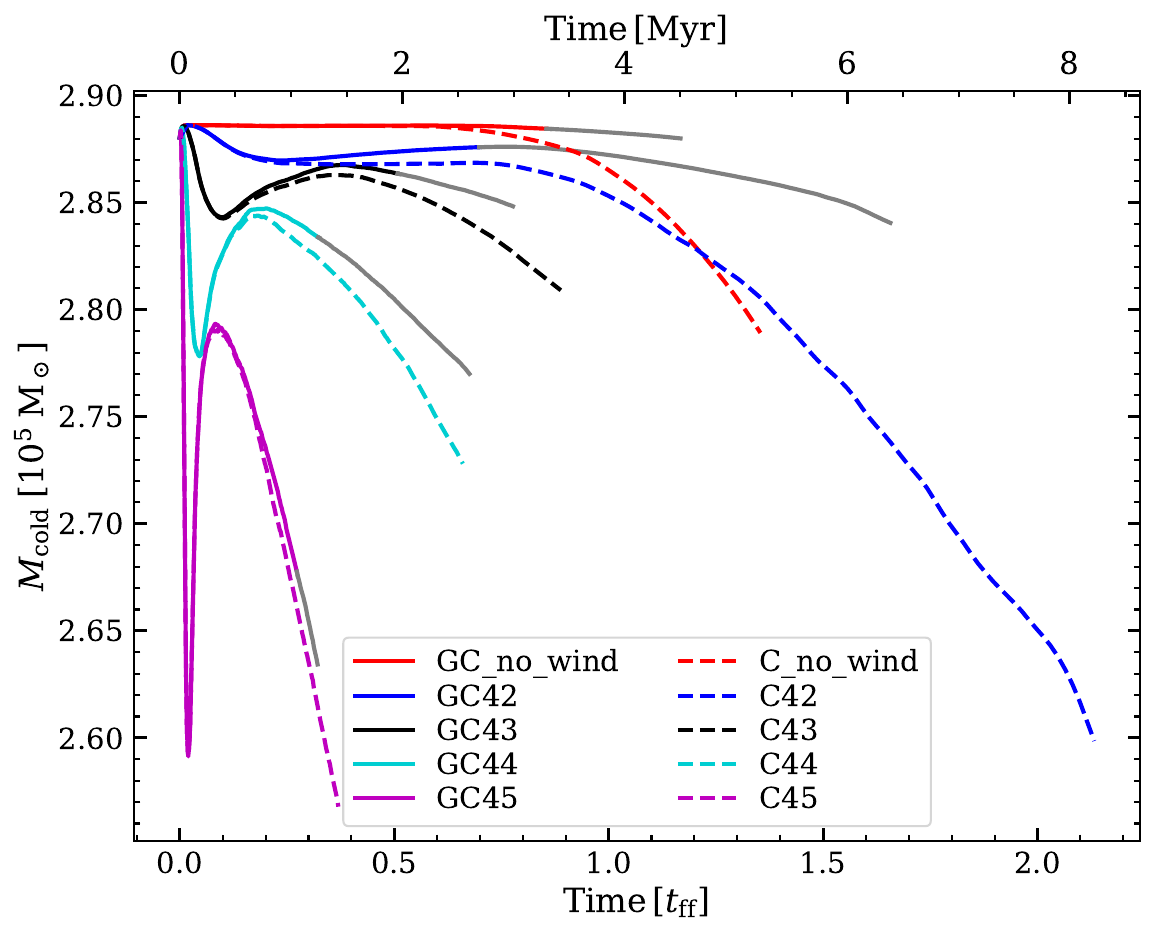}
    \caption{Evolution of the cold gas mass defined in Eq.~\ref{eq:cold_gas} as a function of time for simulations with different wind power, with and without self-gravity. The solid lines correspond to simulations with self-gravity, while the dashed lines with the same colour represent the same wind-power case, but without self-gravity. The red lines are the result of a simulation without any wind. The grey-out portions of the lines indicate the regime in which the Jeans length of the highest-density cell in the simulations is not resolved by at least 4~computational cells.}
    \label{fig:mass_evolve}
\end{figure}

One common convention in all the line plots presented in this section is that the greyed-out parts of the lines (if they exist) correspond to the regime where the Jeans length of the highest density cell is not resolved by four resolution elements (e.g., see Fig.~\ref{fig:mass_evolve}).
This occurs exclusively in the self-gravitating simulations, where few cells can reach very large densities, whose Jeans length can not be resolved by at least 4 computational cells, which is necessary to avoid artificial fragmentation \citep{Truelove_1997}. 

\begin{figure*}
\centerline{
\def\arraystretch{1.0}
\setlength{\tabcolsep}{0.0pt}
\begin{tabular}{lcr}
  \includegraphics[width=0.5\linewidth]{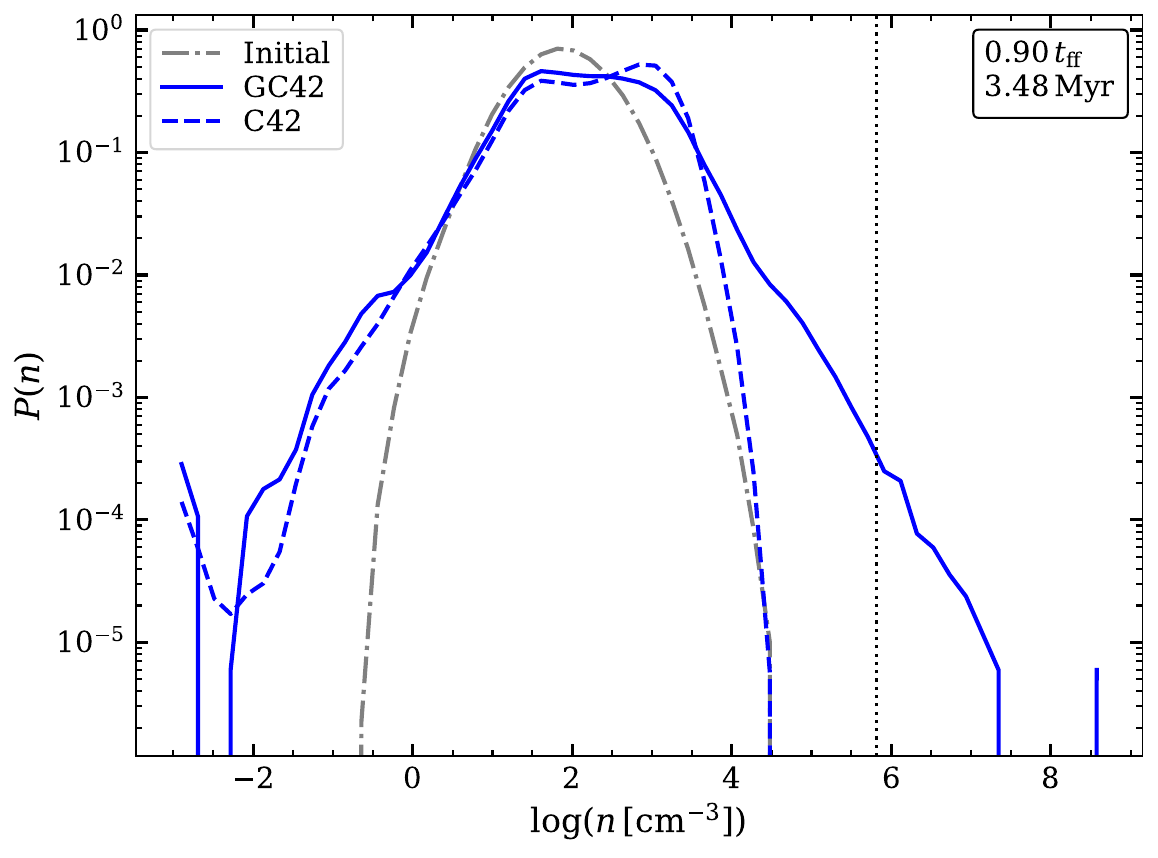} &
  \includegraphics[width=0.5\linewidth]{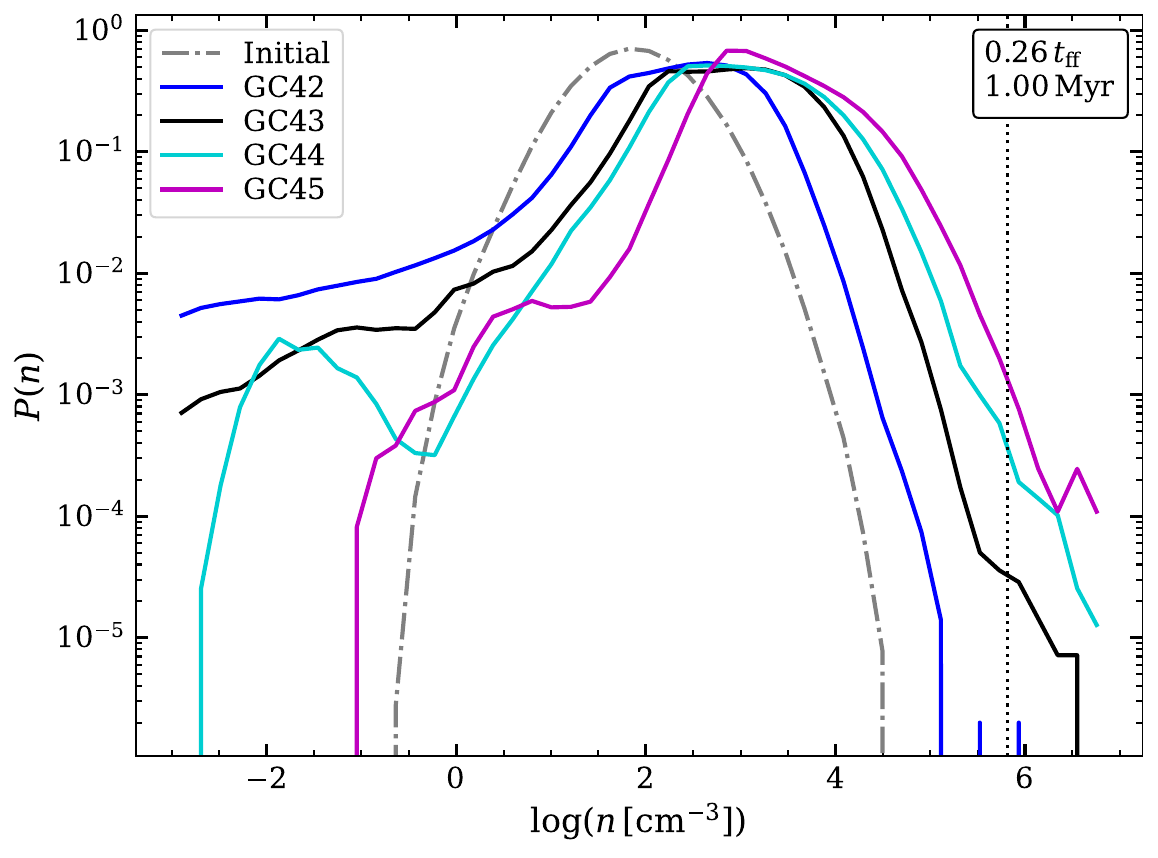}
\end{tabular}}
  \caption{Probability distribution functions (PDF) of log-density of the cloud material for different simulations. Left: PDFs in the simulations with (solid) and without (dashed) self-gravity for the wind power of $\pow{42}\,\ergs$ at $0.9\,\tff$=3.48~Myr. Right: The PDFs of the cloud material in self-gravitating simulations with different wind power at $t=0.26~\tff\approx 1~{\rm Myr}$. The initial density PDF of the cloud material is shown in both panels by the grey dashed-dotted lines. The vertical black dotted lines in both panels mark the density value beyond which the Jeans length cannot be resolved by at least four grid cells with the current computational setup.}
  \label{fig:density_PDF}
\end{figure*}

\subsubsection{Density PDF}\label{sec:density_PDF}
In this section, we investigate the influence of self-gravity and wind power on the distribution of cloud density. 
The left panel of Fig.~\ref{fig:density_PDF} illustrates the number density distribution at $t=0.9\,\tff=3.48\,{\rm Myr}$ in simulations with self-gravity (solid line) and without (dashed line), considering a wind power of $\pow{42}\,\ergs$. 
The grey dashed-dotted line represents the initial probability distribution function (PDF) for the cloud number density. 
Notably, we observe an increase in the low-density tail of the PDFs due to the stripping and mixing of the cloud material with the wind. 
This behaviour remains consistent regardless of the presence of self-gravity.
On the other hand, the wind significantly compresses a substantial portion of intermediate-density gas ($\sim 10-100\,\cc$) to higher densities. 
Interestingly, the highest-density regions, which are surrounded by comparatively lower-density gas, exhibit minimal influence from the wind, as the high-density tail of the PDF in the simulation without self-gravity at 3.48~Myr (blue dashed line) coincides with the initial PDF.
However, in the presence of self-gravity, the densities of these self-gravitating cores increase by accreting lower-density gas from the surroundings, forming an extended, high-density power-law (PL) tail (solid line), which is a prominent feature in turbulent clouds when self-gravity becomes important \citep[e.g.,][]{Klessen_2000,Collins_2011,Collins_2012,Kritsuk_2011,Federrath_2013,Girichidis_2014,Jaupart_2020,Khullar_2021,Appel_2022}.

The right panel of Fig.~\ref{fig:density_PDF} depicts the density PDF of the self-gravitating cloud at $t=0.26\tff=1\,{\rm Myr}$, for simulations with different wind power. 
This time approximately corresponds to the stopping point of the highest power simulation (GC45), where due to increased pressure, the simulation necessitates substantially lower time steps, rendering computations beyond this time prohibitively resource-intensive. 
Nonetheless, due to the shorter cloud-crushing timescales (see Tab.~\ref{tab:sim_list}), all the major evolutionary stages of the clouds are well captured even for the relatively short duration of the simulations.

Notably, as the wind power increases, the compression timescale by the wind decreases, resulting in more rapid growth of the high-density tail of the PDF.
However, at this early stage of the simulations, the influence of self-gravity on the cloud's evolution has not yet given rise to the emergence of the PL tail in the density PDF (see Fig.~\ref{fig:density_PDF_tcc} where we show the density PDF at the same cloud-crushing time, which corresponds to different absolute times for different simulations).
Nonetheless, the mean of the PDF shifts to higher values as the wind power increases due to elevated compression ratios.
Interestingly, for the higher power winds (GC44 and GC45), a secondary peak in the low-density regime of the PDF can be observed (at $n\sim 10^{-2}$ and $1\,\cc$ for GC\_44 and GC\_45, respectively), owing to the presence of a significant portion of hot, diffuse and mixed cloud material which has been stripped off from the cloud by the wind.

%

\subsubsection{Evolution of the cold gas mass}\label{sec:mass_loss}
A commonly employed method for understanding cloud survival is to calculate the evolution of the cloud mass with $n\geq \Bar{n}_{\rm c}/3$, where $\nc$ is the mean number density of the cloud \citep[e.g.,][]{Scannapieco_2015,Gronke_2018}. 
However, as we are interested in the evolution of the cloud gas fraction (which is essential for potential star formation of the cloud), this particular definition might encompass certain portions of mixed and shock-heated dense gas. 
Thus, we define the cold gas mass as
\begin{equation}\label{eq:cold_gas}
    M_{\rm cold} = \int_{T<10^4\,{\rm K}} \rho C dV.
\end{equation}
In Fig.~\ref{fig:mass_evolve}, we show the evolution of the cold gas mass for simulations with different wind power. 
The dashed (solid) lines correspond to simulations without (with) self-gravity.
For comparison, we also present the results of simulations without a wind (red lines).

A general pattern for all the simulations is an initial rapid decline followed by subsequent growth over a brief time span and finally a gradual decline.
During the initial interaction, the wind transfers a significant fraction of thermal energy into the cloud, resulting in an increase in the overall temperature of the cloud, which reduces the amount of cold gas.
The fall-off becomes more pronounced with increasing wind power, as a larger amount of energy is transported into the cloud.
It is important to note that, this decrease in cold gas mass is not due to cloud ablation, which is not significant during this phase.

After the initial interaction, when the shock-compressed gas cools efficiently, the excess internal energy is removed via radiative losses, resulting in the decrease of the mean temperature and the cold gas mass increases to aid the growth phase. 
Nonetheless, the value of the cold cloud mass to which a particular simulation reaches after the initial impact with the cloud decreases with increasing wind power.

Subsequent to the growth phase, the cold gas mass in all the simulations decreases, primarily due to ablation, mixing and removal of the gas from the computational domain by the wind.
As the wind continues to strip the cloud material due to various instabilities and cooling-driven pressure gradients \citep{Gronke_2020}, the cloud material gets mixed and heated by the wind. 
The effectiveness of mixing escalates in tandem with increasing wind power, leading to a quicker decline in the amount of surviving cold gas under the influence of the hot wind.
However, in the simulations without a wind (red lines), the late time decrease in the cold gas mass is due to the escape of a fraction of mass through the computational boundary, which is stirred by the initial random velocity field. 
This is more prominent in the case without self-gravity (red dashed line) due to the absence of attractive gravitational forces.

Interestingly, the simulations with self-gravity (solid lines) retain more cold gas compared to the cases without self-gravity (dashed lines) at the same wind power.
The presence of self-gravity renders the shock-compressed clouds gravitationally bound and more compact, thus, reducing wind's ability to ablate the cloud and induce subsequent mixing. 
Furthermore, the compact cloud provides less surface area for the wind to interact with, compared to that of an expanded cloud in the absence of self-gravity. 
This combination of factors contributes to a higher fraction of cold gas in simulations with self-gravity.
However, the difference between the ensuing evolution in simulations with and without self-gravity becomes less pronounced for high-power winds $(\Pout \gtrsim \pow{44}~\ergs)$. due to the shorter cloud-crushing timescales compared to the freefall time of the cloud.
\begin{figure*}
\centerline{
\def\arraystretch{1.0}
\setlength{\tabcolsep}{0.0pt}
\begin{tabular}{lcr}
  \includegraphics[width=0.5\linewidth]{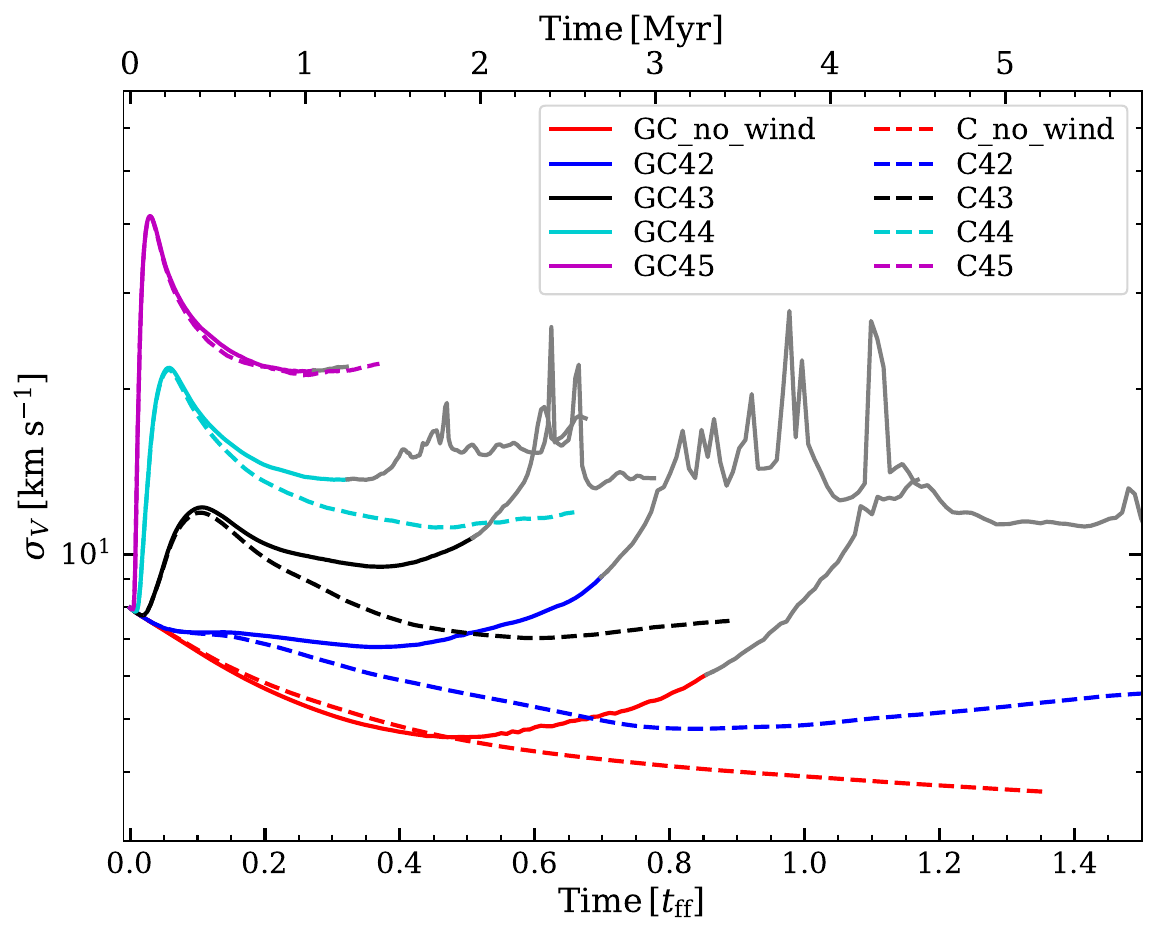} &
  \includegraphics[width=0.5\linewidth]{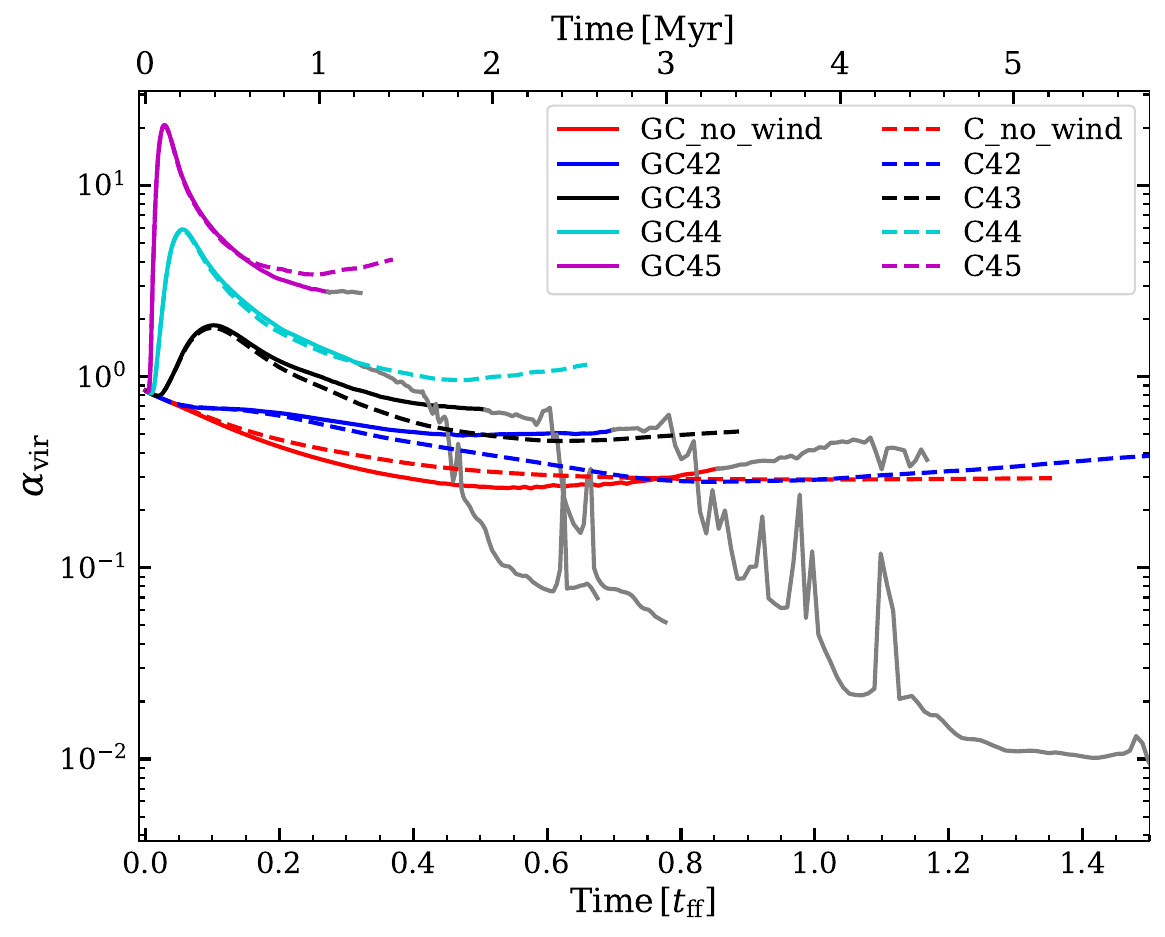}
\end{tabular}}
  \caption{Time evolution of the velocity dispersion (left) and virial parameter (right) as for different simulations. The grey-out parts of the plots correspond to the regime where the Jeans length of the highest-density cell in the simulations is not resolved by at least 4~computational cells.}
  \label{fig:sigma_alpha}
\end{figure*}
%


\subsubsection{Gas turbulence}\label{sec:turbulence}
Quantifying the generation of turbulence inside a cloud impacted by the wind is one of the important factors because this can regulate star formation \citep[e.g.,][]{Federrath_2018}.
As the wind progresses through the fractal cloud, various shear layers are created inside the inter-cloudlet medium, depending on the local density, which leads to the generation of vorticity ($\boldsymbol{\omega}=\curl{v}$).
Additionally, when acting on an inhomogeneous medium, the self-gravitational force causes internal motion between the clumps, creating vorticity inside the cloud \citep{Federrath_2011}, which can further be amplified by the wind.
In order to quantify the turbulence inside the cloud, we define the 3D mass-weighted velocity dispersion ($\sigmav$) as
\begin{equation}\label{eq:sigma_v}
    \sigmav = \sqrt{\sum_{i=1}^3\sigma_{v_i}^2},
\end{equation}
with
\begin{equation}\label{eq:dispersion}
    \sigma_{v_i} = \sqrt{\langle v_i^2 \rangle - \langle v_i \rangle ^2},
\end{equation}
where $\sigma_{v_i}$ is the 1D velocity dispersion along each axis.

The left panel of Fig.~\ref{fig:sigma_alpha} shows the evolution of $\sigmav$ for different simulations with and without self-gravity as a function of $\tff$ (see Fig.~\ref{fig:turbulence_tcc} for the evolutionary trends in terms of the cloud-crushing time in Appendix~\ref{sec:plots_tcc}).
Irrespective of the wind power, the velocity dispersion in all the wind simulations increases due to energy transfer from the wind to the cloud material.
Furthermore, the magnitude of the enhancement increases with increasing wind power.
In all cases, there is an initial enhancement of the velocity dispersion during the compression phase. 
Consequently, in the simulations without self-gravity, when the wind disperses the cloud, it becomes comparatively easier for the wind to traverse through the inter-cloudlet channels. 
Therefore, the velocity dispersion settles to an approximately constant value, depending on the strength of the wind.
In the cases where self-gravity is included, the evolution deviates when gravity starts to dominate. 
In these instances, the fragmented cloudlets are pulled to the central gravitational potential well, towards the core, which induces additional motions between the cloudlets, resulting in a higher velocity dispersion compared to the cases without self-gravity.
It is worth mentioning that, as the wind power increases, the disparity in velocity dispersion evolution between simulations with and without self-gravity diminishes, as a stronger wind disperses the cloud before gravity can significantly influence its evolution.

To thoroughly gauge the wind's influence on star formation, it's inadequate to solely consider velocity dispersion. 
The crucial factor lies in the balance between turbulent kinetic energy ($E_{\rm kin}$) and gravitational binding energy ($E_{\rm grav}$), which determines whether a gas cloud experiences runaway collapse or maintains stability through turbulent support.
Thus, we consider the virial parameter ($\alphavir$) of the cloud, defined as the ratio between the kinetic and gravitational energy \citep{Bertoldi_1992,Federrath_2012},
\begin{equation}\label{eq:alpha_vir}
    \alpha_{\rm vir} = \frac{2 E_{\rm kin}}{|E_{\rm grav}|},
\end{equation}
where $E_{\rm kin}$ and $E_{\rm grav}$ are defined as
\begin{gather}
    E_{\rm kin} = \frac{1}{2} M_{\rm cold} \sigmav^2 \\
    E_{\rm grav} = \int_{T<10^4\,{\rm K}} \Phi \rho C dV,
\end{gather}
where $\Phi$ is the gravitational potential.
In the simulations without self-gravity, we also solve for the gravitational potential at runtime, but do not couple it to the hydrodynamics.

The right-hand panel of Fig.~\ref{fig:sigma_alpha} shows the evolution of the virial parameter with time.
In most of the simulations, except for the cases with wind power $\pow{42}\,\ergs$, the virial parameter initially increases beyond unity because of the infusion of kinetic energy from the wind, which initially surpasses the overall gravitational energy. 
For the highest power simulation (GC45), this increase in $\alphavir$ is more than one order of magnitude ($\sim 20$) higher than the initial value of $\sim 0.9$.
Nevertheless, following the compression phase, there is a rise in the average cloud density, intensifying the gravitational potential well, ultimately leading to a reduction in the value of $\alphavir$.
In the simulations without self-gravity, the subsequent evolution becomes nearly steady-state. 
This suggests that the kinetic and gravitational energy maintain a delicate balance, remaining relatively constant over time.
In the presence of self-gravity, the highly dense collapsing cores give rise to a very deep gravitational potential, leading to the domination of gravitational binding energy over the kinetic energy. 
This results in a very low value of $\alphavir$ below 1, which is particularly prominent in low-power wind cases.

It is important to acknowledge that due to the absence of a sink particles algorithm \citep{Federrath_2010} in our simulations, we cannot take advantage of removing gas from a cell whose local Jeans length becomes unresolved by the computational grid \citep{Truelove_1997}.
Therefore, in scenarios where a region experiences runaway collapse, the density within the specific cell at the bottom of the gravitational potential well becomes exceptionally high as it accumulates matter.
This situation results in a violation of the \citet{Truelove_1997} criterion, which suggests that the Jeans length of any region should be resolved by at least 4~grid cells to prevent artificial fragmentation.
Nonetheless, even if we do not resolve the small-scale structures, the qualitative behaviour in the virial parameter will not change significantly as the potential well created by this very high-density region will still be very deep, even in higher-resolution simulations.

\begin{figure}
    \centering
    \includegraphics[width=\linewidth]{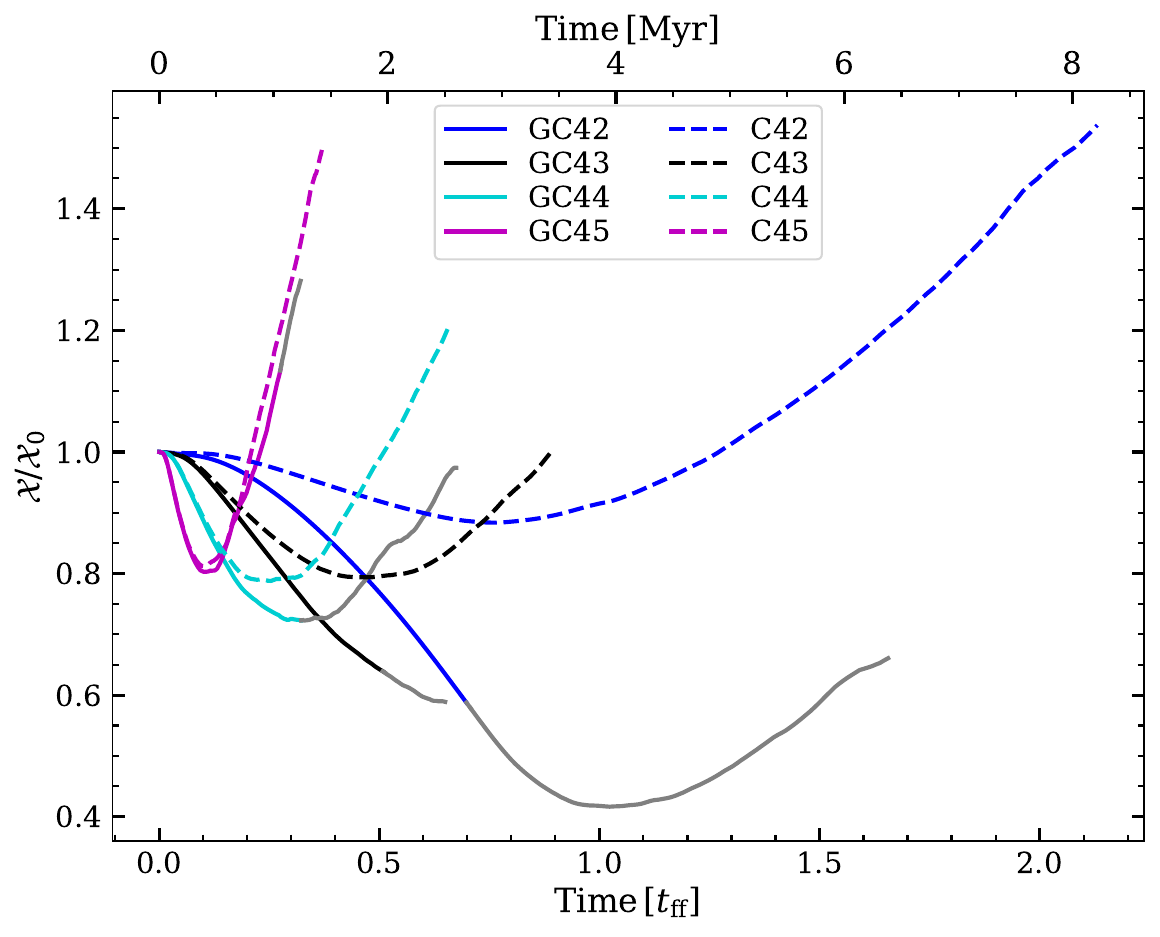}
    \caption{Time evolution of the longitudinal extent (Eq.~\ref{eq:effective_rad}, normalized to the initial value) of the cloud for different simulations. We see that, in the presence of self-gravity, the clouds are more compact.}
    \label{fig:size_evolve}
\end{figure}
%

\subsubsection{Cloud elongation}\label{sec:elongation}
Here, we discuss the role of self-gravity on the compression and subsequent elongation of the cloud, impacted by the wind, from a quantitative point of view.
We calculate the effective length ($\mathcal{X}$) of the cloud along the $x\mbox{-}$direction (i.e., along the direction of the wind) as follows \citep[see][]{Klein_1994, Banda_2018}:
\begin{equation}\label{eq:effective_rad}
    \mathcal{X} \propto \sqrt{\left( \langle x^2 \rangle - \langle x \rangle^2 \right)},
\end{equation}
where Eq.~\eqref{eq:def_mass_weight} is used to calculate the average quantities.
Fig.~\ref{fig:size_evolve} shows the evolution of the cloud length (normalized to the initial value), parallel to the wind ($x$-direction), for different simulations.
The solid and dashed lines correspond to simulations with and without self-gravity, respectively.
We see that in all scenarios, the cloud width initially decreases as a result of the compression. 
Consequently, the cloud undergoes elongation due to the combined effects of shock- and turbulence-induced expansion.
In the absence of self-gravity, this elongation is much more prominent for the lower power winds, as depicted by the blue dashed line (C\_42) in Fig.~\ref{fig:size_evolve}.
In contrast, due to the global collapse of the clouds in the presence of self-gravity in cases with lower wind power, the clouds are more compact leading to a smaller longitudinal width as can be seen from the solid blue line (GC\_42).
Nevertheless, as the wind power increases, the distinction between these two scenarios becomes less pronounced, as the wind initiates the stripping of the cloud before gravity has a significant influence.
\begin{figure}
    \centering
    \includegraphics[width=\linewidth]{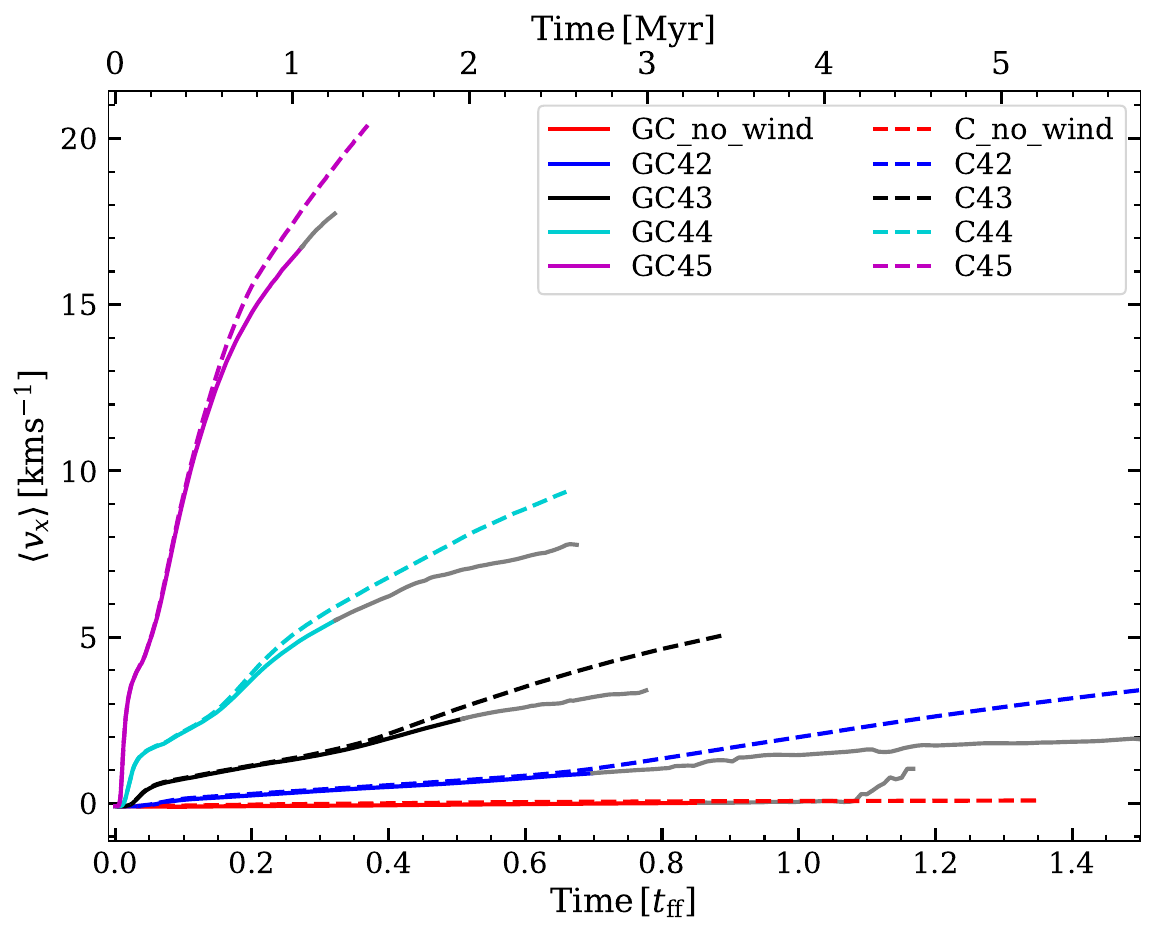}
    \caption{Time evolution of the $x$-component of the centre of mass velocity of the cloud in different simulations as indicated in the legend.}
    \label{fig:vCOM_evolve}
\end{figure}
%

\subsubsection{Cloud acceleration}\label{sec:acceleration}
Fig.~\ref{fig:vCOM_evolve} illustrates the evolution of the $x$-component of the cloud's centre of mass velocity across various simulations. It's evident that higher-powered winds quickly accelerate the clouds due to a greater net momentum transfer from the wind to the cloud.
Notably, despite a similar initial evolution, clouds in simulations incorporating self-gravity at the same wind power begin to decelerate (solid lines) as gravitational forces toward the central potential well counteract the wind-induced acceleration.
It's essential to note that, within the chosen simulation parameters, acceleration timescales significantly exceed the duration of the simulations themselves. 
As a result, the net velocity achieved by the cloud's centre of mass remains significantly lower than the flow velocity. 
Nevertheless, the discernible impact of self-gravity on the cloud's acceleration is clearly evident.

\begin{figure}
    \centering
    \includegraphics[width=\linewidth]{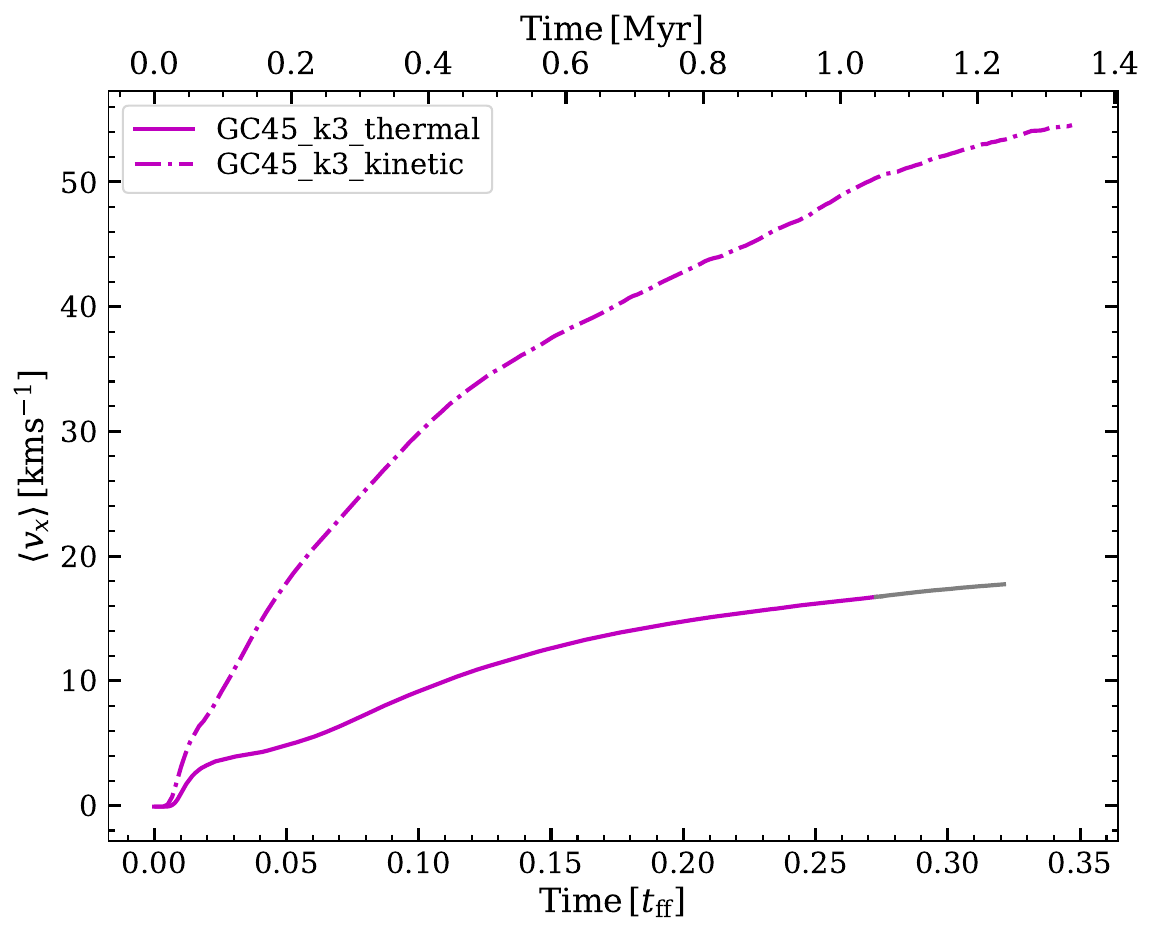}
    \caption{Same as Fig.~\ref{fig:vCOM_evolve} but for a simulation with a thermal wind (GC45\_k3\_thermal; solid line) compared to a kinetic wind (GC45\_k3\_kinetic; dashed-dotted line), with the same wind power ($\pow{45}\,\ergs$).}
    \label{fig:vCOM_kin_therm}
\end{figure}

In Fig.~\ref{fig:vCOM_kin_therm}, we depict the cloud's net centre of mass velocity for GC45\_k3\_thermal (solid line) and GC45\_k3\_kinetic (dashed-dot line) to analyze the impact of wind energy composition on cloud acceleration. As anticipated, the cloud exposed to the kinetic wind demonstrates significantly higher velocities compared to the thermal wind of equal power. This difference arises from the more efficient direct momentum transfer in the kinetic wind, contrasting with the thermal wind where injected internal energy dissipates rapidly due to strong radiative loss, diminishing momentum transfer efficiency.

\begin{figure*}
  \centering
  \includegraphics[width=\linewidth]{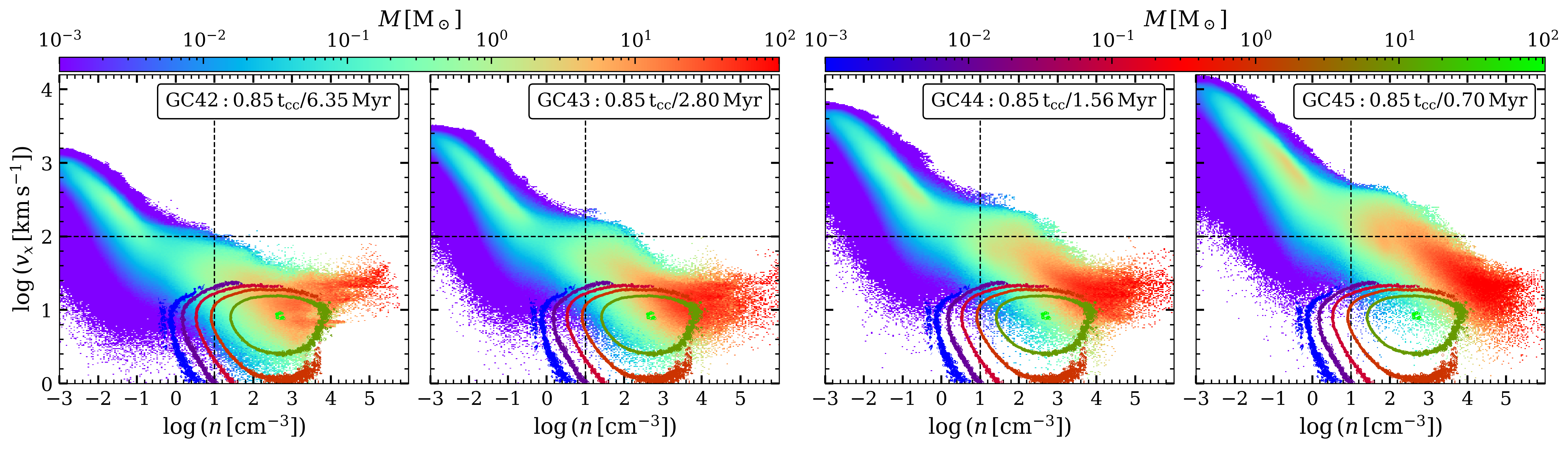}
  \caption{ass-weighted 2D histogram of velocity vs.~density at $0.85~\tcc$ for self-gravity simulations with different wind power. The physical time for each simulation is indicated in the legends. The contours in each panel represent the initial distribution. The vertical dashed line indicates the gas density of $10~\cc$, while a horizontal line marks $v_x = 100~\kms$.}
  \label{fig:n_vel}
\end{figure*}
\begin{figure*}
    \centering
    \includegraphics[width=\linewidth]{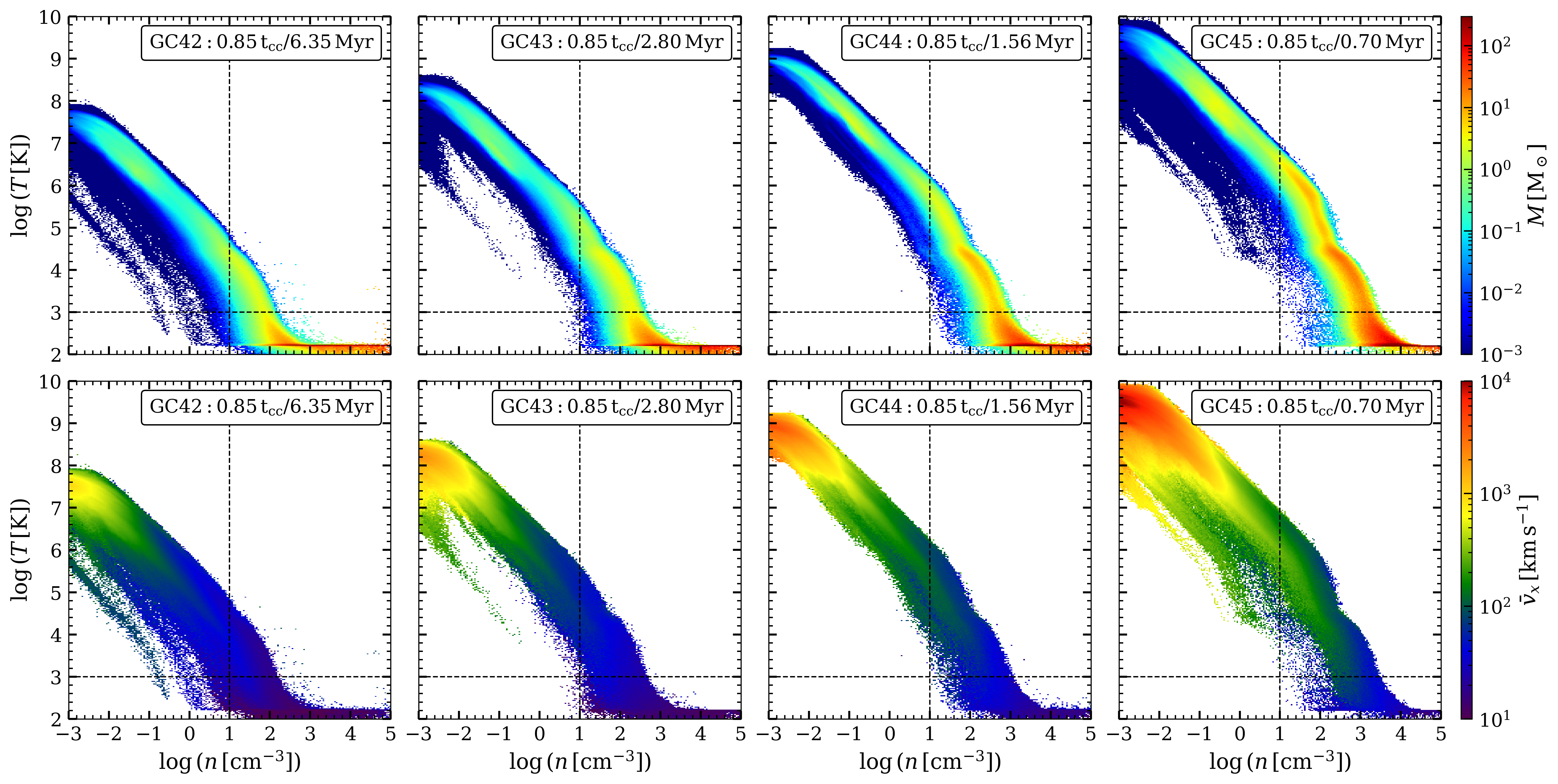}
    \caption{2D histograms of temperature vs.~density of the cloud material for the self-gravitating simulation of different powers at $0.85~\tcc$, which corresponds to different physical times for different wind power as indicated in the legends. The colour bars in the top and bottom rows represent respectively, the mass and mass-weighted mean velocity of the cloud material along the direction of the wind. The vertical dashed line indicates the gas density of $10~\cc$, while a horizontal line marks $T=1000~{\rm K}$.}
    \label{fig:phase_mass_vel}
\end{figure*}

\begin{figure*}
\centerline{
\def\arraystretch{1.0}
\setlength{\tabcolsep}{0.0pt}
\begin{tabular}{lcr}
  \includegraphics[width=0.5\linewidth]{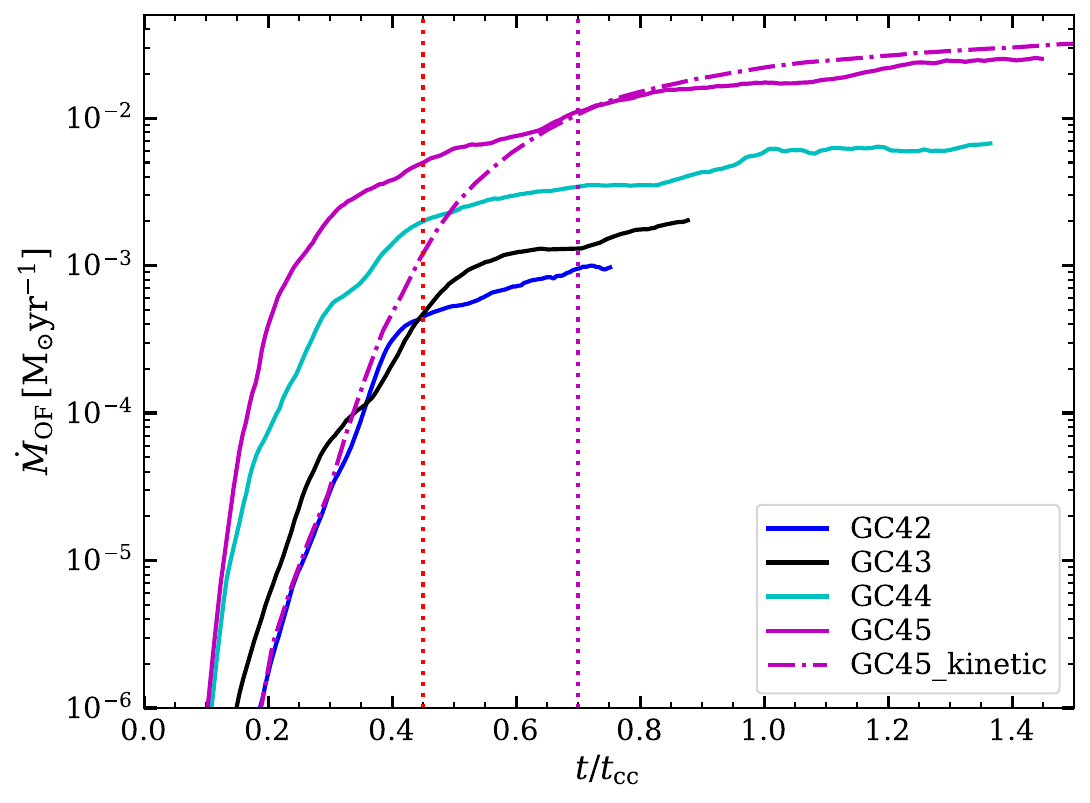} &
  \includegraphics[width=0.5\linewidth]{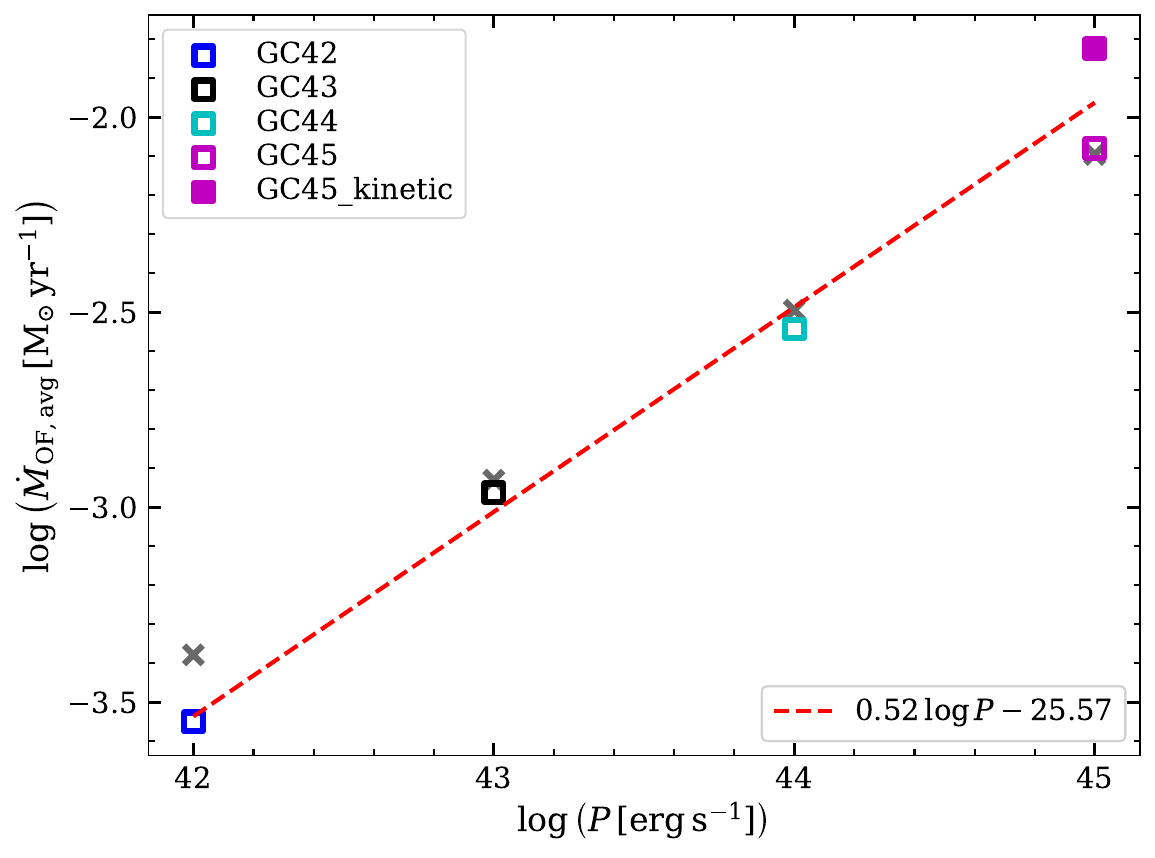}
\end{tabular}}
  \caption{Left: The time evolution of mass outflow rates ($\Mdot$) through $x = 75~\pc$ surface for self-gravitating simulations with different powers as indicated in the legend. Right: The average mass outflow rate ($\Mdotavg$) as a function of wind power. These $\Mdotavg$ values are determined after reaching a near-steady state, marked by vertical dotted lines in the left panel (red for thermal winds and magenta for kinetic winds). The grey cross marks denote values for non-self-gravitating simulations at corresponding wind powers. Additionally, the red dashed line represents a power law fit to $\Mdotavg$ obtained from the self-gravitating runs.}
  \label{fig:outflow_scaling}
\end{figure*}

\subsection{Multi-phase outflow}\label{sec:outflow}
While the findings presented in Sec.~\ref{sec:acceleration} provide insights into the acceleration of the ``cloud as a whole'' by the wind, it becomes increasingly evident that the definition of the cloud as a single object diminishes as the wind fragments the initial cloud within a spherical zone into numerous smaller pieces. 
This effect is particularly prominent for a fractal cloud, which is the case in this study.
The fragmented and stripped cloud material with different densities and temperatures are entrained in the hot wind, resulting in a multiphase outflow that can extend up to a few kpc from the centre of a galaxy, with velocities from a few 100 to several 1000~$\kms$.
These kinds of ionized \citep{Nesvadba_2006,Nesvadba_2008,Holt_2008,Westmoquette_2012,Harrison_2014,Rupke_2017}, neutral atomic \citep{Morganti_2005,Nesvadba_2010,Guillard_2012,Cazzoli_2016,Morganti_2016}, and molecular \citep{Feruglio_2010,Cicone_2014,Fiore_2017,Fluetsch_2019} outflows in galaxies, resulting from AGN, have been observed and characterized by a plethora of observational studies \citep[see][for a review]{Morganti_2017}.
Moreover, numerical simulations of the interaction between relativistic jets and interstellar clouds have demonstrated that atomic gas can form in-situ inside the post-shock material which cools rapidly, at least for low-power jets \citep[e.g.,][]{Perucho_2024}.
Thus, from a theoretical point of view, it is important to investigate and quantify how AGN-driven winds with different powers lead to such multi-phase outflows. 

\subsubsection{Velocity distribution}
Fig.~\ref{fig:n_vel} presents the mass-weighted 2D histogram of velocity vs.~number density for the cloud material, for four self-gravitating simulations with different wind power at $0.85~\tcc$, which corresponds to different physical times for different power as indicated in the legends.
The contours in each panel represent the initial distribution of the cloud material in the density-velocity plane.
The impact of the wind on the cloud can be readily seen from the diagram, particularly in the high-power cases (GC44, GC45), where a significant portion of the cloud material has been compressed to higher densities and accelerated to higher velocities. 
The velocities of the dense gas with number densities of $\sim 10-100\,\cc$ (right side of the vertical line) is enhanced up to $\lesssim 400~\kms$, which constitutes a considerable fraction of the cloud's mass.
A notable amount of diffuse gas with densities of $\sim 0.01\mbox{-} 1\,\cc$ is accelerated to velocities of several $1000~\kms$.
However, the velocities of the very dense gas ($n\gtrsim 10^3\,\cc$) are only mildly affected by the wind, as the gas with larger column densities is more resistant to acceleration through direct momentum transfer.

\subsubsection{Multiphase structure}
To analyze the gas phases of the outflowing material, we have constructed phase diagrams representing gas temperature versus number density in Figure~\ref{fig:phase_mass_vel}, where the colourbar in the top and bottom rows show the total mass and the mass-weighted mean velocity of the cloud material, respectively.
From the top row, it is evident that most of the cloud material is in the cold phase ($T\lesssim10^{3}~{\rm K}$) for the winds with lower powers, whereas there exists a notable amount of moderately dense gas ($10~\cc \lesssim 100~\cc$) in warm ($10^{3}~{\rm K}\lesssim T \lesssim 10^{4}~{\rm K}$) and hot ($10^{4}~{\rm K}\lesssim T \lesssim 10^7~{\rm K}$) phases in the higher-power cases.  Cloud materials in these phases experience acceleration up to $400~\kms$. Interestingly, comparatively diffused gas with densities ($0.1-10~\cc$), which notably contributes to the ionized outflows, exists mostly in the hot phase, with velocities up to several $1000~\kms$. On the other hand, gas in the dense and cold phase exhibits very low velocities, less than $100~\kms$, even in the highest power case. This suggests that accelerating dense, cold interstellar gas to high velocities ($\sim 1000~\kms$) via AGN-driven winds is exceedingly challenging.

In the simulations featuring lower-power winds (GC42 and GC43) shown in Fig.\ref{fig:n_vel}, the presence of high-velocity gas is notably limited, due to the low ram pressure of the wind, which primarily accelerates the cloud.

\subsubsection{Outflow rate scaling with wind power}
In order to quantify the outflow driven by the winds and its dependence on the wind power, we calculate the total mass-outflow rates ($\Mdot$) of the cloud material as a function of time for self-gravitating simulations with different wind power, including the simulation with the kinetic wind (GC45\_kinetic), which is illustrated in the left panel of Fig.~\ref{fig:outflow_scaling}. 
The outflow rates are calculated across the $y\mbox{-}z$ plane at $x=75\,{\rm pc}$. 
In the global picture, this corresponds to the outflow rate through an area of $A = 100~\pc\times100~\pc$ from a single cloud situated at a distance $\ROF\sim 1.075\,{\rm kpc}$ from the AGN as per our configuration.
Notably, the mass outflow initiates after the compression phase when the ablation and stripping of the cloud start to dominate the evolution and progressively approaches a near-steady state value over time.
The amount of outflow increases with increasing wind power, as expected.
Furthermore, during the steady-state phase, the outflow rate induced by a kinetic wind (dashed-dotted magenta) surpasses that of a thermal wind (solid magenta) with equivalent power due to the higher momentum transfer efficiency.
It is important to note that, the mass-outflow rates, calculated at $x=75~\pc$, are entirely contributed by the stripped and entrained materials from the cloud.

To establish the relationship between the mass outflow rate and wind power, we determine the average mass outflow rate ($\Mdotavg$) in each simulation after $\Mdot$ reaches a near-steady state, as indicated by the dotted lines (red for thermal and magenta for kinetic wind) in the left panel, which is plotted against the wind power in the right panel of Fig.~\ref{fig:outflow_scaling}. 
The open squares correspond to $\Mdotavg$ values for self-gravitating simulations with different powers, as indicated in the legend.
Additionally, the grey cross marks represent values obtained from non-self-gravitating simulations with corresponding wind powers.

From the right panel of Fig.~\ref{fig:outflow_scaling}, it is evident that the winds with power in the range $\pow{42}-\pow{45}\,\ergs$ cause mass-outflow rates of $\sim 10^{-4}-10^{-2}\,{\rm \Msun\,yr^{-1}}$ from a single cloud through an area of $A = 100\pc\times 100\pc$.
If we consider multiple clouds are distributed around the AGN globally, the global mass-outflow rates ($\Dot{M}_{\rm OF, glob}$) at a distance of $\ROF$ from the AGN can be calculated as,
\begin{equation}\label{eq:mdot_glob}
    \Dot{M}_{\rm OF, glob} \approx f_{\rm V} \Mdotavg\times \frac{4\pi \ROF^2}{A}\left(\frac{\Omega}{4\pi}\right),
\end{equation}
where $f_{\rm V}$ is the volume filling factor of the dense cloud, typically of the order of $0.01-0.1$ \citep{Avillez_2004,Mukherjee_2018} and $\Omega$ is the solid angle covered by the outflowing region. For a spherical outflowing region with a maximum outflow rate $\Omega = 4\pi$, and $f_{\rm V} = 1$. Thus, Eq.~\ref{eq:mdot_glob} becomes $\Dot{M}_{\rm OF, glob} \approx \Mdotavg\times 4\pi R_{\rm OF}^2/A$. 
Therefore, assuming a simple spherical arrangement of the clouds, the global mass outflow rates at a distance of $\ROF\sim 1.1\kpc$ from the AGN are to the order of $0.1-10\,\rm{\Msun\,yr^{-1}}$  for the considered wind powers in this study.
These values are similar to what has been found in observations \citep[see][and references therein]{Fiore_2017}.

A noteworthy feature in Figure~\ref{fig:outflow_scaling} is the tight correlation between the mass-outflow rate and the wind power. 
As the power of the wind increases, there is a corresponding increase in the mass-outflow rate, a trend that is in agreement with our intuitive expectations.
Fitting a simple power law ($\Mdotavg\propto P^\kappa$) to the results from the self-gravitating simulations (indicated by the red dashed line) yields a power-law exponent of $\kappa \sim 0.52$.
However, from the left panel, it is important to note the outflow rates for low-power winds (GC42 and GC43) have not reached a steady state by the end of the simulation and are expected to rise further before becoming constant.
Therefore, the calculated average mass outflow rates for the lower power winds presented in the right panel serve as lower bounds. 
Thus, the value of $\kappa \sim 0.52$ represents the upper bound of the power-law exponent estimated in this study.

We present a simple complementary picture, where the mass outflow, driven by the AGN, is due to the stripped-off material from the embedded clouds inside the flow, rather than the swept-up shell model \citep[e.g.,][]{FGQ_2012,King_2015}.
We assume that the mass loss due to the ablation of the cloud by the hot wind is driven by the pressure gradient at the interface \citep[see][]{Harquist_1986}.
For a purely hydrostatic ablation, the ablation rate ($\Dot{M}_{\rm abl}$) of one cloud embedded in a hot wind of Mach number $\MachWind$ is given as \citep{Harquist_1986,Wagner_2012},
\begin{equation}\label{eq:mdot_ablation_mach}
    \Dot{M}_{\rm abl} = \alpha \min \left(1,\MachWind^{4/3}\right)(M_{\rm c}c_{\rm s,c})^{2/3}(\rho_{\rm w} \vwind)^{1/3},
\end{equation}
where $M_{\rm c}$ and $c_{\rm s,c}$ are the mass and internal isothermal sound speed of the cloud and $\alpha$ is a constant factor of the order of unity for a spherical cloud.
Keeping the cloud configuration identical, the mass ablation rate depends on the wind velocity as 
\begin{equation}\label{eq:mdot_abalation_velocity}
    \Dot{M}_{\rm abl} \propto \vwind^{5/3}.    
\end{equation}
Thus, by combining Eq.~\eqref{eq:mdot_abalation_velocity} and Eq.~\eqref{eq:wind_velocity_norm}, we can show that,
\begin{equation}
    \Dot{M}_{\rm abl} \propto \Pout^{5/9},
\end{equation}
where $\Pout$ is the power of the wind.
Therefore, if we assume the ablated mass is the source of the majority of the mass-loading in the outflow, then the outflow rate depends on the wind power as $\Mdot \propto \Pout^{5/9} \propto \Pout^{0.55}$, which is close to the value we are estimating from the simulations.

It is worth noting that similar correlations between the outflow rates and the bolometric luminosity of the AGN have been consistently reported in numerous observational investigations, spanning both the ionized and cold molecular phases. 
For instance, studies focusing on ionized outflows have documented a power-law exponent of about $\kappa\sim 1.29$ \citep{Fiore_2017}, whereas the exponent for cold molecular outflows varies between $\kappa\sim 0.68-0.76$ in different instances \citep{Cicone_2014,Fiore_2017,Fluetsch_2019}.

While our analysis does not differentiate between different phases of the outflow, the scaling relation we estimate for the overall mass-outflow rate, encompassing both the hot, warm, and cold phases, as a function of wind power, is fairly close to the scaling relations to the molecular outflow.
However, it's crucial to emphasize that despite the observed strong correlation between the average mass outflow rate ($\Mdotavg$) and wind power as found in this study, there exist several sources of scatter contributing to this correlation. 
For instance, a cloud impacted by a kinetic wind tends to show a higher mass outflow rate than in the case of a thermal wind with similar power, as illustrated by the magenta squares in the right panel (filled for kinetic and open for thermal). 
Additionally, factors such as the morphology (i.e., whether the cloud is porous or compact) of the cloud and whether self-gravity plays a dominant role in the cloud's evolution at the impact stage can introduce considerable scatter to the correlation.
Moreover, on a global scale the geometric distribution (i.g., spherical, disk-like, etc.) of clouds around the driving source, as represented by the $f_{\rm V}$ factor in Eq.~\eqref{eq:mdot_glob}, can also cause significant scatter in the global mass outflow rate.
Furthermore, comparing outflow rates for different systems with the same power but at different evolutionary stages can also introduce scatter as the temporal evolution of $\Mdot$ can vary significantly (left panel of Fig.~\ref{fig:outflow_scaling}).
Therefore, acknowledging these factors while interpreting the correlation between the mass outflow rate and wind power is crucial. However, it is noteworthy that our simplified estimates do present qualitative similarities to the observed results, reinforcing the idea that an ablation-based model of mass loss can likely explain the observed correlations.

\begin{figure*}
\centerline{
\def\arraystretch{1.0}
\setlength{\tabcolsep}{0.0pt}
\begin{tabular}{lcr}
  \includegraphics[width=0.5\linewidth]{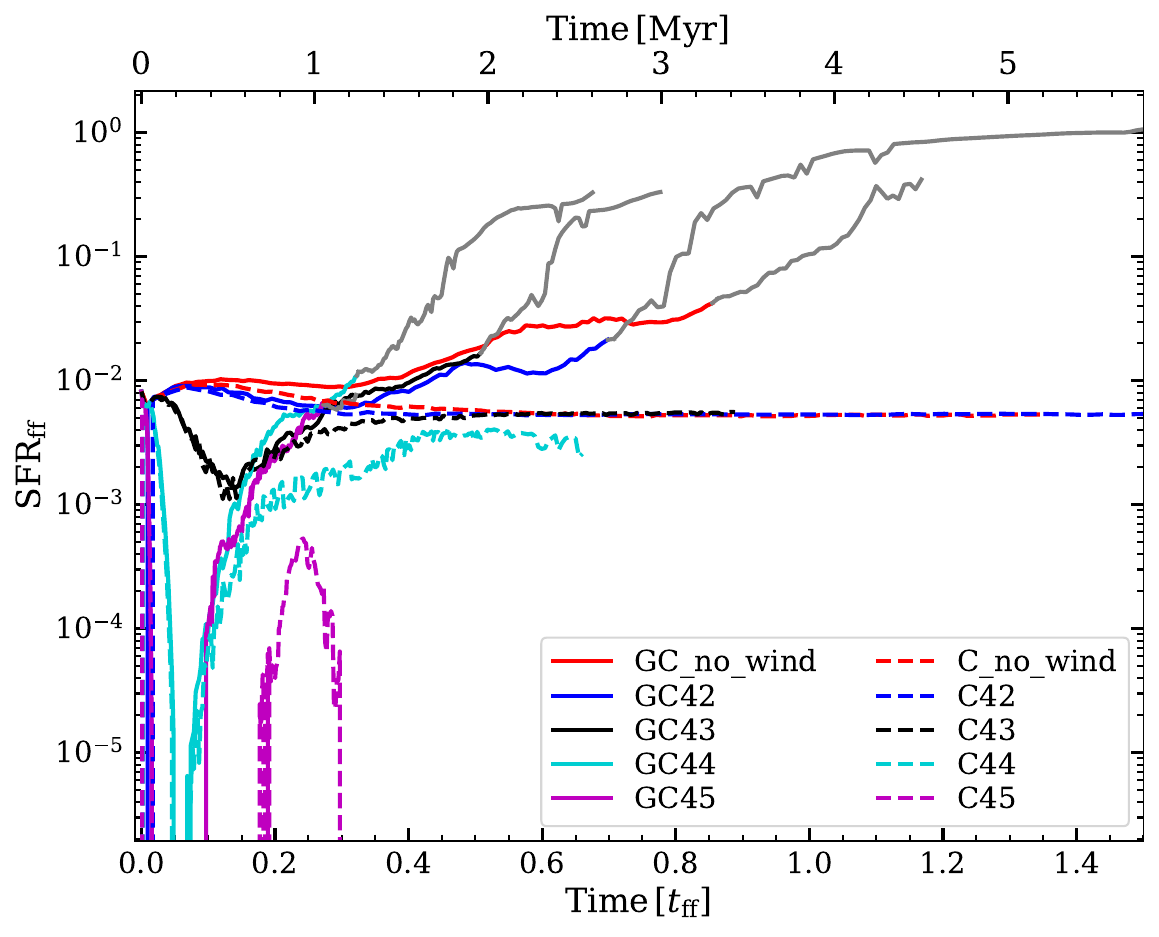} &
  \includegraphics[width=0.5\linewidth]{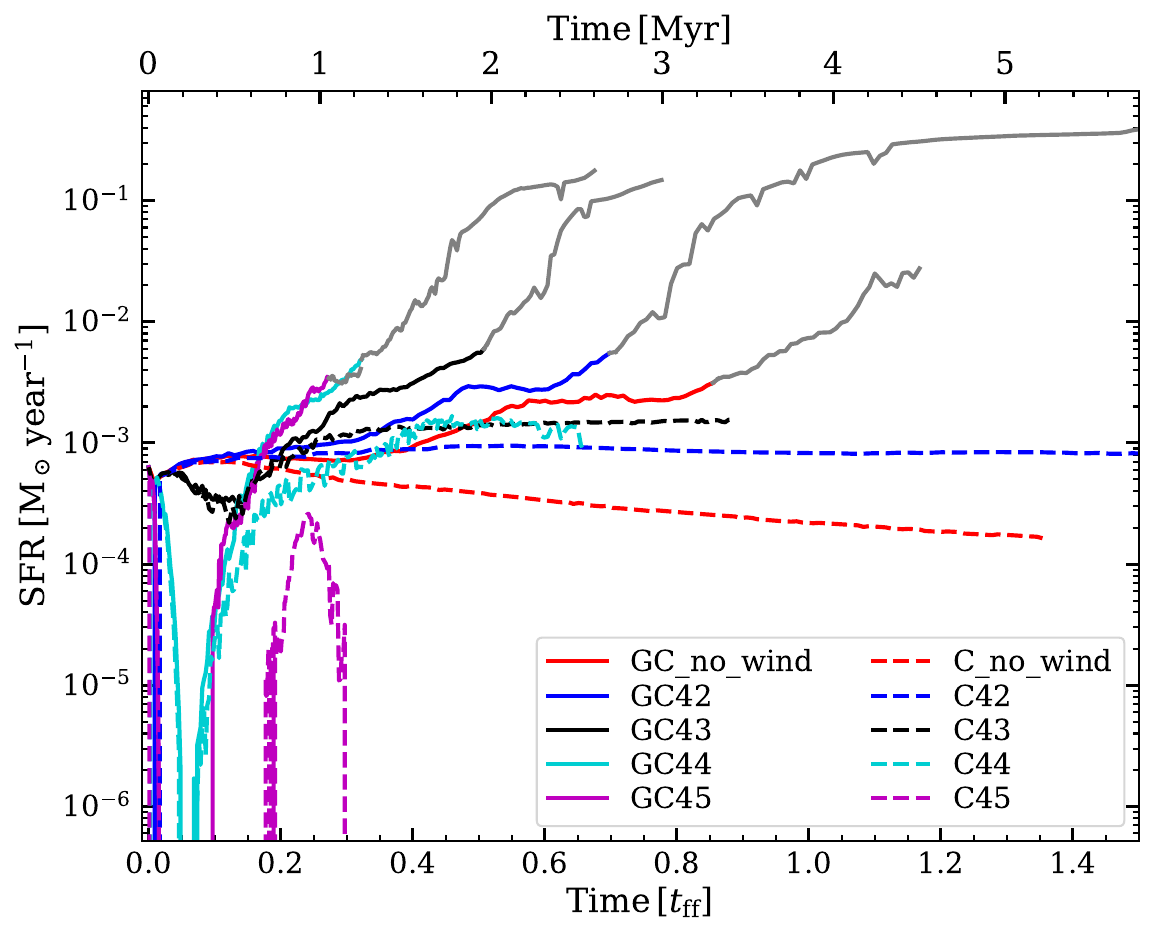}
\end{tabular}}
  \caption{Time evolution of the star-formation rate per freefall time (left) and the absolute star formation rate (right), for different simulations as indicated in the legend. For comparison, the results for simulations without a wind are also shown (red lines). The grey-out parts of the plots in self-gravitating simulations correspond to the regime where the Jeans length of the highest-density cell in the simulations is not resolved by at least 4~computational cells.}
  \label{fig:SFR}
\end{figure*}
%
\subsection{Star formation rate}
Outflows from AGNs are believed to be one of the major drivers behind the star formation activity in galaxies.
The local input of energy and momentum from these outflows into the star-forming regions can induce turbulence, which may have a dual role. 
On the one hand, the induced turbulence increases the stability of the clouds against gravitational forces preventing global collapse, which can reduce the star formation inside such clouds.
On the other hand, it can promote over-densities via shock compression \citep[e.g., see][]{MacLow_2004}, which may result in an enhanced star formation.
Therefore, the effect of these AGN-driven winds on the star formation activity is determined by a complex interplay between different physical processes acting on the cloud scale.

We estimate the star formation rate in the simulation using a semi-analytical model of turbulence-regulated star formation \citep[see][]{Krumholz_2005,Padoan_2011,Hennebelle_2011,Federrath_2012,Burkhart_2019}, which take into account the local physical quantities such as the velocity dispersion, virial parameter, local sound speed, turbulence driving mode, and magnetic field (if present) to estimate the star formation rate. 
In this framework, the star formation rate (SFR) is calculated by integrating the density PDF from a critical density. The SFR of a cloud of mass $M_{\rm c}$ is calculated as,
\begin{equation}\label{eq:SFR}
    {\rm SFR} = \SFRff\frac{M_{\rm c}}{\tff(\rho_0)},
\end{equation}
where $\tff(\rho_0)$ is the freefall time at the mean density ($\rho_0$) of the cloud. The quantity $\SFRff$, defined as the star formation rate per freefall time, is given by \citep{Federrath_2012},
\begin{equation}\label{eq:SFRff}
    \SFRff = \frac{\epsilon}{\phi_t}\int_{s_{\rm crit}}^{\infty}\frac{\tff(\rho_0)}{\tff(\rho)}\frac{\rho}{\rho_0}p(s)ds,
\end{equation}
where $s=\ln(\rho/\rho_0)$ is the logarithmic density contrast, and $p(s)$ is the density PDF expressed in terms of $s$. 
On the GMC scale, the parameter $\epsilon$ is interpreted as the fraction of the global mass of the whole cloud that eventually turns into stars, and is typically $1-2\%$ \citep[see Section 3.1.1 of][for a discussion]{Mandal_2021}. 
Therefore, we set $\epsilon=0.015$ in the calculation of $\SFRff$.
The $\phi_t$ parameter is a numerical factor of the order of unity to account for uncertainties in the integral and is set to $\phi_t = 2.04$ as calibrated in \citet{Federrath_2012}.

The critical logarithmic density ($s_{\rm crit}$) of collapse, used as the lower limit of the integral of Eq.~\eqref{eq:SFRff}, is estimated by comparing the Jeans length to the sonic length (where the turbulent velocity dispersion is of the order of the local sound speed) and is given by \citep{Krumholz_2005,Federrath_2012},
\begin{equation}\label{eq:s_crit}
    s_{\rm crit} = \ln\left[\left(\frac{\pi^2}{5}\right)\phi_x\alphavir\Mach^2\right],
\end{equation}
where $\phi_x$ is a numerical factor of the order of unity to account for slight differences in the exact equality between the Jeans length and the sonic scale \citep{Krumholz_2005,Federrath_2012} and set to $0.19$ in this study \citep[calibrated in][]{Federrath_2012}. 
For the detailed theoretical background of the turbulence-regulated star formation model, we refer interested readers to \citet{Krumholz_2005,Federrath_2012,Mandal_2021}, and references therein.

In order to estimate the SFR in our simulations, we use Eq.~\eqref{eq:sigma_v} and Eq.~\eqref{eq:alpha_vir} for calculating the Mach number ($\Mach$) and virial parameter ($\alphavir$), which gives us the value of $s_{\rm crit}$ via Eq.~(\ref{eq:s_crit}). 
We then construct the density PDF $p(s)$ from the simulation data, which is subsequently integrated according to Eq.~\eqref{eq:SFRff} to obtain $\SFRff$.
Finally, we calculate the SFR from Eq.~\eqref{eq:SFR}, where Eq.~\eqref{eq:cold_gas} is used to estimate the cloud mass.

\subsubsection{Dependence on wind power}
Fig.~\ref{fig:SFR} shows the time evolution of $\SFRff$ (left panel) and SFR (right panel) for different simulations, as denoted in the legend. 
During the initial stages, as the wind begins to interact with the cloud, it injects a substantial amount of kinetic and thermal energy into it, leading to an increase in the velocity dispersion and virial parameter, as depicted in Fig.~\ref{fig:sigma_alpha}. 
Thus, the turbulent motions of the gas impede the gravitational collapse of many regions, resulting in a decreased SFR compared to the simulation without a wind.
This behaviour is consistent among all the wind simulations; however, the degree of initial suppression in $\SFRff$ increases with rising wind power. 
In fact, for the most powerful wind (magenta), the star formation is completely quenched.
This is because higher wind power leads to a greater transfer of energy into the cloud, resulting in more pronounced turbulent motion within the gas.

Subsequent to this phase, as the wind fragments the cloud, it becomes easier for the wind to pass through the inter-cloudlet channels due to the smaller column depth. 
Hence, the strength of the interaction between the wind and cloud reduces.
Additionally, the dense cloudlets embedded within the hot wind are more resilient to propagation of the transmitted shocks into the cores, thus, weakening the effect of the wind within the dense cores.
Furthermore, the cloudlets that have been shock-compressed achieve significantly higher densities, resulting in a deeper gravitational potential. 
Therefore, the gravitational energy becomes dominant over the kinetic support, leading to an increased $\SFRff$ after the initial suppression phase.
For simulations without self-gravity, except for the C\_45 case, the rebound reaches a level comparable to that in simulations without the wind, as depicted by the dashed lines. For the C\_45 case, the cloud is destroyed and entrained by the wind before reaching that value of $\SFRff$.

However, in the presence of self-gravity, the shock compression causes the cloudlets to attain a much higher density. 
Additionally, the cloudlets merge due to the attractive gravitational force, forming a massive and highly dense central core, which undergoes runaway collapse.
As a consequence, the $\SFRff$ experiences a rapid increase following the initial decline (solid lines), to the extent that it surpasses the $\SFRff$ value in the simulation without wind (red solid line). 
It is important to note that the stronger the wind, the earlier and the more rapid this increase in $\SFRff$ occurs, as the higher-power winds quickly compress the cloud, effectively reducing the free-fall time.
Therefore, in all the simulations with self-gravity, the wind triggers the collapse of the cloud after the initial suppression.
While the morphological features in Fig.~\ref{fig:proj_diff_power} show that a high-power wind significantly disrupts the cloud, we note that a substantial portion of cloud material still undergoes compression and experiences rapid collapse.

It is crucial to emphasize that our simulations do not include self-consistent modelling of star formation using sink particles, which we aim to explore in a forthcoming study.
Therefore the reported SFR values in this context should be interpreted as a qualitative trend between various wind parameters.


\begin{figure*}
    \centering
    \includegraphics[width=\linewidth]{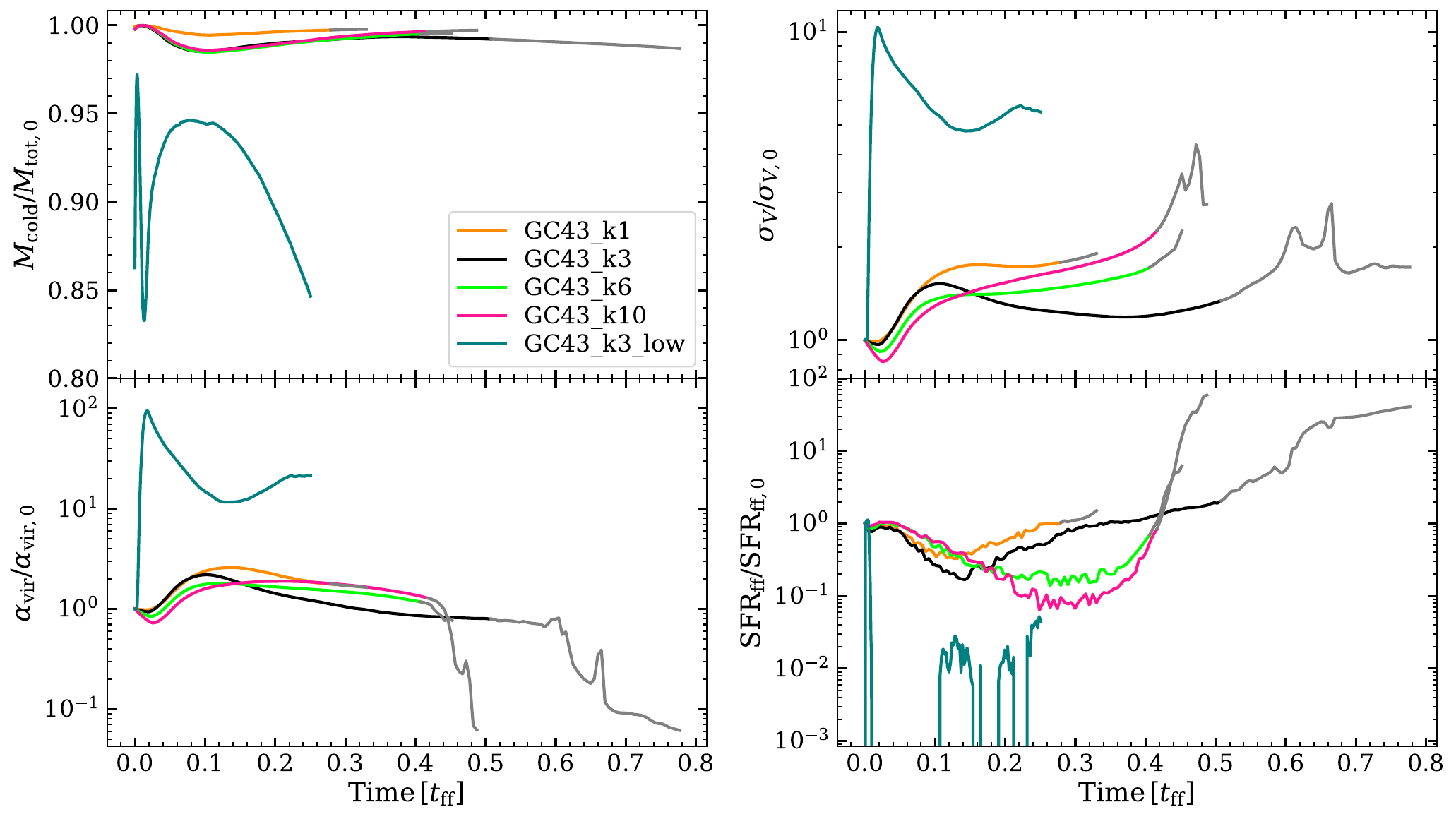}
    \caption{Time evolution of the cold gas fraction (top-left), velocity dispersion (top-right), virial parameter (bottom-left) of the cloud material and the star formation rate per freefall time (bottom-right) in the simulations with same wind power of $\pow{43}\,\ergs$ but different cloud wavenumber (GC43\_k\{1,3,6,10\}) and lower mean density (GC43\_k3\_low).}
    \label{fig:plots_diff_k}
\end{figure*}

\subsubsection{Dependence on cloud properties}
The results discussed in the earlier sections focused on the impact of winds with different powers on the same initial cloud structures and density. 
However, clouds with different densities and morphology can react differently to the wind. 
Therefore, to investigate this, in this section, we present the results for the wind-cloud interaction for a fixed value of wind power ($\pow{43}~\ergs$) but varying cloud properties, viz. different cloud wavenumber (GC43\_k\{1,3,6,10\}) and mean density (GC43\_k3\_low).

The top-left panel of Fig.~\ref{fig:plots_diff_k} shows the evolution of the cold gas fraction with time. 
Notably, the evolution appears similar for high-density clouds. 
Initially, there is a decrease in the cold gas fraction attributed to energy injection from the wind, followed by a subsequent increase due to the efficient cooling of the compressed gas. 
However, for the low-density cloud (GC43\_k3\_low), the initial cold gas fraction is lower compared to the other cases. 
This is because, initially, the clouds in our simulations are in pressure equilibrium with the ambient medium, as described in Sec.~\ref{sec:cloud_setup}. 
Due to the lower mean density, the equilibrium condition leads to a higher mean initial temperature ($\sim 2700~{\rm K}$) of the low-density cloud, resulting in a lower fraction of cold gas mass compared to the high-density clouds. 
However, as the evolution progresses, the gas rapidly cools, and the cold gas fraction increases. 
Nonetheless, as the wind impacts the cloud, the fraction decreases, and the subsequent evolution closely resembles that of the high-density cases. 
The major difference lies in the fact that the cold gas fraction remains low and decreases rapidly due to higher degrees of ablation and mixing. 
However, until the end of the simulation ($\sim 3~{\rm Myr}$), a significant amount of cold gas survives.

The top-right and bottom-left panels of Fig.~\ref{fig:plots_diff_k} illustrate the time evolution of velocity dispersion and virial parameter, respectively.
All the values are normalized to the initial values in order to examine the effect of the wind on the evolution of the parameters with respect to the initial values.
The qualitative evolution of both parameters is similar in all the simulations to what we observe in Fig.~\ref{fig:sigma_alpha}.
Interestingly, the evolution for $\kmin=1, 6$ and $10$ is quite similar apart from some initial deviation due to different initialization.
As discussed in Sec.~\ref{sec:diff_k}, while in $\kmin=1$ case the single large clump experiences rapid global collapse, for $\kmin=6,10$ cases, the narrow inter-cloudlets channels prevent the wind from penetrating and dispersing the cloud significantly, eventually leading to global collapse.
Despite different reasons, the ultimate fates of the clouds in these cases are similar.
In contrast, for the $\kmin=3$ case, the inter-cloudlets channels are wide enough for the wind to penetrate into the cloud, which directly compresses the cloudlets locally but prevents the global collapse for a longer period.

For the cloud with a lower mean density (GC43\_k3\_low), the velocity dispersion increases by more than an order of magnitude as the wind induces stronger turbulent motion due to a lower density.
The virial parameter also increases by two orders of magnitude in this case by the impact of wind. 
However, here we observe a second increase of the viral parameter at $\sim 0.25~\tff$.
This is primarily a result of the increase in turbulent velocity together with the decrease in mean density due to the disintegration of the cloud, which contributes to the decrease in gravitational binding energy.

The bottom-right panel of Fig.~\ref{fig:plots_diff_k} shows the evolution of the normalized star formation rate per freefall time ($\SFRff$) for all these simulations.
Similar to Fig.~\ref{fig:SFR}, the initial increase of the virial parameter reduces the $\SFRff$ for all simulations. 
However, the consequent rebound of $\SFRff$ depends on the cloud properties.
For the case with $\kmin=1$, one big clump (see top panel of Fig.~\ref{fig:slice_diff_k}) rapidly collapses via compression, leading to an early growth of $\SFRff$.
For $\kmin=6$ and $10$ cases, the initial small-scale velocity field of the cloud, characterized by the same values of $\kmin$, causes an enhanced amount of inter-cloud motion. 
This motion persists for a longer period before decaying, keeping the star formation rate relatively low during this time.
However, as the wind compresses the cloud globally, the majority of the cloudlets merge, effectively forming a large, compact clump, which undergoes rapid collapse globally. 
As a result, the star formation rate increases rapidly after the initial suppression.

In the case of the cloud with a lower mean density, we observe a complete cessation of star formation for approximately $0.1~\tff$ ($\sim 1.3~{\rm Myr}$) due to the initial rise of the virial parameter by two orders of magnitude.
However, as some of the dense clump gets compressed by the wind, we observe some intermittent star formation activity after the initial quenching.
Nonetheless, the SFR values during these intervals remain more than one order of magnitude lower than the initial value until the end of the simulation.

\section{Discussion}\label{sec:discussion}
\subsection{Comparison to previous studies}\label{sec:multiphase}
There have been a few studies concerning the effect of highly pressurized winds possibly originating from AGN on the interstellar clouds \citep{Mellema_2002,Zubovas_2014b,Dugan_2017a}.
The results obtained in the present work add to the previous studies of AGN-driven wind-cloud interaction by including additional physics such as self-gravity and more realistic cloud morphologies (fractal structures) and carrying out simulations over a wide range of parameters for both winds and clouds.
Some of the key findings from our study align with the results from previous works in the context of shock/wind-cloud interaction.
Our results reveal the emergence of dense filaments due to ram pressure stripping and KH instabilities acting upon the interface between the cloud and the wind, similar to the results from previous studies \citep{Mellema_2002,Orlando_2005,Cooper_2009,Pittard_2010,Dugan_2017a,Li_2020,Banda_2019,Banda_2020,Banda_2021}.

However, it is worth noting that in our simulations, these filaments display a more clumpy structure as opposed to the extended, continuous filaments observed in earlier studies. 
This difference arises from our incorporation of a fractal medium characterized by the presence of dense cores and a diffuse inter-core medium.
This configuration facilitates the infiltration of the wind into the cloud by clearing out low-density channels and enveloping the cores, effectively breaking the cloud into numerous cloudlets.
Moreover, as the KH instability grows faster at smaller wavelengths, the fragmentation of the cloud is much higher compared to an uniform or smooth cloud \citep{Cooper_2009}.
Therefore these fragmented cloudlets undergo acceleration to shape the clumpy filaments observed in our simulations.
In this way, the morphology of the filaments in our study resembles more the findings by \citet{Cooper_2009,Banda_2020,Banda_2021}, where a similar fractal cloud configuration is considered.

Another interesting phenomenon to be compared with previous studies is the lifetime of the dense clumps that are generated in the process of fragmentation.
\citet{Cooper_2009} concluded that radiative cooling plays a significant role in prolonging the survival of dense clumps entrained within a hot wind. 
This cooling process leads to the creation of a dense protective layer around the clump, enhancing the cloud's lifetime. 
In such scenarios, these clouds can survive for more than $1~{\rm Myr}$, which aligns with our results.
In all simulations performed in this study, the majority of the fragmented clumps survive until the end of the simulations ($1.2-5.5\,{\rm Myr}$, depending on the wind power).
Additionally, as demonstrated previously in Sec.~\ref{sec:diff_model} and \ref{sec:density_PDF}, self-gravity further increases the density of the shock-compressed cloudlets, and since the growth rate of instabilities is inversely proportional to the density contrast (see Eq.~\ref{eq:t_KH}), the growth of instabilities is diminished, therefore, further prolonging the lifetime of the cloudlets. Additionally, the increased density contrast due to self-gravity reduces the velocity of the transmitted shock ($v_{\rm ts}\sim v_{\rm w}/\chi^{1/2}$) into the cloudlets, consequently reducing the overall heating of the cores.

The effect of self-gravity has been considered in a few previous studies \citep{Zubovas_2014b,Dugan_2017a,Falle_2017,Li_2020,Girichidis_2021,Kupilas_2022} in the context of the cloud wind/shock interaction. 
All of these studies have demonstrated that when self-gravity is taken into account, the compression resulting from the wind contributes to an increased rate of fragmentation and eventual gravitational collapse \citep{Zubovas_2014b}, which agrees with our findings.
Furthermore, \citet{Dugan_2017a} has shown that under specific conditions, a sufficiently high ram pressure in the wind can lead to the complete destruction of the cloud. 
However, our study diverges from this result as we do not observe any instance of a cloud being entirely destroyed by the wind, even when subjected to the most powerful winds in our simulations.
Moreover, despite the substantial disruption of the cloud by the wind in the simulation with an initial lower mean density (GC43\_k3\_low), pockets of dense material persist, fostering intermittent star formation activity (see bottom panel of Fig.~\ref{fig:plots_diff_k}).  
This discrepancy can be attributed to two major factors. 
Firstly, in order to obtain that large value of ram pressure of the wind, \citet{Dugan_2017a} consider quite higher values of the wind density.
Therefore, the momentum imparted by the wind into the cloud is orders of magnitude higher than in our highest-wind-power case. As a result, the cloud in their study undergoes rapid acceleration, and the combination of Rayleigh-Taylor instabilities and ram pressure stripping leads to the cloud's disintegration.
Secondly, we consider a fractal cloud in our simulations, which has many dense gravitationally-bound cores at the initiation of the simulations.
These cores are further compressed by the wind, rendering them self-shielded and resistant to ablation, which contributes to the cloud's survival, despite the wind's powerful effects.

\subsection{Implication for AGN feedback}\label{sec:feedback}
The impact of AGN feedback on star formation activity remains a complex and elusive process in current astrophysical research. 
While numerous studies support the idea of negative feedback, attributed to the turbulence and thermal energy enhancement induced by AGN winds or the jet-inflated cocoon, there are also proponents of positive feedback. 
In this scenario, the over-pressurized wind associated with AGN activity has the effect of compressing and fragmenting star-forming clouds. 
This compression leads to an increased star formation rate (SFR) by triggering the collapse of gas clouds into new stars.

While there is little consensus on negative feedback from the theoretical point of view, a few studies have shown that AGN activity can indeed suppress the star formation rate globally inside the host galaxy by overall induction of turbulence and thermal energy \citep{Mandal_2021,Felize_2023}.
Furthermore, \citet{Wagner_2012} have argued that the gas within AGN-driven outflows can become unbound, effectively escaping the host galaxy's potential and thereby reducing the available star-forming fuel. 
This long-term process may contribute to a negative feedback mechanism.

Numerous theoretical studies have presented scenarios in which the triggering of star formation by AGN activity can be a viable mechanism \citep[e.g.][]{Wagner_2011,Nayakshin_2012,Gaibler_2012,Dugan_2017a,Dugan_2017b,Mukherjee_2018,Mandal_2021}. 
For instance, \citet{Gaibler_2012,Mukherjee_2018} have shown that the compression resulting from the high-pressure bubble inflated by an AGN jet can enhance the star formation rate (SFR) in a disc galaxy by a factor of $2-3$.
Similarly, \citet{Zubovas_2014b} arrived at a similar conclusion, demonstrating that external pressurization within the ISM can confine and compress star-forming regions and accelerate the onset of gravitational collapse and subsequent star formation.
Moreover, the study by \citet{Dugan_2017a} revealed that the star formation efficiency depends on the ram pressure of the wind. 
They identified a critical threshold value of the ram pressure, below which the clouds experience rapid collapse, leading to an enhancement in SFR. 
However, above this threshold, ram pressure becomes strong enough that the clouds are ablated before gravitational forces can significantly influence their evolution.

The result we present in this study also leans toward a positive feedback scenario.
While there exists a short period of suppressed/quenched star formation during the initial interaction between the wind and the cloud, the compression due to the over-pressurized wind dominates in the long term. 
The radiative shocks compress many massive cloudlets to significantly higher densities, and the presence of self-gravity intensifies this process.
As a result, the shock propagating into these cores decelerates swiftly as the gas densities progressively increase, and the cloudlets effectively become self-shielded from the wind, such that the cloudlets can survive for a long time in the hot wind. 
Without the support from internal turbulence inside these cores, which is crucial for their stability, they eventually collapse to form stars, which will stop once all the gas in the core is consumed.

Therefore, from our present results, positive feedback by the AGN is inevitable even for high-power winds. 
Indeed, some observational studies support this scenario. 
For instance, \citet{Maiolino_2017,Gallaghe_2019} detected a substantial amount of star formation ($\sim 15\,{\rm \Msun\, yr^{-1}}$) inside a galactic outflow, with a significant young ($\sim 10\,{\rm Myr}$) stellar population and higher stellar velocities ($\lesssim 100\,\kms$). 
This indicates that the stars have been formed inside the outflows, which have been triggered by the compression induced by the out-flowing gas. 
A recent investigation by \citet{Duggal_2023} identified a young ($1-10\,{\rm Myr}$) stellar population inside young compact steep spectrum (CSS) radio galaxies. 
Notably, the dynamical age of the radio sources in these galaxies corresponds well to the stellar ages, suggesting that star formation was triggered by jet activity.
There are various other observational studies where enhanced star formation by AGN activity (jets or winds) has been reported \citep[e.g.,][]{Bicknell_2000,Zinn_2013,Salome_2015,Salome_2017,Bernhard_2016}, in agreement with our findings.

Moreover, as outlined in Sec.\ref{sec:wind_parameters}, the wind velocity and pressure experienced by the cloud at a distance of $1~\kpc$ for different powers can be mapped to various distances from the AGN for a specific wind power.
For example, the wind parameters corresponding to powers of $\pow{42}, \pow{43}, \pow{44}$ and $\pow{45}~\ergs$ at $1\kpc$ (refer to Tab.\ref{tab:sim_list}) can be replicated by considering a wind with a power of $\pow{43}\ergs$ at distances of $3.2, 1, 0.3, 0.1~\kpc$ from the AGN. 
Therefore, our findings hold relevance for comprehending the impact of AGN-driven winds on a galactic-scale environment. 
Importantly, even though a lower power wind (e.g., $\pow{42}, \pow{43}~\ergs$) may not appear to significantly affect the cloud at a distance of $1\kpc$ according to our study, the closer environment of the AGN on a scale of hundreds of parsecs will likely be strongly influenced by the wind, akin to the outcomes observed in higher power simulations (e.g., $\pow{44}, \pow{45}~\ergs$).
Similarly, for a higher-power wind, the outskirts of the galaxy will be mildly affected, mirroring the outcomes of simulations with lower-power winds presented in this study.
In any case, the presence of self-gravity is likely to trigger star formation inside the clouds on a wide length scale while impacted by AGN-driven winds, albeit at different timescales, as observed in simulations with different powers (Fig.~\ref{fig:SFR}).

The reported suppression of star formation by AGN activity, as observed in several studies \citep[e.g.][]{Ogle_2007,Ogle_2010,Nesvadba_2010,Alatalo_2014,Alatalo_2015,Lanz_2016}, continues to await theoretical confirmation.
Our study, which incorporates comprehensive modelling of the interaction between a star-forming cloud and an AGN-driven wind, does not reveal any significant long-term suppression of star formation activity over the duration of an AGN lifetime (typically a few Myr). 
Based on our results, it appears that the only plausible means of reducing the star formation would be to employ mechanisms capable of disrupting the dense cores from the inside.
Stellar feedback in the form of winds, jets, radiation pressure, and photo-ionization appears to be a viable mechanism \citep{Menon_2023}, at least on the scale of a few parsecs, which corresponds to the typical size of the dense cloudlets we have identified in our study.
Indeed, previous studies on star-cluster formation in clouds of size $\sim 1-10\,\pc$ have demonstrated that stellar feedback can slow down the star formation rate \citep{Federrath_2015,Grudic_2018}.
Therefore, the underlying concept is that following the initial burst of star formation induced by wind-driven compression, the feedback from young and massive stars acts to disrupt the cloud, which in turn makes wind material entrainment more favourable.
Therefore, in tandem with stellar feedback, the AGN wind could potentially contribute to the destruction of dense cores, breaking them into even smaller structures that are no longer prone to gravitational collapse due to their reduced size.
This presents an intriguing scenario that we plan to investigate in future studies.

Additionally, the wind parameters examined in this study represent fast and hot AGN-driven outflows. However, the impact of slow and relatively cold winds, which are typically found at greater distances from the AGN, is not considered here and should be investigated in future.

\section{Conclusion}\label{sec:conclusion}
In this study, we have performed a series of three-dimensional hydrodynamical simulations of the interaction between AGN-driven winds and star-forming interstellar clouds, including radiative cooling and realistic cloud morphology such as fractal geometry.
We have conducted two sets of simulations with and without self-gravity to examine the effect of self-gravity on the evolution of the clouds.
We consider a large range of parameter space for investigating various aspects of the evolution process, including the power of the wind, the mean density of the cloud, the fractal density distribution of the cloud, and whether the wind is dominated by kinetic or thermal energy.
In the following, we summarize the main results of this study:
\begin{enumerate}
    \item \textbf{\textit{Interaction of the wind with the fractal cloud}:}
    When the wind interacts with the fractal cloud consisting of dense cores separated by low-density channels, it rapidly erases the low-density areas, resulting in the formation of numerous dense cores. Subsequently, these dense cores undergo compression due to radiative shocks.
    
    \item \textbf{\textit{Effect of self-gravity}:}
    While the cloudlets get compressed by the wind irrespective of the presence of self-gravity with the same wind power, the cloudlets formed in the self-gravitating simulations attain much higher densities and become gravitationally bound, compared to the cloudlets in the simulations without self-gravity. In the absence of self-gravity, after attaining the maximum possible compression, the clouds start to disintegrate and expand as a result of the momentum transfer from the wind to the cloud material.

    \item \textbf{\textit{Dependence on wind power}:}
    The amount of cloud material that is retained and accreted by the gravitationally bound cloudlets depends on the strength of the wind. For lower-power winds, the momentum transfer is reduced, and as a result, a significant portion of the cloud material does not gain enough acceleration to overcome the gravitational potential of the cloud. Consequently, this material falls back into the potential well, forming a central, massive clump that undergoes rapid gravitational collapse. Conversely, for the higher-power winds, although some initially high-density clumps collapse under self-gravity, a significant number of relatively low-density clumps are accelerated and dispersed by the wind prior to self-gravity becoming a significant factor.

    \item \textbf{\textit{Effect of cloud morphology}:}
    The size of the cloudlets (determined by the minimum wave number $\kmin$ used to generate the fractal density distribution) inside the cloud significantly affects the evolution when impacted by a wind with the same power. For the $\kmin=1$ case, the whole cloud roughly contains a single large dense clump and is therefore large enough to prevent the shock from penetrating into its cores before it undergoes global gravitational collapse due to external pressurization. In contrast, for $\kmin=6$ and $\kmin=10$, the cloud is characterized by numerous dense cores, separated by narrow channels. However, when the initial shock sweeps across the cloud, it effectively blocks these channels with swept-up material, preventing the wind material from infiltrating the cloud. As a result, in the absence of significant dispersal of the cloudlets, the cloudlets accumulate near the centre by the gravitational force, ultimately experiencing global collapse. Thus, there exists a narrow range of $\kmin$, for which the inter-clumps channels are wide enough for the wind to penetrate into the cloud and provide stability against global collapse by transferring energy and momentum, which is the case for $\kmin=3$.

    \item \textbf{\textit{Relative effect of thermal vs.~ kinetic wind}:} 
    The total energy partition of the wind in the thermal and kinetic components has a major effect on the evolution of the cloud. A thermal-energy-dominated wind primarily affects the cloud material by increasing its internal energy, which in turn is converted into kinetic energy of the gas. However, in the presence of strong radiative cooling, the internal energy of the gas quickly dissipates, therefore the effective momentum transfer from the wind to the cloud is reduced for a thermal wind. A kinetic-energy-dominated wind directly transfers momentum to the cloud material, resulting in a higher level of expansion and acceleration.

    \item \textbf{\textit{Evolution of the velocity dispersion}:}
    The velocity dispersion inside the cloud in all the wind simulations increases due to energy transfer from the wind to the cloud material. This effect is more pronounced with stronger winds. In cases without self-gravity, the velocity dispersion stabilizes at a roughly constant value as the cloud disperses. With self-gravity, the velocity dispersion starts to increase again when gravity becomes dominant as it pulls fragmented cloudlets towards the core, inducing additional motions between the cloudlets.

    \item \textbf{\textit{Effect on the virial parameter}:}
    In all the simulations, an initial increase in the virial parameter occurs as the wind imparts kinetic energy, initially surpassing the gravitational energy. As the cloud compresses and the gas density rises, the gravitational potential deepens, reducing $\alphavir$. Simulations without self-gravity reach a steady state with balanced kinetic and gravitational energy. On the other hand, with self-gravity, the cloud becomes gravitationally bound and undergoes collapse, and $\alphavir$ drops significantly, especially in cases with low-power winds, due to the dominant gravitational binding energy.

    \item \textbf{\textit{Generation of multiphase outflow}:}
    The ablation of the cloud material by the wind can give rise to multi-phase outflows with velocities from a few $100\,\kms$ to several $1000\,\kms$ over a huge range of temperatures ($10^2-10^7\,{\rm K}$), consisting of cold, warm and hot gas. The calculated mass-outflow rates correlate tightly with the wind power ($\Mdot\propto P^\kappa$). We find a power-law exponent of $\kappa\approx 0.52$.

    \item \textbf{\textit{Impact on the star formation rate}:}
    In the presence of self-gravity, which is very important within the environment we are interested in, our results favour a positive feedback scenario triggered by the AGN-driven winds, at least within the parameter space we consider in this study. Even though the wind can suppress or quench the star formation for about $1~{\rm Myr}$ during the initial interaction, a substantial number of shock-compressed, dense cloudlets manage to shield themselves from the wind's influence and subsequently undergo rapid gravitational collapse. This process ultimately leads to an increased star formation rate.
\end{enumerate}
%
\section*{Acknowledgements}
We thank the referee, Dr.~Tiago Costa, for his thorough and insightful comments, which have improved the quality of the paper significantly. A.~Mandal would like to thank Moun Meenakshi and Kavita Kumari for useful discussions about the effect of the AGN radiation field on interstellar gas. A.~Mandal further thanks Moun Meenakshi for the help with the radiative transfer calculation using \textsc{cloudy}.
C.~F.~acknowledges funding provided by the Australian Research Council (Discovery Project DP230102280), and the Australia-Germany Joint Research Cooperation Scheme (UA-DAAD). We gratefully acknowledge high-performance computing resources provided by the Australian National Computational Infrastructure (grant~n72 and ek9) and the Pawsey Supercomputing Centre (project~pawsey0810) in the framework of the National Computational Merit Allocation Scheme and the ANU Merit Allocation Scheme, the Pegasus\footnote{\url{http://hpc.iucaa.in}} high-performance computing facilities of IUCAA, and the Leibniz Rechenzentrum and the Gauss Centre for Supercomputing (grants~pr32lo, pr48pi and GCS Large-scale project~10391). D.~Mukherjee and N.~Nesvadba acknowledge support from the IFCPAR/CEFIPRA (project no. 6504-2) for collaborative research. This work has received funding from the European High Performance Computing Joint Undertaking (JU) and Belgium, Czech Republic, France, Germany, Greece, Italy, Norway, and Spain under grant agreement No 101093441.
This work is supported by "Italian Research Center on High Performance Computing Big Data and Quantum Computing (ICSC)", project funded by European Union - NextGenerationEU - and National Recovery and Resilience Plan (NRRP) - Mission 4 Component 2.  Spoke 3, Astrophysics and Cosmos Observations.

\section*{Data Availability}
Data related to this work will be shared upon reasonable request to the corresponding author.



\bibliographystyle{mnras}
\bibliography{manuscript} 



\appendix
\section{Non-reflecting boundary condition for subsonic inflow}\label{sec:NRBC}
In this section, we present a simplified approximation of the non-reflecting boundary conditions for subsonic inflow for one-dimensional flow.

\subsection{Characteristic form of the conservation equations}
The one-dimensional (1D) conservation equations in the primitive variable are given as
\begin{gather}
  \del{\rho}{t} + u\del{\rho}{x} + \rho\del{u}{x} = 0, \label{eq:density}\\
  \del{u}{t} + u\del{u}{x} + \frac{1}{\rho}\del{p}{x} = 0, \label{eq:velocity}\\
  \del{p}{t} + \rho a^2\del{u}{x} + u\del{p}{x} = 0, \label{eq:pressure}
\end{gather}
where $a$ is the sound speed and defined as $a = \sqrt{\gamma p/\rho}$. Eqs.~\eqref{eq:density}-\eqref{eq:pressure} can be written in terms of the primitive variable vector $\mathbf{Q}$ as
\begin{gather}
    \del{\mathbf{Q}}{t} + \mathbf{A(Q)}\del{\mathbf{Q}}{x} = 0,\label{eq:privimitive_eq}
\end{gather}
where $\mathbf{Q}$ and $\mathbf{A(Q)}$ are defined as
\begin{gather}\label{eq:A}
    \renewcommand\arraystretch{1.5}
    \mathbf{Q} = 
    \begin{bmatrix}
        \rho \\
        u \\
        p
    \end{bmatrix},
    \quad
    \mathbf{A(Q)} = 
    \begin{bmatrix}
        u & \rho & 0\\
        0 & u & \dfrac{1}{\rho} \\
        0 & \rho a^2 & u
    \end{bmatrix}.
\end{gather}

Applying a similarity transformation to $\mathbf{A}$, one obtains
\begin{gather}\label{eq:similarility_tras}
    \mathbf{A} = \mathbf{S}\mathbf{\Lambda}\mathbf{S}^{-1},
\end{gather}
where $\mathbf{S} = [\mathbf{K}^1,\mathbf{K}^2,\mathbf{K}^3]$ is the matrix consisting of the right eigenvectors of $\mathbf{A}$ such that $\mathbf{AK}^{(i)} = \lambda_{i}\mathbf{K}^{(i)}$, where $\lambda_i$ are the eigenvalues corresponding to $\mathbf{K}^{(i)}$. $\mathbf{\Lambda}$ is a diagonal matrix with its diagonal elements being $\lambda_{i}$.
Replacing Eq.~\eqref{eq:similarility_tras} in Eq.~\eqref{eq:privimitive_eq}, we obtain
\begin{equation}\label{eq:euler_charac1}
    \del{\mathbf{Q}}{t}  + \mathbf{S}\Lambda\mathbf{S}^{-1}\del{\mathbf{Q}}{x} = 0.
\end{equation}
Multiplying Eq.~\eqref{eq:euler_charac1} with $\mathbf{S}^{-1}$ from the left, we obtain
\begin{equation}\label{eq:inverse_equation}
    \mathbf{S}^{-1}\del{\mathbf{Q}}{t} + \mathbf{\Lambda S}^{-1}\del{\mathbf{Q}}{x} = 0.
\end{equation}
If we approximate $\mathbf{A}(\mathbf{Q})$ to be locally constant, so is $\mathbf{S}^{-1}$. Therefore, defining
\begin{equation}\label{eq:char_variable}
    \mathbf{W} = \mathbf{S}^{-1}\mathbf{Q},    
\end{equation}
Eq.~\eqref{eq:inverse_equation} reduces to
\begin{equation}\label{eq:char_variable_vec}
    \del{\mathbf{W}}{t} + \mathbf{\Lambda}\del{\mathbf{W}}{x} = 0, \\
\end{equation}
which is the characteristic form of the conservation equations, and $\mathbf{W} = [w^1, w^2, w^3]^T$ is the characteristic variable vector.
In component form, Eq.~\eqref{eq:char_variable_vec} can be written as
\begin{equation}
    \del{w^i}{t} +\lambda_i\del{w^i}{x} = 0,
\end{equation}
which is a set of wave equations with characteristic velocities $\lambda_i$.

In order to compute the characteristic variables, first we find the eigenvalues ($\lambda_i$) and right eigenvectors ($\mathbf{K}^i$) of $\mathbf{A}$, which are given by
\begin{gather}
    \lambda_1 = u-a,\\
    \lambda_2 = u, \\
    \lambda_3 = u+a,
\end{gather}
and the corresponding eigenvectors are given as
\begin{equation}
    \renewcommand\arraystretch{1.5}
    \mathbf{K}^1 = \alpha_1
    \begin{bmatrix}
        1 \\
        -\dfrac{a}{\rho} \\
        a^2
    \end{bmatrix},
    \qquad
    \mathbf{K}^2 = \alpha_2
    \begin{bmatrix}
        1\\
        0\\
        0
    \end{bmatrix},
    \qquad
    \mathbf{K}^3 = \alpha_3
    \begin{bmatrix}
        1\\
        \dfrac{a}{\rho}\\
        a^2
    \end{bmatrix},
\end{equation}
where $\alpha_i$ are the scale factors. Assuming $\alpha_1 = \rho/2a$, $\alpha_2=1$, and $\alpha_3 = \rho/2a$, the $\mathbf{S}$ matrix and its inverse is calculated as
\begin{equation}\label{eq:S}
    \renewcommand\arraystretch{2}
    \mathbf{S} = 
    \begin{bmatrix}
        \dfrac{\rho}{2a} & 1 & \dfrac{\rho}{2a}\\
        -\dfrac{1}{2} & 0 & \dfrac{1}{2} \\
        \dfrac{\rho a}{2} & 0  &\dfrac{\rho a}{2}
    \end{bmatrix},
    \qquad
    \mathbf{S}^{-1} =
    \begin{bmatrix}
        0 & -1 & \dfrac{1}{\rho a}\\
        1 & 0 & -\dfrac{1}{a^2} \\
        0 & 1 & \dfrac{1}{\rho a}
    \end{bmatrix}.
\end{equation}
Therefore, from Eqs.~\eqref{eq:A}, \eqref{eq:char_variable} and \eqref{eq:S}, the form of characteristic variables ($\mathbf{W}=[w^1,w^2,w^3]^T$) for the 1D Euler equations are given as
\begin{align}
    w^1 &= \frac{p}{\rho a} - u, \quad \text{corresponding to $\lambda_1 = u-c$}\label{eq:w1}\\
    w^2 &= \rho - \frac{p}{a^2}, \quad \text{corresponding to $\lambda_2 = u$}\label{eq:w2}\\
    w^3 &= \frac{p}{\rho a} + u, \quad \text{corresponding to $\lambda_3 = u+c$}.\label{eq:w3}
\end{align}

\subsection{Boundary conditions}
The primary idea behind the non-reflecting boundary condition is that any outgoing wave ($w^{\rm out}$) that is leaving the computational domain through the boundary should leave the domain without getting reflected back into the interior region, which means the wave amplitude associated with the outgoing wave is constant at the boundary, i.e.,
\begin{equation}\label{eq:nr_bc}
    w^{\rm out} = \mathrm{const}, \qquad {\rm or} \qquad \Delta w^{\rm out} = 0,\,\, \text{at the boundary},
\end{equation}
and the boundary condition associated with the outgoing characteristic variables should be extrapolated from the interior solution. The other waves that are incoming into the computational domain require physical boundary conditions.

\begin{figure*}
    \centerline{
    \def\arraystretch{1.0}
    \setlength{\tabcolsep}{0.0pt}
        \begin{tabular}{lcr}
            \includegraphics[width=0.5\linewidth]{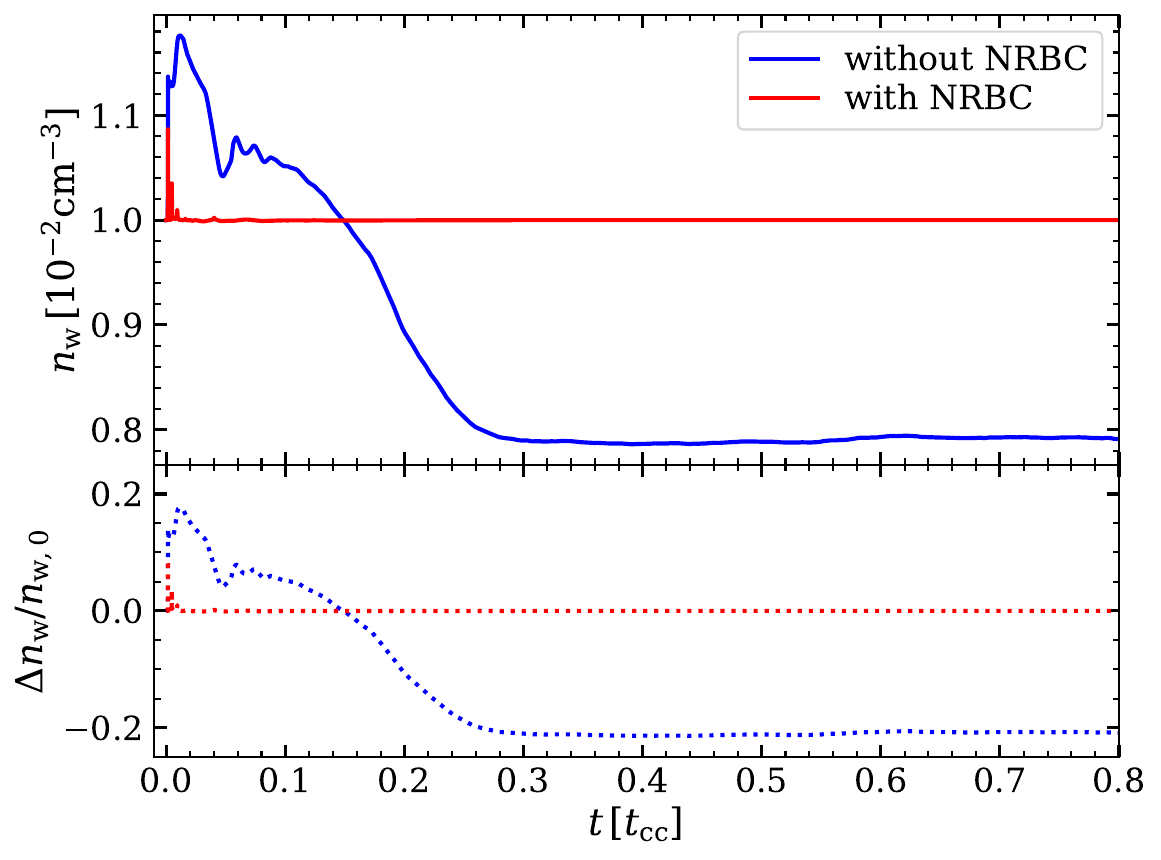} &
            \includegraphics[width=0.5\linewidth]{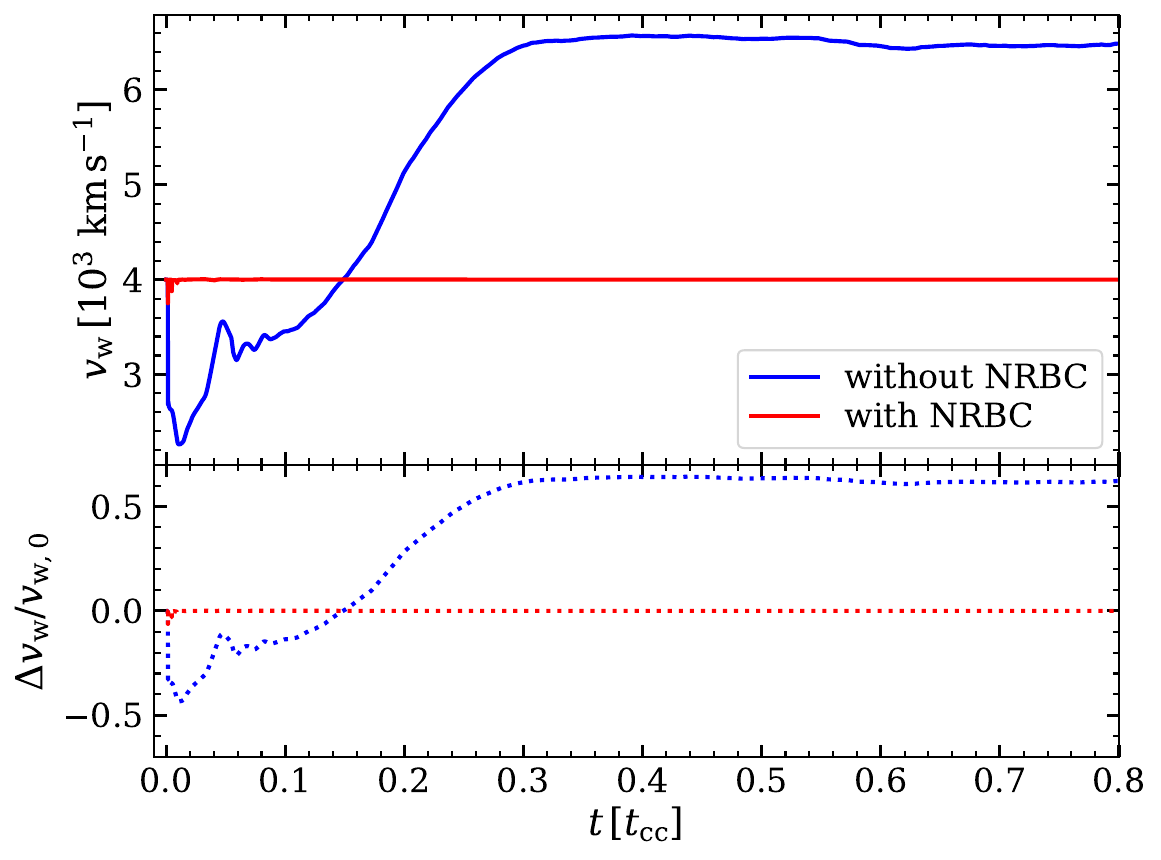} \\
            \includegraphics[width=0.5\linewidth]{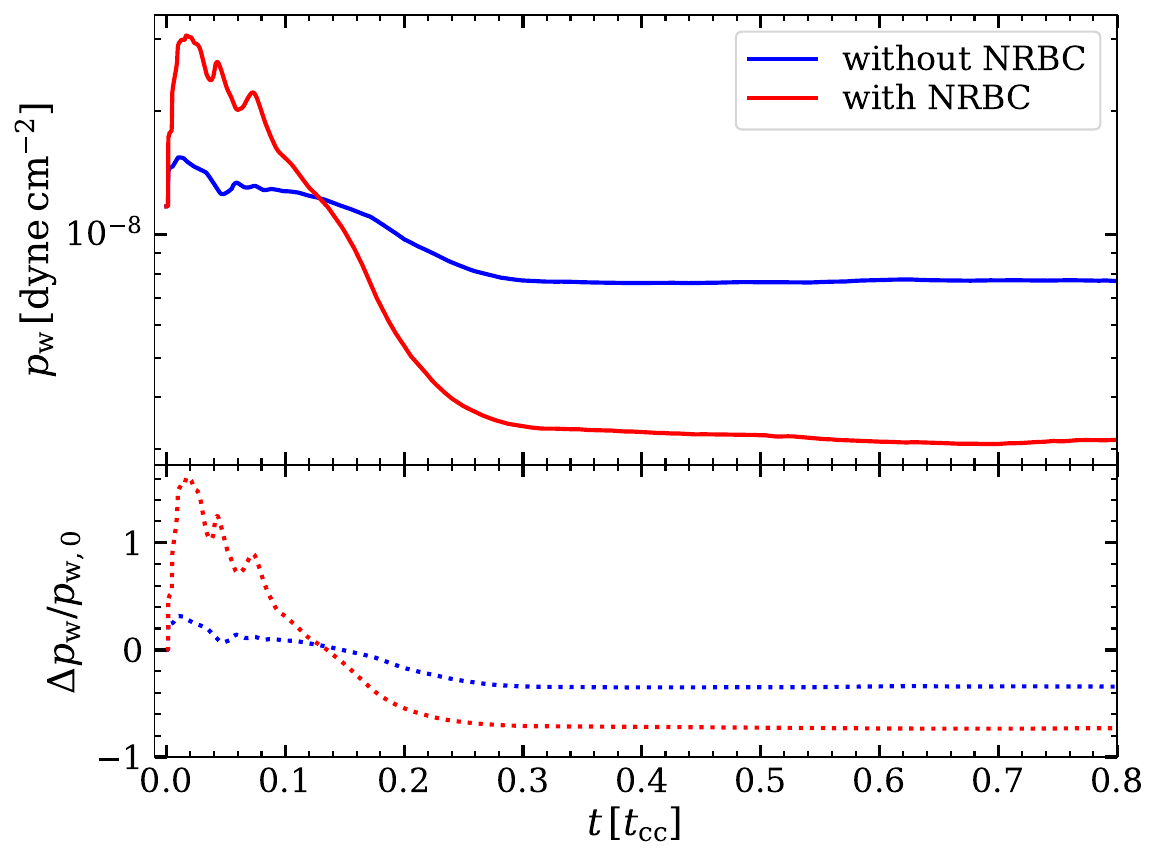} &
        \includegraphics[width=0.5\linewidth]{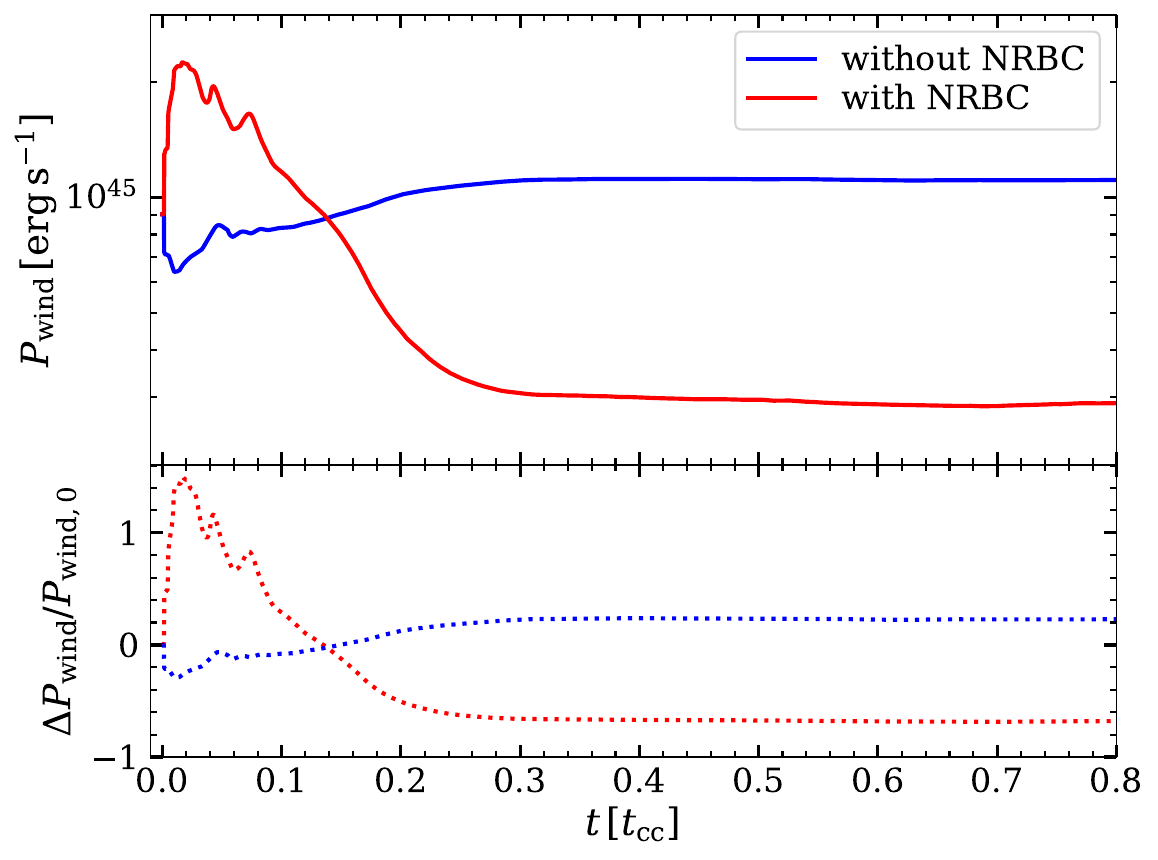}
        \end{tabular}}
    \caption{Time evolution of the wind parameters for a 3D model problem of wind-cloud interaction without (blue) and with (red) a non-reflecting boundary condition (NRBC) at the inflow boundary. The bottom section of each panel shows the fractional deviation from the injected values at the ghost zone.}
    \label{fig:input_parameter_comparison}
\end{figure*}

For a subsonic inflow boundary (let us assume $x$-beg), one wave ($w^1$) is leaving the domain ($\lambda_1 = u - a < 0$), and the other two ($w^2, w^3$) are incoming, which means only two out of three primitive variables (ideally ($\rho, u$) or ($\rho, p$) for well-poised conditions, e.g., see  Sec.~19.3 of \citealp{Laney_1998}) can be specified physically at the boundary. The remaining variable has to be set numerically from the interior solution in such a way that $\Delta w^1 = 0$ at the boundary. Let's say, we want to specify the value of $\rho_{\rm B}$ and $u_{\rm B}$ physically, then the pressure value ($p_{\rm B}$) at the boundary is calculated by setting
\begin{gather}
    \Delta w^1 = 0,\\
    w^1_{\rm B} -w^1_{\rm I} = 0, \\
    p_{\rm B} = \gamma \rho_{\rm B}(w^1_{\rm I} + u_{\rm B})^2,
\end{gather}
where $w^1_{\rm I}$ is the value of $w^1$ at the first active zone ($i=1$) of the computational domain, which is to be calculated from the interior solution. Therefore, for target density and velocity values of $\rho_{\rm t}$ and $u_{\rm t}$, the ghost cell values of the primitive variables are given by
\begin{align}
    \rho_{\rm B} &= \rho_{\rm t}, \label{eq:rho_BC}\\
    u_{\rm B} &= u_{\rm t}, \label{eq:u_BC}\\
    p_{\rm B} &= \gamma \rho_{\rm t}(w^1_{\rm I} + u_{\rm t})^2. \label{eq:p_BC}
\end{align}
Eqs.~\eqref{eq:rho_BC}-\eqref{eq:p_BC} ensure that $\Delta w^1= 0$ at the boundary, and the wave-reflection is minimised.

\subsection{Test and comparison}
In order to demonstrate how the wind emerges in the interior solution with and without NRBC, we perform one simulation with NRBC implemented at the subsonic inflow boundary with the same setup as described in Sec.~\ref{sec:method}. The wind and cloud initialisations are the same as in GC45\_k3 in Table~\ref{tab:sim_list}, with half the resolution of the fiducial runs. We compare the evolution of the wind parameters with a simulation that uses the same initial condition and resolution, but using the method of wind injection adopted in this paper (Sec.~\ref{sec:BC}). In the simulation with NRBC, we specify the density and velocity values at the ghost zone, while the pressure of the wind is calculated using Eq.~\eqref{eq:p_BC}. 
    
In Fig.~\ref{fig:input_parameter_comparison}, we present the time evolution of the wind density (top-left), velocity (top-right), pressure (bottom-left), and power (bottom-right), at the first active cell in simulations with (red lines) and without (blue lines) NRBC. The lower section of each panel illustrates the fractional deviation of each parameter from its intended value.
As observed, the evolution of the physically specified wind parameters---namely density and velocity---remains consistent throughout the simulations with NRBC. However, the pressure of the injected wind shows a significant deviation compared to the simulation without NRBC. Consequently, the deviation of wind power from its intended value is substantial ($\sim 90\%$).
On the other hand, although all wind parameters in the simulation without NRBC deviate from their intended values (blue lines), the cumulative effect results in more stable wind power (with a variation of $\sim 30\%$) compared to the scenario with NRBC.
Moreover, we have performed additional simulations using the saturated values of the density and velocity for the GC45 case (see Tab.~\ref{tab:saturated_values}) with the NRBC. We confirmed that, although the quantitative results differ from the result presented in this study during the transient phase, they closely resemble the results (e.g. the diagnostics presented in Fig.~\ref{fig:mass_evolve}, \ref{fig:sigma_alpha}, \ref{fig:SFR}) at the saturation phase ($t\gtrsim 0.4~\tcc$). Therefore, the qualitative results presented in this study are robust.

\section{Density PDF of the cloud material at the same cloud-crushing time for different power}

\begin{figure}
    \centering
    \includegraphics[width=\linewidth]{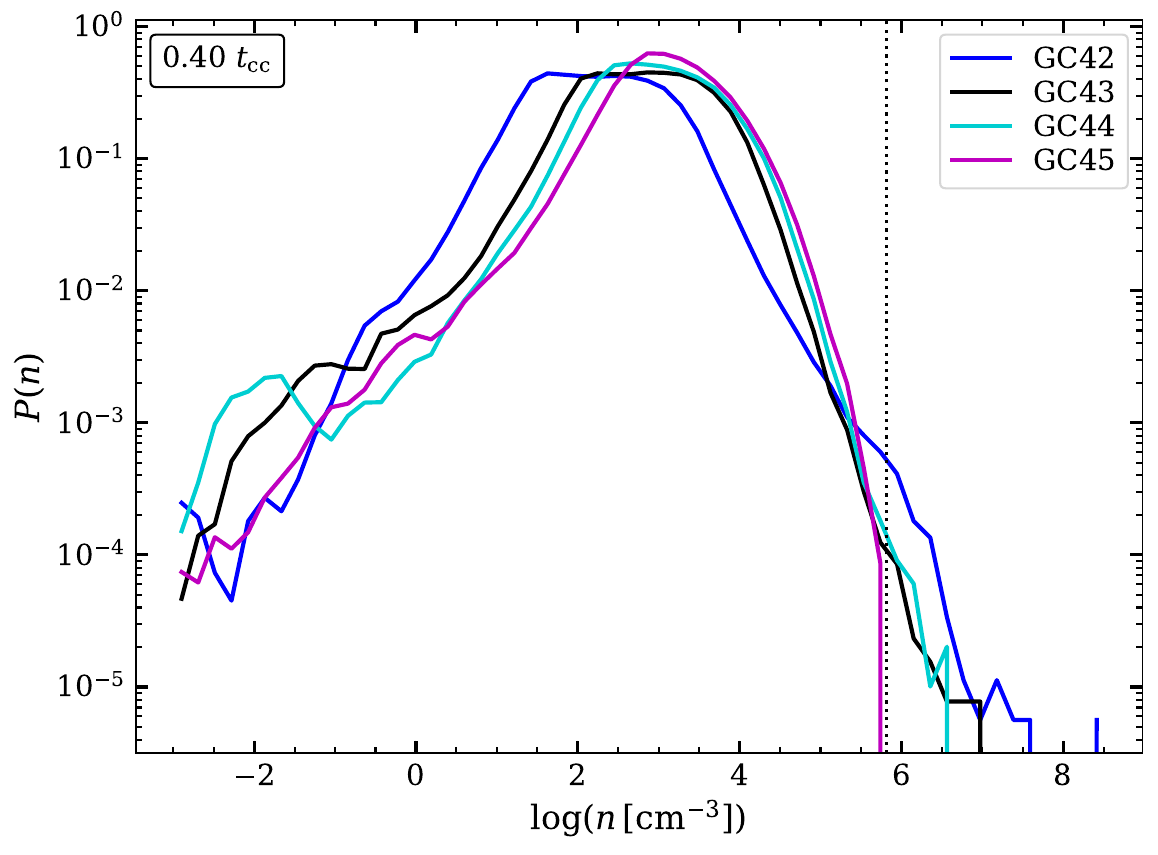}
    \caption{The PDFs of the cloud material in self-gravitating simulations with different wind power as indicated in the legend at $t=0.4\tcc$. The vertical black dotted lines in both panels mark the density value beyond which Jeans length cannot be resolved by at least four grid cells with the current computational setup.}
    \label{fig:density_PDF_tcc}
\end{figure}

Fig.~\ref{fig:density_PDF_tcc} depicts the density PDF of the self-gravitating cloud at $t=0.4\tcc$ for simulations with different wind power. This time corresponds to an absolute time of 3.46, 1.38, 0.78 and 0.34~Myr for GC42, GC43, GC44 and GC45, respectively.

We notice a distinctly contrasting trend in the high-density tail of the PDF compared to what is depicted in Fig.~\ref{fig:density_PDF}.
In this case, the cloud in the lowest power simulation (GC42, blue) has the highest fraction of cloud material in the power-law tail. As wind power increases, this fraction as well as the maximum cloud density decreases due to an increasingly smaller absolute time for self-gravity to influence the evolution. Notably, in the highest-power simulation (GC45), there is no evident signature of the power-law tail at $0.4\tcc$. Corresponding to an approximate time of $0.4\tcc = 0.34{\rm Myr} \approx 0.1~\tff$ in this instance, the evolution is at a very early stage in terms of the impact of self-gravity.

\section{Turbulent property evolution in terms of cloud-crushing time}\label{sec:plots_tcc} 
\begin{figure}
\centerline{
\def\arraystretch{1.0}
\setlength{\tabcolsep}{0.0pt}
\begin{tabular}{lcr}
  \includegraphics[width=\linewidth]{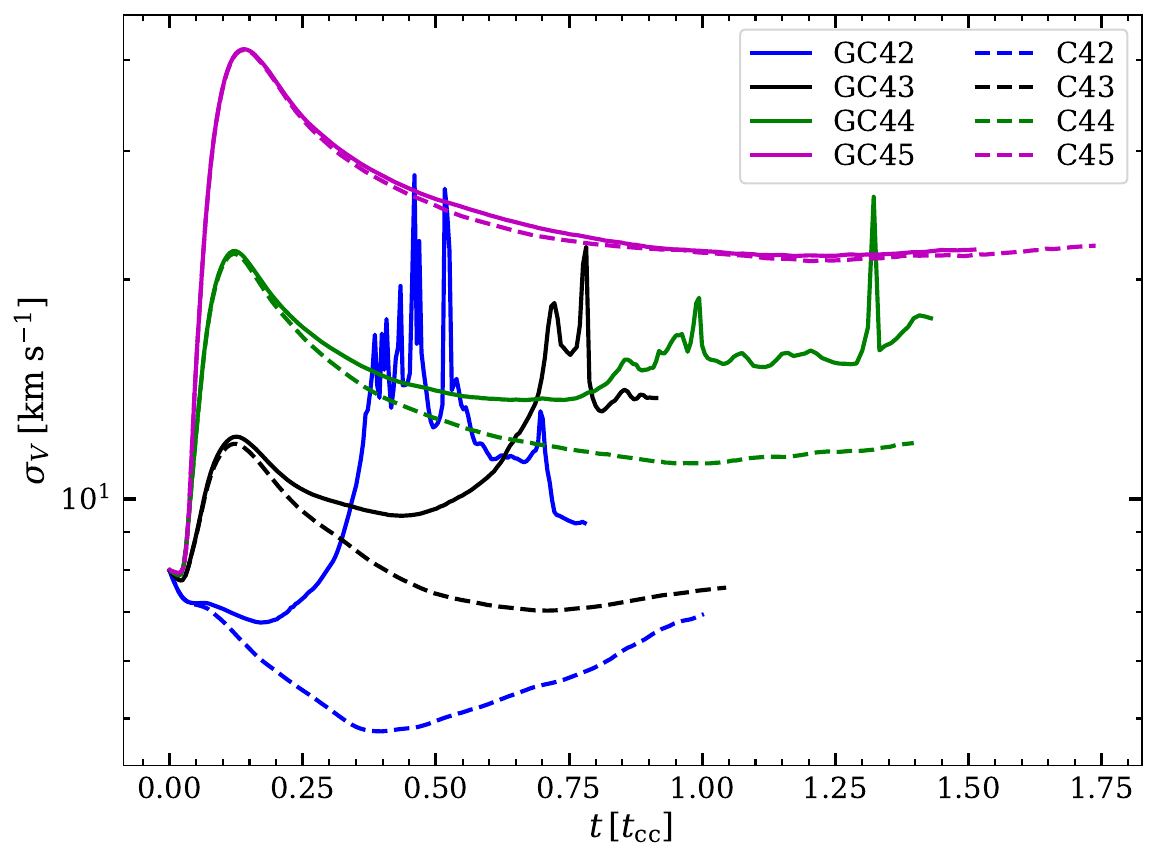} \\
  \includegraphics[width=\linewidth]{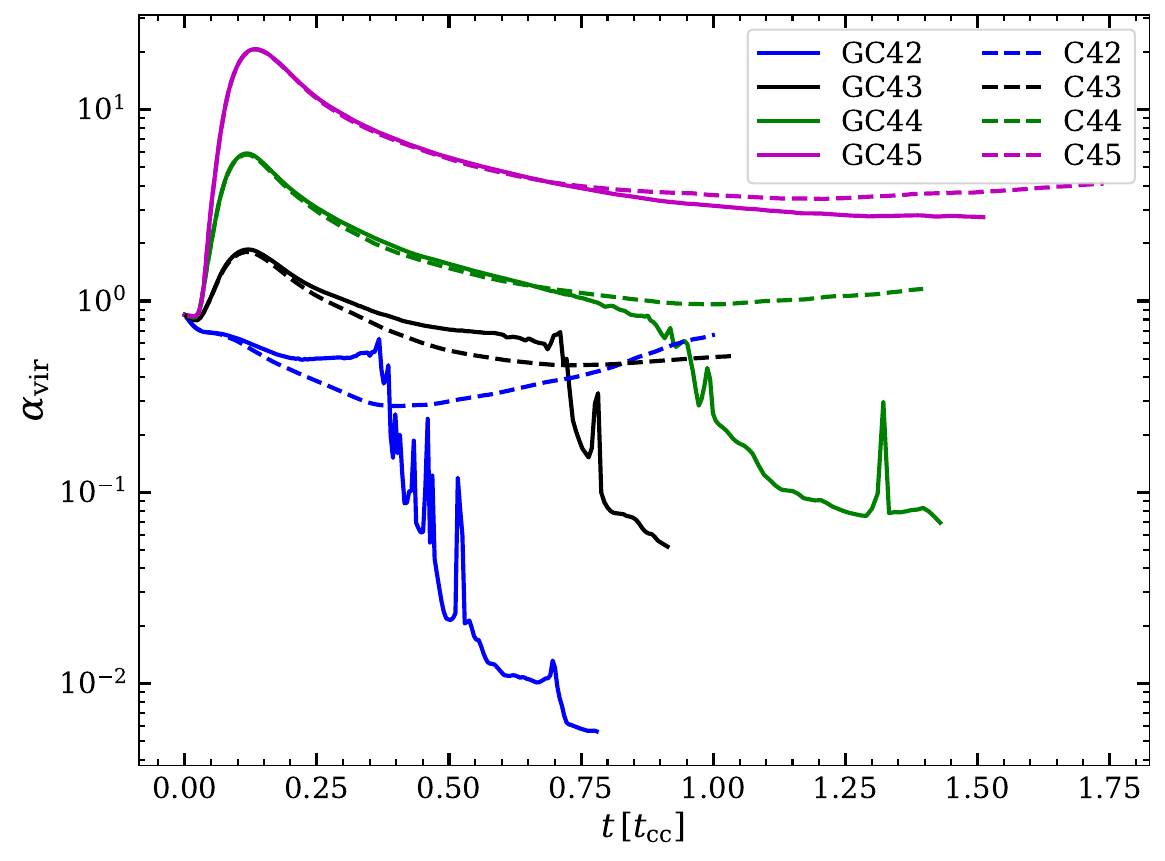}
\end{tabular}}
  \caption{Same as Fig~.\ref{fig:sigma_alpha} but expressed in terms of $\tcc$.}
  \label{fig:turbulence_tcc}
\end{figure}

Here we examine how the turbulent properties of the clouds in the simulations with different wind power evolve with time in unit of the cloud-crushing timescale.

Fig.~\ref{fig:turbulence_tcc}, shows the evolution of velocity dispersion (left) and the virial parameter (right) of the cloud material as a function of cloud-crushing time for the fiducial cloud setup with different wind power.
It is evident that in the absence of self-gravity (dashed line), the evolutionary trends are similar for all the powers. The velocity dispersion as well as the virial parameter peaks at $\sim 0.2~\tcc$, although the magnitudes are different.
However, in the self-gravity runs, the freefall time ($\tff$) is a more relevant timescale to compare to. As the freefall time of the cloud in the lower-power case (GC\_42) is shorter than $\tcc$, the cloud collapses early at around $\sim 0.5~\tcc$. With increasing power, $\tcc$ of the cloud becomes comparable or shorter than $\tff$.
Therefore, the collapse of the clouds gets delayed in terms of $\tcc$ with increasing wind power.

\section{Effect of numerical resolution}\label{sec:resolution_study}
\begin{figure*}
    \centerline{
    \def\arraystretch{1.0}
    \setlength{\tabcolsep}{0.0pt}
        \begin{tabular}{lcr}
            \includegraphics[width=0.5\linewidth]{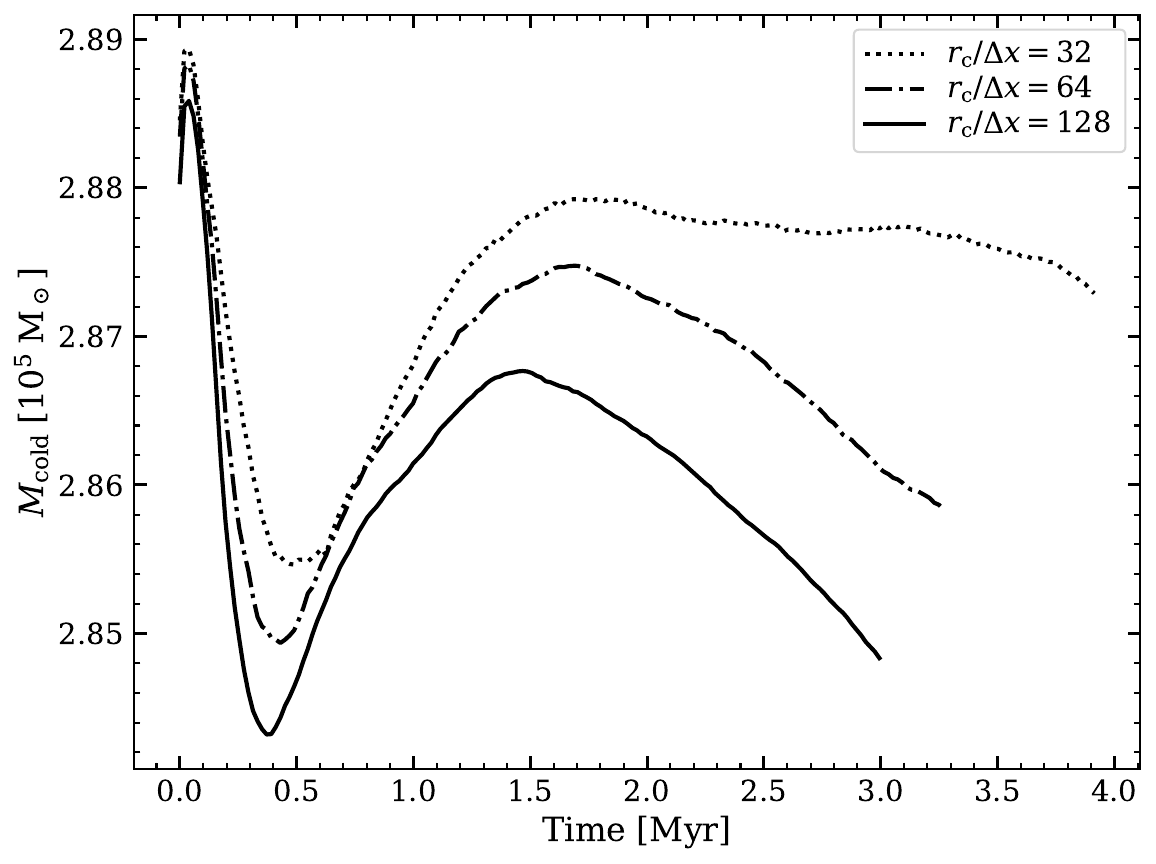} &
            \includegraphics[width=0.5\linewidth]{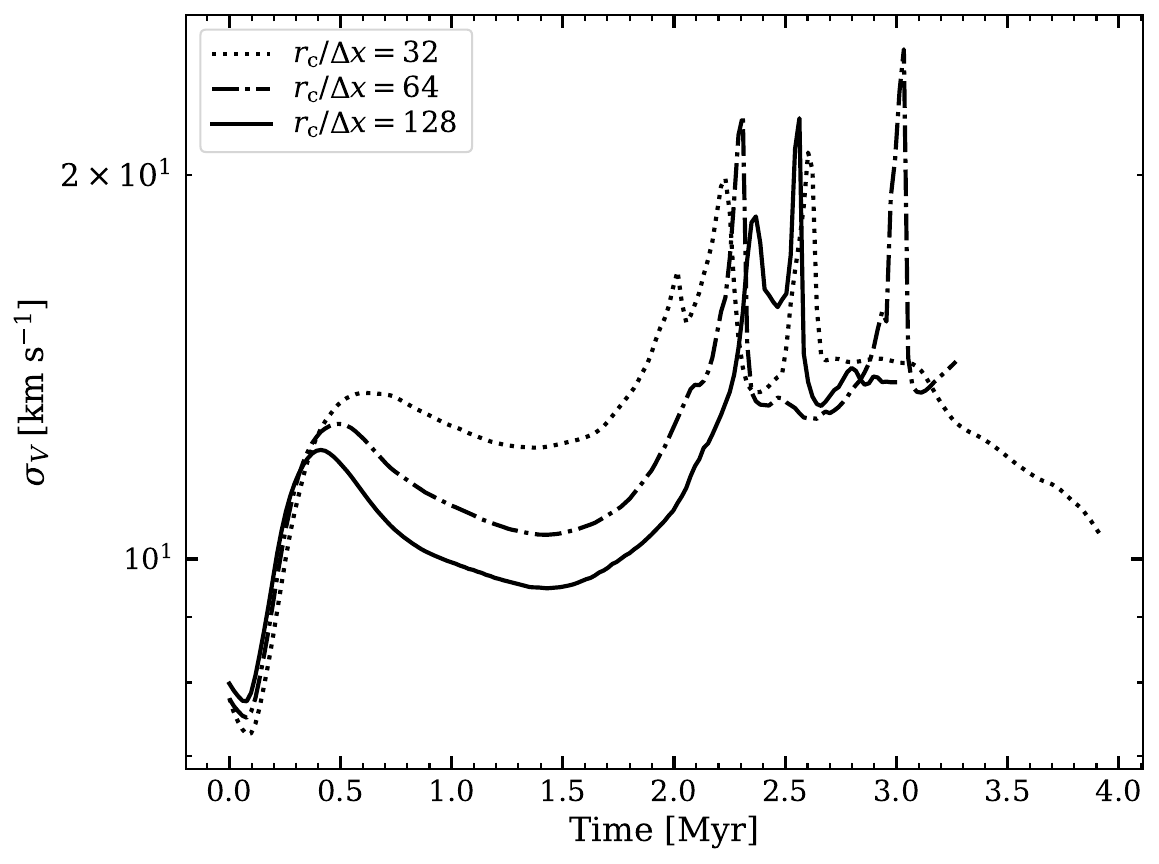} \\
            \includegraphics[width=0.5\linewidth]{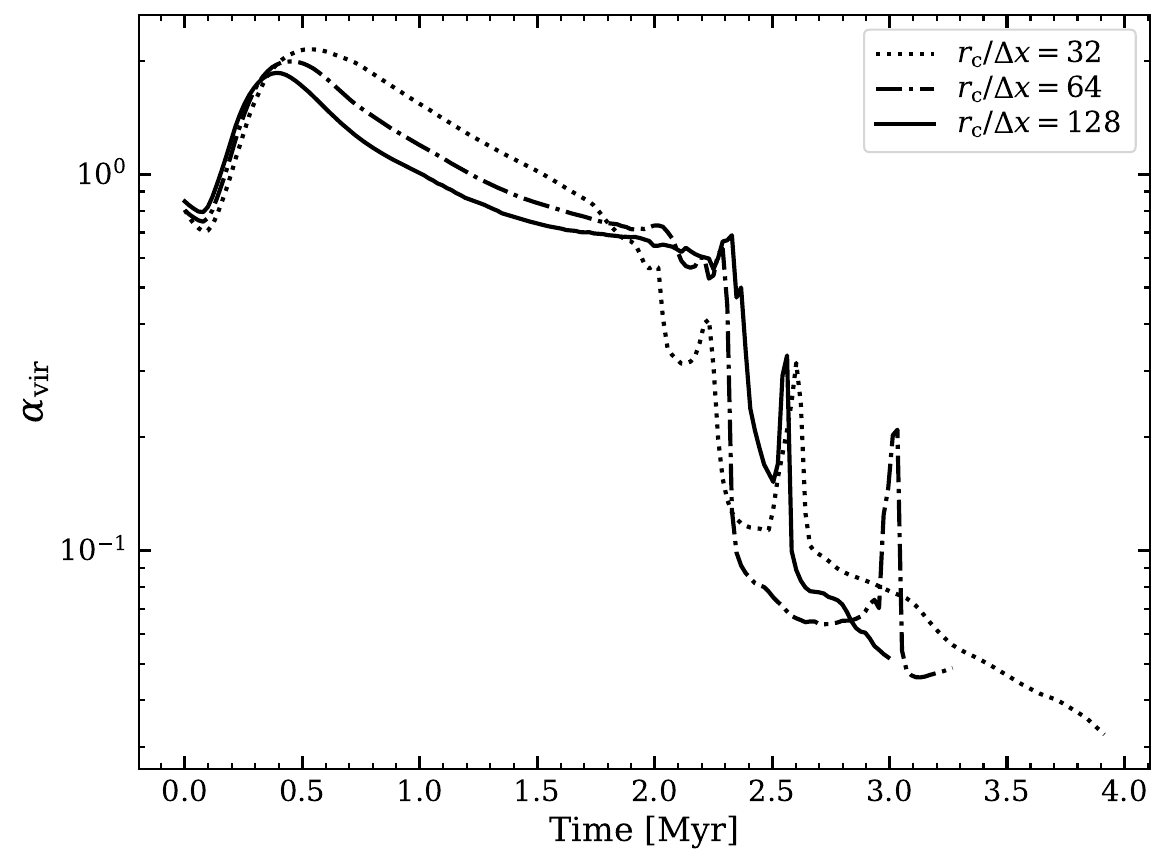} &
        \includegraphics[width=0.5\linewidth]{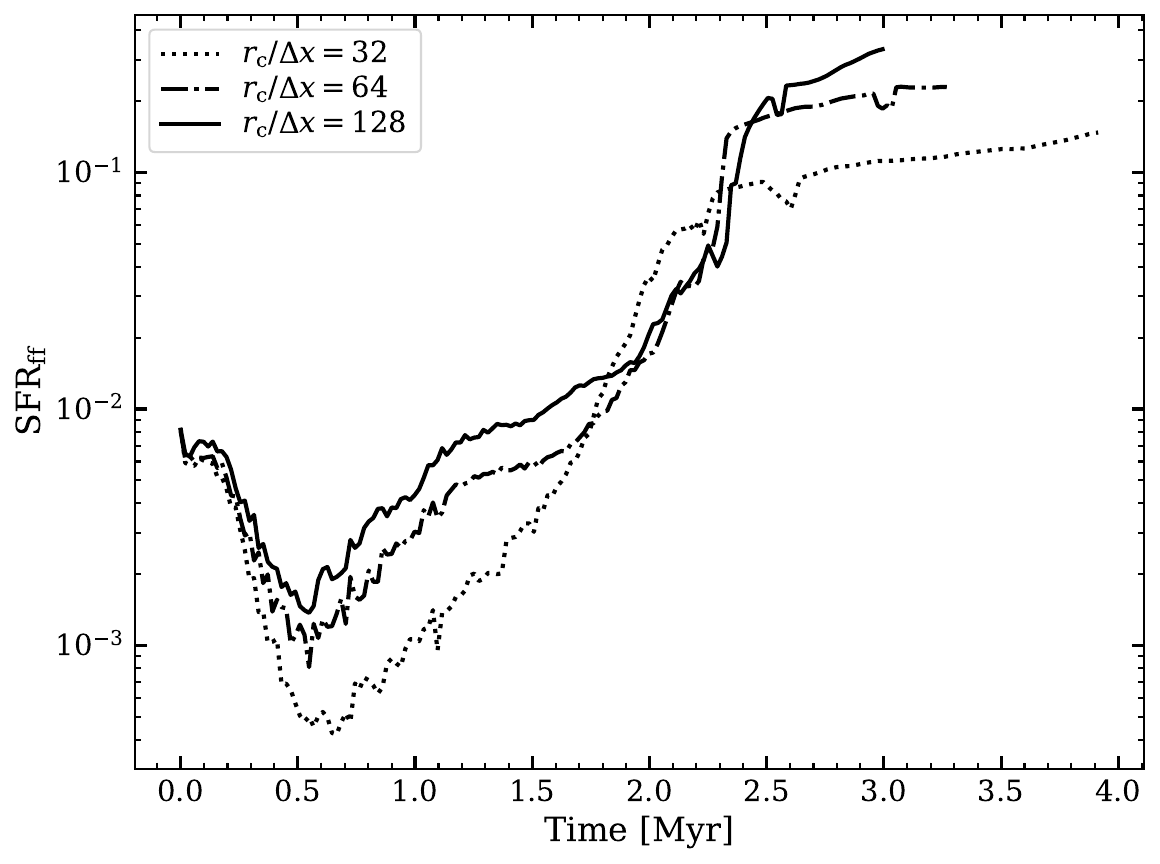}
        \end{tabular}}
    \caption{Time evolution of the cold gas mass (top-left), velocity dispersion (top-right), virial parameter (bottom-left), and star formation rate per freefall time (bottom-right), for simulations with wind power of $\pow{43}~\ergs$, but with different numerical grid resolutions. The widths of the cells in these three simulations are $\Delta x = $ 0.19~pc~(solid), 0.39~pc (dashed-dottet), and 0.78~pc (dotted), and the initial cloud radius ($r_{\rm c}$) is resolved by 128, 64 and 32 cells, respectively.}
    \label{fig:resolution_study}
\end{figure*}

In order to quantify the numerical resolution dependence of the results presented in this study, we perform two simulations with the same wind and cloud initialisation as in GC43\_k3, but with decreasing resolution. The widths of a cell in the considered simulations are $\Delta x = 0.19, 0.39$ and $0.78~\pc$ and the initial cloud radius ($r_{\rm c}$) is resolved by 128~(high), 64~(medium) and 32~(low)~cells, respectively. 

Fig.~\ref{fig:resolution_study} illustrates the time evolution of the cold gas mass (top-left), velocity dispersion (top-right), virial parameter (bottom-left), and star formation rate per freefall time (bottom-right), for simulations with $r_{\rm c}/\Delta x = $~128~(solid), 64~(dashed-dotted), and 32~(dotted). 
Across all panels, there is a clear dependence on resolution regarding the evolution of these quantities. 
While the evolution in the high (solid) and medium (dashed-dotted) resolution simulations closely align, the low (dotted) resolution one exhibits more significant deviations from the medium resolution case. 
This systematic dependence on resolution in the scenario of shock-cloud interaction directly stems from how well the instabilities acting on the cloud surface are resolved \citep{Klein_1994,Cooper_2009,Yirak_2010,Banda_2016,Banda_2018,Banda_2020}. 
As resolution increases, we can resolve perturbations with smaller wavelengths ($\lambda$), i.e., higher wavenumber ($k$). 
Since the growth rates for the Kelvin-Helmholtz and Rayleigh-Taylor instabilities are directly proportional to the wavenumber of the perturbation, the clouds in higher-resolution simulations are more susceptible to instabilities and get ablated faster and mixed with the wind. 
This systematic variation affects all derived quantities.
For instance, the amount of cold gas in the top-left panel of Fig.~\ref{fig:resolution_study} in the higher-resolution simulation is consistently lower compared to the lower-resolution case, due to the increased amount of ablation and mixing. This systematic variation is reflected in all the other panels as well.
However, we find that the differences of the results between $r_{\rm c}/\Delta x$ = 128 and~64 are much less compared to the $r_{\rm c}/\Delta x = 64$ and~32 pair, implying that the resolution of the fiducial simulations ($r_{\rm c}/\Delta x=128$) presented in this paper is adequate, as found by various previous studies \citep[e.g.,][]{Klein_1994,Fujita_2009,Banda_2018}.


\bsp	
\label{lastpage}
\end{document}